\documentclass[reprint,
 amsmath,amssymb,
 aps,
 longbibliography,
 pre,
floatfix,
]{revtex4-1}

\usepackage[dvipsnames]{xcolor}
\usepackage{xcolor}
\usepackage{graphicx}
\usepackage{dcolumn}
\usepackage{bm}
\usepackage{hyperref}
\usepackage[normalem]{ulem} 
\usepackage{cancel}

\begin{document}

\preprint{}

\title{Synchronization and exceptional points in nonreciprocal active polar mixtures}

\author{Kim L. Kreienkamp}
\email{k.kreienkamp@tu-berlin.de}
\author{Sabine H. L. Klapp}
\email{sabine.klapp@tu-berlin.de}
\affiliation{%
 Institute for Physics and Astronomy, Technische Universit\"at Berlin, Berlin, Germany
}%


\begin{abstract}
Many active matter systems consist of different particle types that interact via nonreciprocal couplings. Such nonreciprocal couplings can lead to the spontaneous emergence of time-dependent states that break parity-time symmetry. On the field-theoretical level, the transition to these states is marked by so-called exceptional points. However, their precise impact on observable particle dynamics remains poorly understood. In this study, we address this gap by providing a scale-bridging view of a minimal active mixture with nonreciprocal polar interactions. We find that nonreciprocity induces chiral motion on the particle level, yet no full, homogeneous synchronization. Instead, we observe various behaviors, ranging from fully synchronized clusters to chimera-like states. The nonreciprocity-induced spontaneous chirality increases with the degree of nonreciprocity and peaks at coupling strengths associated with exceptional points.
\end{abstract}

\maketitle


\section*{\label{sec:introduction}Introduction}
Nonreciprocity is ubiquitous in heterogeneous nonequilibrium systems and significantly impacts their dynamics \cite{Ivlev_2015_statistical_mechanics_where_newtons_third_law_is_broken,meredith_zarzar_2020_predator-prey_droplets,saha_scalar_active_mixtures_2020,frohoff_thiele_2023_nonreciprocal_cahn-hilliard,fruchart_vitelli_2023_odd_viscosity_and_elasticity,mandal_sollich_2022_robustness_travelling_states_generic_non-reciprocal_mixture}. Notable examples include chase-and-run behavior in bacterial predator-prey systems \cite{xiong_tsimring_2020_flower-like_patterns_multi-species_bacteria,theveneau_mayor_2013_chase-and-run_cells} and catalytic colloids \cite{saha_golestanian_2019_pairing_waltzing_scattering_chemotactic_active_colloids}, robots following chiral trajectories due to opposing alignment goals  \cite{chen_zhang_2024_emergent_chirality_hyperuniformity_active_mixture_NR}, and demixing driven by nonreciprocal torques in mixtures of differently sized Quincke rollers \cite{maity_morin_2023_spontaneous_demixing_binary_colloidal_flocks}. More generally, nonreciprocity has shown to be crucial in neural \cite{sompolinsky_kanter_1986_temporal_association_asymmetric_neural_networks,rieger_zittartz_1989_glauber_dynamics_asymmetric_SK_model} and social \cite{helbing_molnar_1995_social_force_model_pedestrian_dynamics,puy_romanczuk_2024_selective_social_interactions_schooling_fish} networks, systems with vision cones \cite{barberis_peruani_2016_minimal_cognitive_flocking_model,lavergne_bechinger_2019_group_formation_visual_perception-dependent_motility,loos_martynec_2023_long-range_order_non-reciprocal_XY_model}, and quantum optics \cite{metelmann_2015_nonreciprocal_photon_transmission,zhang_2018_non-reciprocal_quantum_optical_system,mcdonald_2022_nonequilibrium_quantum_non-Hermitian_lattice,reisenbauer_delic_2024_non-hermitian_dynamics_optically_coupled_particles,livska_brzobohat_2024_pt-like_phase_transition_optomechanical_oscillators}.

Among the range of phenomena induced by nonreciprocity, one particularly striking effect is the emergence of time-dependent states under certain conditions \cite{You_Baskaran_Marchetti_2020_pnas,fruchart_2021_non-reciprocal_phase_transitions,suchanek_loos_2023_irreversible_fluctuations_emergence_dynamical_phases,suchanek_loos_2023_time-reversal_PT_symmetry_breaking_non-Hermitian_field_theories,suchanek_loos_2023_entropy_production_nonreciprocal_Cahn-Hilliard,avni_vitelli_2023_non-reciprocal_Ising_model}. Recently, the transition to these time-dependent states accompanied by the occurrence of exceptional points (EPs) has gained much interest \cite{suchanek_loos_2023_time-reversal_PT_symmetry_breaking_non-Hermitian_field_theories,suchanek_loos_2023_irreversible_fluctuations_emergence_dynamical_phases}. EPs are often discussed in the context of non-Hermitian quantum mechanics \cite{fruchart_2021_non-reciprocal_phase_transitions,miri_alu_2019_exceptional_points_optics_photonics}.
For nonreciprocal systems and non-Hermitian field theories of classical systems, this framework is equally relevant. For example, in scalar nonreciprocal Cahn-Hillard models, EPs mark the transition from a static demixed state to a traveling demixed state with a phase shift, which breaks parity-time (PT) symmetry \cite{You_Baskaran_Marchetti_2020_pnas,suchanek_loos_2023_irreversible_fluctuations_emergence_dynamical_phases}.

Our focus is on nonreciprocal \textit{polar} active fluids, composed of motile particles of different species with competing goals regarding their mutual orientation. As shown in continuum-theoretical studies \cite{fruchart_2021_non-reciprocal_phase_transitions,kreienkamp_klapp_2022_clustering_flocking_chiral_active_particles_non-reciprocal_couplings}, such systems can feature not only \mbox{(anti)}flocking, i.e., coherent motion in constant direction, but also exceptional transitions towards chiral states, where the polarization direction rotates over time without intrinsic torques.

In contrast to these continuum approaches, which often base on mean-field-like assumptions, the microscopic, i.e., particle, dynamics is only partly understood.
Several important questions remain: How are nonreciprocity-induced chiral states characterized on the particle level? And how does the concept of EPs, widely discussed in field theory, relate to observable particle dynamics?

To address these issues, we perform a combined continuum and particle-level analysis of a minimal model for a binary mixture of active Brownian particles \cite{romanczuk_schimansky-geier_2012_active_brownian_particles} with additional nonreciprocal torques between particles of different species plus mutual repulsion. The system exhibits two qualitatively distinct regimes of nonreciprocity-induced dynamics:
At low intraspecies coupling strengths, nonreciprocal interspecies alignment leads to asymmetric density dynamics, where predominantly one of the two species forms clusters. The asymmetric clustering in the weak-intraspecies-coupling regime has been studied in detail in \cite{kreienkamp_klapp_2024_non-reciprocal_alignment_induces_asymmetric_clustering,kreienkamp_klapp_2024_dynamical_structures_phase_separating_nonreciprocal_polar_active_mixtures}. Here, we focus on the regime of strong intraspecies coupling strengths, where the polarization dynamics play a dominant role. This leads to significantly different, time-dependent collective behavior, which is not present in the system with weaker intraspecies couplings. In particular, in the here considered strong-intraspecies-coupling regime, the corresponding field theory predicts spontaneous time-dependent dynamics and exceptional transitions that have been previously associated with the emergence of spontaneous chirality \cite{fruchart_2021_non-reciprocal_phase_transitions}. The latter work suggests that these chiral states are characterized by a homogeneous density distribution and full synchronization of the rotating particle orientations for very large interaction radii. Our particle simulations, conducted at smaller interaction radii, confirm the emergence of spontaneous chirality on the particle level for sufficiently strong nonreciprocity. However, a homogeneous state with full synchronization of all particles is \textit{not} observed. Instead, we observe chimera-like states with a coexistence of locally synchronized and disordered regions \cite{abrams_strogatz_2004_states_chimera_coupled_oscillators}, whose appearance is completely eluded by the mean-field continuum theory. The size of the synchronized regions and resulting polarization depend on the strength of nonreciprocity, with the effects becoming more pronounced when nonreciprocity increases. Moreover, at the coupling strengths related to EPs in the continuum theory, the spontaneous chirality is significantly enhanced.

\section*{Results and discussion}
\subsection*{Model}
We consider a binary mixture of circular, self-propelling particles consisting of species $a=A,B$. The particle dynamics are described by overdamped Langevin equations for the positions $\bm{r}_{i}^a$ and the polar angles $\theta_i^a$ of the heading vectors $\bm{p}_i^a=(\cos\theta_i^a,\sin\theta_i^a)^{\rm T}$, given by \cite{kreienkamp_klapp_2022_clustering_flocking_chiral_active_particles_non-reciprocal_couplings,kreienkamp_klapp_2024_non-reciprocal_alignment_induces_asymmetric_clustering,kreienkamp_klapp_2024_dynamical_structures_phase_separating_nonreciprocal_polar_active_mixtures}
\begin{subequations} \label{eq:Langevin_eq}
	\begin{align}
		\dot{\bm{r}}^a_{i} &= v_0\,\bm{p}^a_{i} + \mu_{r} \sum_{j,b} \bm{F}_{\rm{rep}}(\bm{r}^a_{i},\bm{r}^b_{j}) + \sqrt{2\,D'_{\rm t}}\,\bm{\xi}_{i}^a \label{eq:Langevin_r}\\
		\dot{\theta}^a_{i} &= \mu_{\theta} \sum_{j,b \in \partial_i(R_{\theta})} k_{ab}\, \sin(\theta_{j}^b-\theta_{i}^a) + \sqrt{2\,D'_{\rm r}}\,\eta_{i}^a \label{eq:Langevin_theta}.
	\end{align}
\end{subequations}
Both species have the same self-propulsion velocity ($v_0$) and equal mobilities ($\mu_r,\mu_{\theta}$). The particles interact through symmetric steric repulsion ($\bm{F}_{\rm{rep}}$, see ``Steric repulsion'' subsection in the Methods) and are subject to translational ($\bm{\xi}_{i}^a$) and rotational ($\eta_{i}^a$) Gaussian white noise with zero mean and unit variance. The two species differ only in their Vicsek-like torques of strength $k_{ab}$. When $k_{ab}>0$, particles of species $a$ aim to orient parallel (align) with particles of species $b$ within the shell $\partial_i$ of radius $R_{\theta}$. For $k_{ab}<0$, $a$-particles seek to orient antiparallel (antialign) with $b$-particles. The interspecies couplings, $k_{AB}$ and $k_{BA}$, can be either reciprocal ($k_{AB}=k_{BA}$) or nonreciprocal ($k_{AB}\neq k_{BA}$). The model~\eqref{eq:Langevin_eq} includes both, steric repulsion and generic alignment couplings, each driving paradigmatic active matter transitions, namely, motility-induced phase separation \cite{cates_tailleur_2015_mips,Bialke_2013_microscopic_theory_phase_seperation} and flocking \cite{marchetti_simha_2013_hydrodynamics_soft_active_matter,vicsek_1995_novel_type_phase_transition,gregoire_chate_2004_collective_motion}. Importantly, however, the results regarding the synchronization behavior remain qualitatively equivalent in systems without repulsion, see Supplementary Note 6.

We set the particle diameter $\ell = \sigma$ and time $\tau = \sigma^2/D'_{\rm t}$ as characteristic length and time scales. The control parameters are the particle density $\rho_0^a$, the reduced orientational coupling strength $g_{ab}=k_{ab}\,\mu_{\theta}\,\tau$, the P\'eclet number ${\rm Pe}=v_0\,\tau/\ell$, and the rotational noise strength $D_{\rm r}=D'_{\rm r}\,\tau$. For details on corresponding Brownian Dynamics (BD) simulations, see ``Brownian Dynamics simulations'' subsection in the Methods.

To study the impact of (non-)reciprocal torques in an otherwise symmetric system, we assume equal densities for both species, $\rho_0^A=\rho_0^B=\rho_0/2$, and equal intraspecies alignment strengths $g_{AA}=g_{BB}=g>0$. The density ($\rho_0^a = 4/(5\,\pi)$), motility (${\rm Pe}=40$), and rotational noise strength ($D_{\rm r} = 3\cdot 2^{-1/3}$) are chosen to ensure motility-induced phase separation in the absence of alignment couplings ($g_{ab}=0 \ \forall \ ab$) \cite{kreienkamp_klapp_2024_non-reciprocal_alignment_induces_asymmetric_clustering}. The alignment radius is set to $R_{\theta}=10\,\ell$ [for smaller $R_{\theta}$, see Supplementary Note 5].

The alignment couplings between particles can give rise to states with nonzero global polarization. Our system features two types of polarized states with constant flock directions: flocking (parallel orientation of $A$- and $B$-flocks) and antiflocking (antiparallel orientation) \cite{kreienkamp_klapp_2024_non-reciprocal_alignment_induces_asymmetric_clustering,kreienkamp_klapp_2024_dynamical_structures_phase_separating_nonreciprocal_polar_active_mixtures}. In addition to these polarized states with stationary directions, time-dependent chiral states can occur, where the polarization vector rotates over time \cite{fruchart_2021_non-reciprocal_phase_transitions}. This striking phenomenon is solely induced by nonreciprocal couplings, in the absence of any intrinsic chirality \cite{liebchen_2017_collective_behavior_chiral_active_matter_pattern_formation_flocking,levis_liebchen_2018_Brownian_chiral_disks}, and only emerges for sufficiently strong intraspecies alignment $g$. The particle behavior observed in this study therefore differs significantly from that reported in previous works at weaker intraspecies alignment \cite{kreienkamp_klapp_2024_non-reciprocal_alignment_induces_asymmetric_clustering,kreienkamp_klapp_2024_dynamical_structures_phase_separating_nonreciprocal_polar_active_mixtures}.

\subsection*{Mean-field continuum analysis}
To study the emergence of chiral states, we first consider a coarse-grained description of the microscopic model \eqref{eq:Langevin_eq} in terms of density fields $\rho^a(\bm{r},t)$ and polarization densities $\bm{w}^a(\bm{r},t)$ \cite{kreienkamp_klapp_2022_clustering_flocking_chiral_active_particles_non-reciprocal_couplings,kreienkamp_klapp_2024_dynamical_structures_phase_separating_nonreciprocal_polar_active_mixtures,kreienkamp_klapp_2024_non-reciprocal_alignment_induces_asymmetric_clustering}. The full equations are given in the ``Continuum model'' subsection in the Methods. On the continuum level, the alignment strength scales as $g'_{ab} = g_{ab}\,R_{\theta}^2\,\rho_0^b/2$.
We perform linear stability analyses of the homogeneous disordered and homogeneous (anti)flocking states.
We are mostly interested in long-wavelength (wavenumber $k=0$) fluctuations of the polarization fields.
At $k=0$, fluctuations of the densities do not occur due to number conservation. Density fluctuations do occur at $k>0$, but are dominated by those of the polarization at the strong coupling conditions considered here [see Supplementary Note 4]. 

The linear stability analysis of the disordered state $(\rho^a, \bm{w}^a)=(1,\bm{0})$ against $k=0$-perturbations $\sim {\rm e}^{i\sigma^{\rm dis}t}$ reveals the onset of states with non-zero polarization. The corresponding complex growth rates are given by \cite{kreienkamp_klapp_2024_non-reciprocal_alignment_induces_asymmetric_clustering}
\begin{equation}
\label{eq:eigenvalue_disordered}
	\sigma^{\rm dis}_{1/2} = g' - D_{\rm r} \pm \sqrt{g'_{AB}\,g'_{BA}} .
\end{equation}
Non-zero polarization emerges when ${\rm Re}(\sigma^{\rm dis}_{1/2})>0$. Stationary flocking or antiflocking occurs when ${\rm Im}(\sigma^{\rm dis}_{1/2})=0$, with the associated eigenvector determining whether the system flocks or antiflocks. In contrast, ${\rm Im}(\sigma^{\rm dis}_{1/2})\neq 0$ indicates time-dependent oscillatory polarization dynamics. 

\begin{figure}
	\includegraphics[width=1\linewidth]{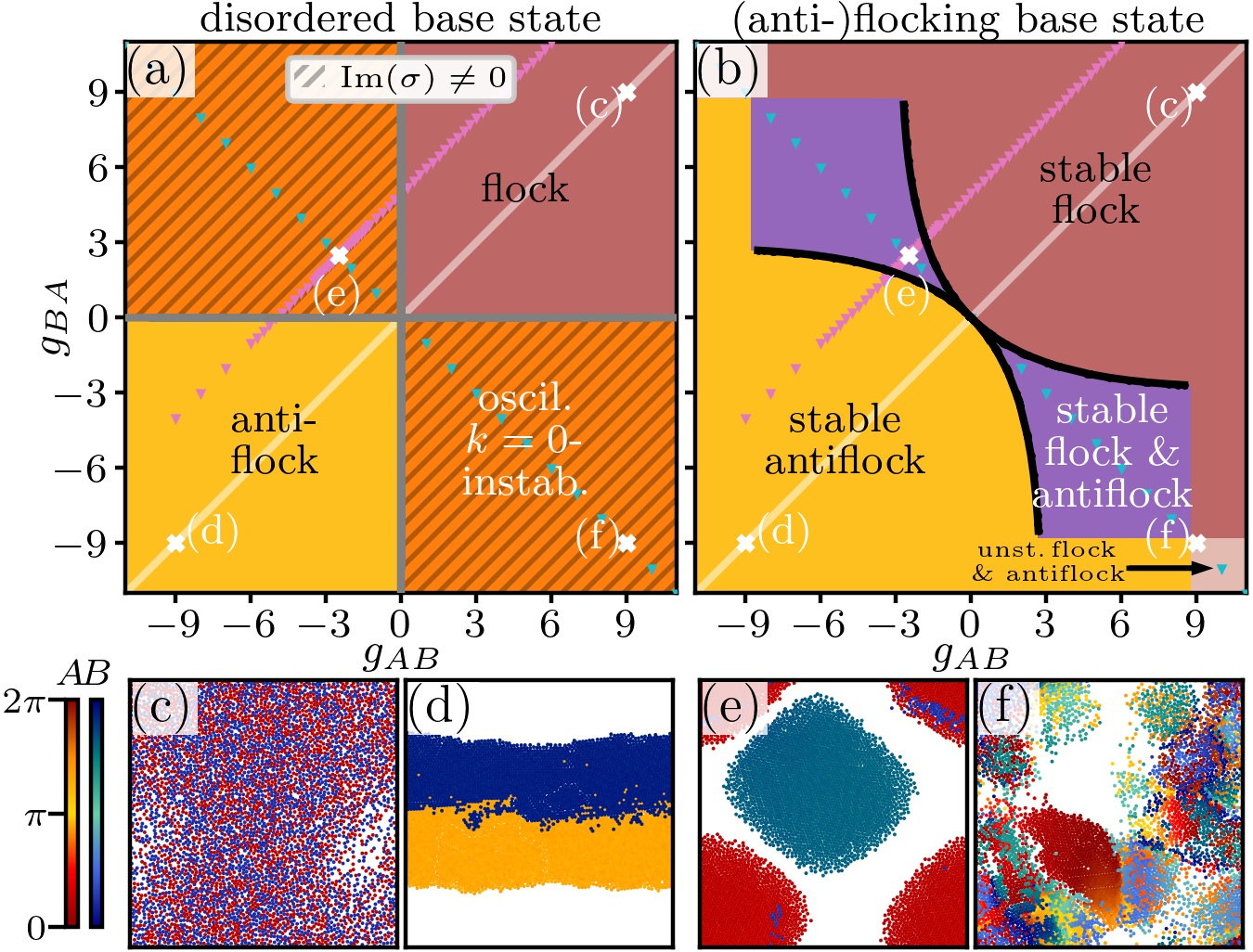}
	\caption{\label{fig:stability_diagram_with_snapshots_Rtheta-10_mean-field}\textbf{Stability diagrams at wavenumber $k=0$ and particle simulation snapshots.} The stability diagrams are obtained from linear stability analyses of the (a) uniform disordered and (b) homogeneous (anti)flocking base states of the continuum Eqs.~\eqref{eq:continuum_eq_density_main},\eqref{eq:continuum_eq_polarization_main} for different interspecies coupling strengths $g_{AB}$ and $g_{BA}$. Exceptional points of the disordered and (anti)flocking base states are indicated as gray and black lines, respectively. The white line indicates reciprocal couplings. Pink and cyan triangles denote data points corresponding to the simulations discussed in the main text. Brownian Dynamics simulation snapshots at (c) $g_{AB}= g_{BA}=9$, (d) $g_{AB}= g_{BA}=-9$, (e) $g_{AB}=-g_{BA}=-2.5$, and (f) $g_{AB}=-g_{BA}=9$. The color code in (c) indicates the particle type and orientation. Other parameters are specified in text.}
\end{figure}

We focus on the ``strong-intraspecies-coupling'' regime, where the term $g'-D_{\rm r}>0$ in Eq.~\eqref{eq:eigenvalue_disordered}, by setting $g=9$. This ensures that non-zero polarization occurs regardless of the values of $g_{AB}$ and $g_{BA}$. The corresponding stability diagram is shown in Fig.~\ref{fig:stability_diagram_with_snapshots_Rtheta-10_mean-field}(a). Flocking and antiflocking in stationary direction occur when both $g'_{AB}$ and $g'_{BA}$ are positive or negative, respectively. However, if the two species have opposing alignment goals, i.e., $g'_{AB}\,g'_{BA}<0$, oscillatory polarization is predicted.

The stability of polarized states can be further analyzed by considering these as base states in a secondary linear stability analysis, shown in the ``Linear stability analysis'' subsection in the Methods. The homogeneous flocking and antiflocking base states are defined by $\rho^a_{\rm b} = 1$ and $\bm{w}_{\rm b}^a = (w_0^a, 0)^{\rm T}$, where $w_0^A$ and $w_0^B$ are obtained as solutions to the continuum equations. Flocking (antiflocking) corresponds to $w_0^A \, w_0^B >(<) 0$. The result is shown in Fig.~\ref{fig:stability_diagram_with_snapshots_Rtheta-10_mean-field}(b). Within the regime of opposing alignment goals there exist regions where both flocking and antiflocking are stable against $k=0$-perturbations [purple regimes in Fig.~\ref{fig:stability_diagram_with_snapshots_Rtheta-10_mean-field}(b)]. No oscillatory instabilities emerge from (anti)flocking base states.
Note that linear stability analyses of both the disordered state [Fig.~\ref{fig:stability_diagram_with_snapshots_Rtheta-10_mean-field}(a)] and the homogeneous (anti)flocking states [Fig.~\ref{fig:stability_diagram_with_snapshots_Rtheta-10_mean-field}(b)] provide complementary insights on the field-theoretical level, but relate to the same behavior on the particle level.

\subsection*{Exceptional points in continuum description}
The oscillatory instabilities of the disordered base states already hint at non-trivial time-dependent collective behavior in the regime of opposing alignment goals. Another indicator for time-dependent states are so-called EPs. In the context of non-Hermitian field theories, they indicate transitions to states with broken (generalized) PT symmetry \cite{miri_alu_2019_exceptional_points_optics_photonics}, including the chiral states found in nonreciprocal polar active matter \cite{fruchart_2021_non-reciprocal_phase_transitions}. At EPs, eigenvalues of the linear stability matrix coalesce and their corresponding eigenvectors become parallel \cite{You_Baskaran_Marchetti_2020_pnas,fruchart_2021_non-reciprocal_phase_transitions,suchanek_loos_2023_time-reversal_PT_symmetry_breaking_non-Hermitian_field_theories,el-ganainy_demetrios_2018_non-Hermitian_physics_PT-symmetry_breaking}. If this happens at a bifurcation -- where the system's dynamical behavior undergoes a qualitative change -- these points correspond to as ``critical exceptional points'' (CEPs) \cite{suchanek_loos_2023_entropy_production_nonreciprocal_Cahn-Hilliard}. In the here considered system EPs (of any type) only occur for $k=0$-perturbations [see Supplementary Note 4]. 

Non-critical EPs emerge from the disordered base state at the lines separating stationary (anti)flocking instabilities and oscillatory instabilities [gray lines in Figs.~\ref{fig:stability_diagram_with_snapshots_Rtheta-10_mean-field}(a)].

For the (anti)flocking base states, we find (only) CEPs. At these points, a previously damped mode coalesces with the Goldstone mode related to the spontaneously broken rotational invariance of steady (anti)aligned states \cite{fruchart_2021_non-reciprocal_phase_transitions}. The CEPs separate regimes where both flocking and antiflocking are stable from those with only stable flocking or only stable antiflocking. These CEPs are indicated as black lines in Figs.~\ref{fig:stability_diagram_with_snapshots_Rtheta-10_mean-field}(b). Note that while the qualitative stability diagram remains unchanged for any intraspecies alignment strength $g'>D_r$, the precise positions of the CEPs depend on the chosen value of $g'$ and are therefore, to some extent, tunable.

\subsection*{Particle dynamics}
\begin{figure}
	\includegraphics[width=1\linewidth]{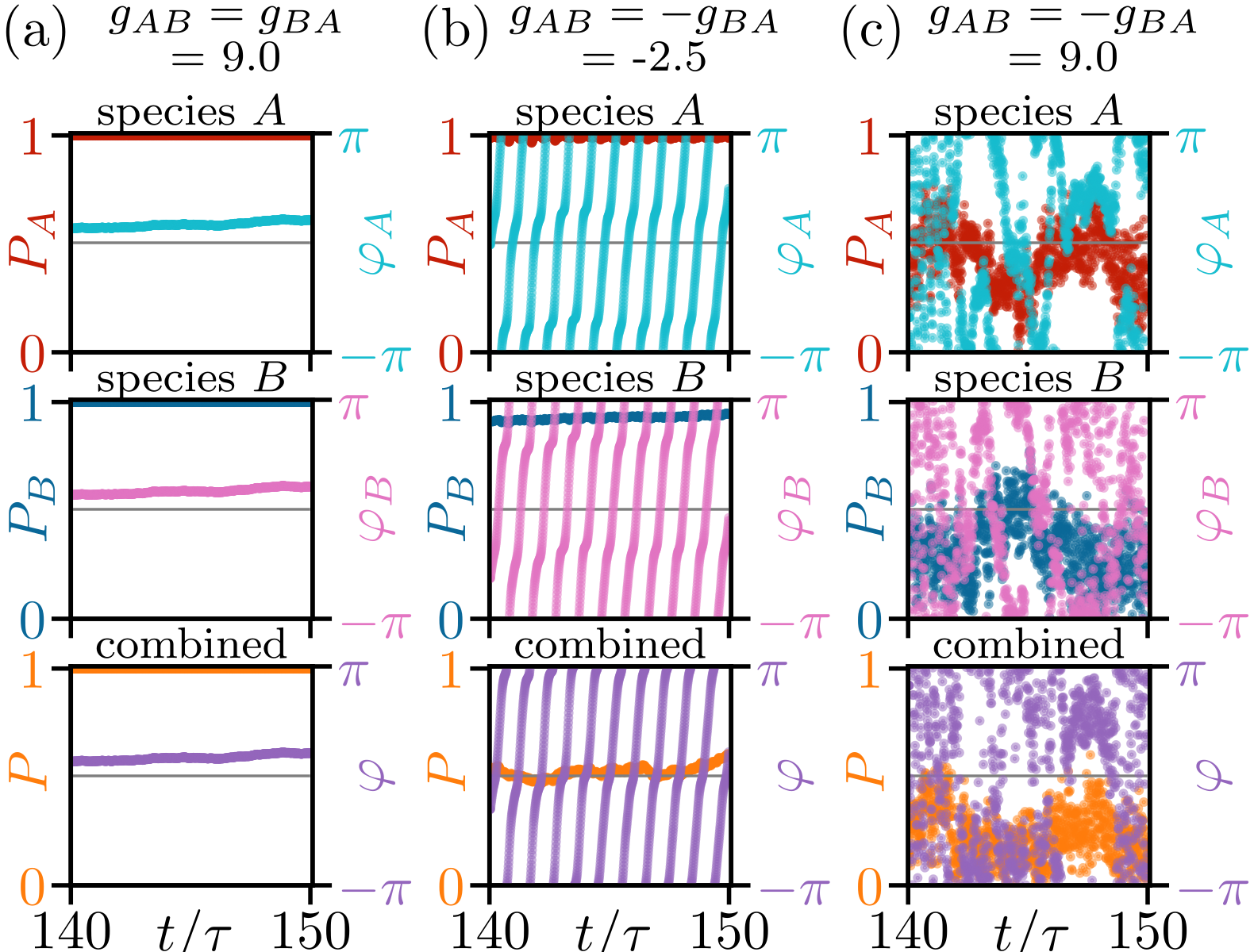}
	\caption{\label{fig:synchronization_over_time_paper}\textbf{Time evolution of the polarization and phase in a single noise realization.} The polarization and phase are shown for (a) reciprocal flocking and nonreciprocal chiral motion of (b) low and (c) high degree of nonreciprocity. The top two rows show the polarizations ($P_{A}, P_B$) and phases ($\varphi_A, \varphi_B$) for each species individually as a function of time $t$. The bottom row displays the combined polarization ($P$) and phase ($\varphi$) for all particles. The time evolutions correspond to the following interspecies coupling strengths ($g_{AB},g_{BA}$) and snapshots: (a) $g_{AB}=g_{BA}=9$ in Fig.~\ref{fig:stability_diagram_with_snapshots_Rtheta-10_mean-field}(c), (b) $g_{AB}=-g_{BA} = -2.5$ in Fig.~\ref{fig:stability_diagram_with_snapshots_Rtheta-10_mean-field}(e), and (c) $g_{AB}=-g_{BA}=-9$ in Fig.~\ref{fig:stability_diagram_with_snapshots_Rtheta-10_mean-field}(f). The characteristic time scale is given by $\tau$.}
\end{figure}

We now turn to the dynamics on the particle scale.
We quantify the time evolution of polarization dynamics in terms of the global polarization $P_a(t)$ of species $a=A,B$, measuring the coherence and synchronization of $a$-particles, and the average phase $\varphi_a(t)$, defined as $P_a(t)\,{\rm e}^{i\varphi_a(t)} = N_a^{-1} \sum_{j}^{N_a} {\rm e}^{i\theta_j(t)}$ \cite{acebron_spigler_2005_kuramoto_model}, see the ``Classification of synchronized states'' subsection in the Methods. The noise- and time-averaged polarizations are denoted by $\langle P_a \rangle$. Snapshots from particle simulations are shown in Figs.~\ref{fig:stability_diagram_with_snapshots_Rtheta-10_mean-field}(c-f). Corresponding simulation videos are provided as Supplementary Movies 1-4. Examples of polarization and average phase evolutions over time for single noise realizations are shown in Fig.~\ref{fig:synchronization_over_time_paper}.

In reciprocal systems, sufficiently strong alignment couplings overcome noise-induced reorientation. The result is \mbox{(anti)}flocking, characterized by \textit{coherent motion} of particles in constant direction [snapshots in Fig.~\ref{fig:stability_diagram_with_snapshots_Rtheta-10_mean-field}(c,d)]. For $g_{AB}=g_{BA}=9$ [Fig.~\ref{fig:synchronization_over_time_paper}(a)], flocking emerges. The two species have the same single-species polarizations $\langle P_{A} \rangle = \langle P_{B} \rangle = 1$ and the same average phases $\varphi_A(t)\approx \varphi_{B}(t)\approx{\rm const.}$, reflecting the stationarity of the flocking direction. The resulting species-combined polarization $\langle P\rangle$, which takes into account the orientations of all particles of both species, is $\langle P \rangle = 1$. On the other hand, anti-flocking is characterized by polarized single-species flocks $\langle P_{A} \rangle = \langle P_{B} \rangle = 1$, whose anti-parallel orientation leads to $\langle P \rangle = 0$ [see Supplementary Note 2].

In contrast to these essentially stationary states, nonreciprocal alignment can induce persistent \textit{rotational} motion with time-dependent $\varphi_a(t)$ \cite{fruchart_2021_non-reciprocal_phase_transitions}. The phenomenon appears in all regimes with opposite alignment goals ($g_{AB}\,g_{BA}<0$), yet the particle dynamics significantly depends on the degree of nonreciprocity. 

To describe the rotational motion, we use the following terminology. A \textit{fully synchronized} state refers to a state where all particles of a \textit{single} species are synchronized with $\langle P_{A} \rangle = \langle P_{B} \rangle = 1$. The average phases, $\varphi_A(t)$ and $\varphi_B(t)$, periodically oscillate in time. If particles of different species rotate with a constant phase shift, their motion is \textit{phase-locked}. In this case, the combined polarization is $\langle P \rangle<1$. The precise value carries information about the phase shift between the species. In a \textit{partially synchronized} state, only some of the particles are synchronized, such that both $\langle P_{A}\rangle,\langle P_{B}\rangle<1$. This comprises \textit{chimera-like} states, where spatially separated synchronized and disordered regions coexist \cite{abrams_strogatz_2004_states_chimera_coupled_oscillators}. To quantify the rotational motion on an individual particle level, we further calculate the rate of phase differences of particles $i$,
\begin{equation}
	\Omega^i_{\rm s}(t) =  \frac{\theta_i(t) - \theta_i(t-\Delta t)}{\Delta t},
\end{equation}
where we set $\Delta t = 0.01\,\tau$. In the absence of any alignment couplings, the distribution $\mathcal{P}(\Omega^i_{\rm s}\,\tau)$ is Gaussian with zero mean and variance $2\,D'_{\rm r}/\Delta t$.

The dependence of the dynamics on nonreciprocity strength becomes clear when comparing the cases of weak and strong antisymmetric couplings, $g_{AB}=-g_{BA}=\delta$. The parameters related to such antisymmetric couplings do \textit{not} coincide with EPs, but lie in-between the EPs within the regime of oscillatory instabilities (see Fig.~\ref{fig:stability_diagram_with_snapshots_Rtheta-10_mean-field}). Note that reversing the sign of $\delta$ yields the same dynamics, with the roles of $A$- and $B$-particles exchanged.

\begin{figure}
	\includegraphics[width=1\linewidth]{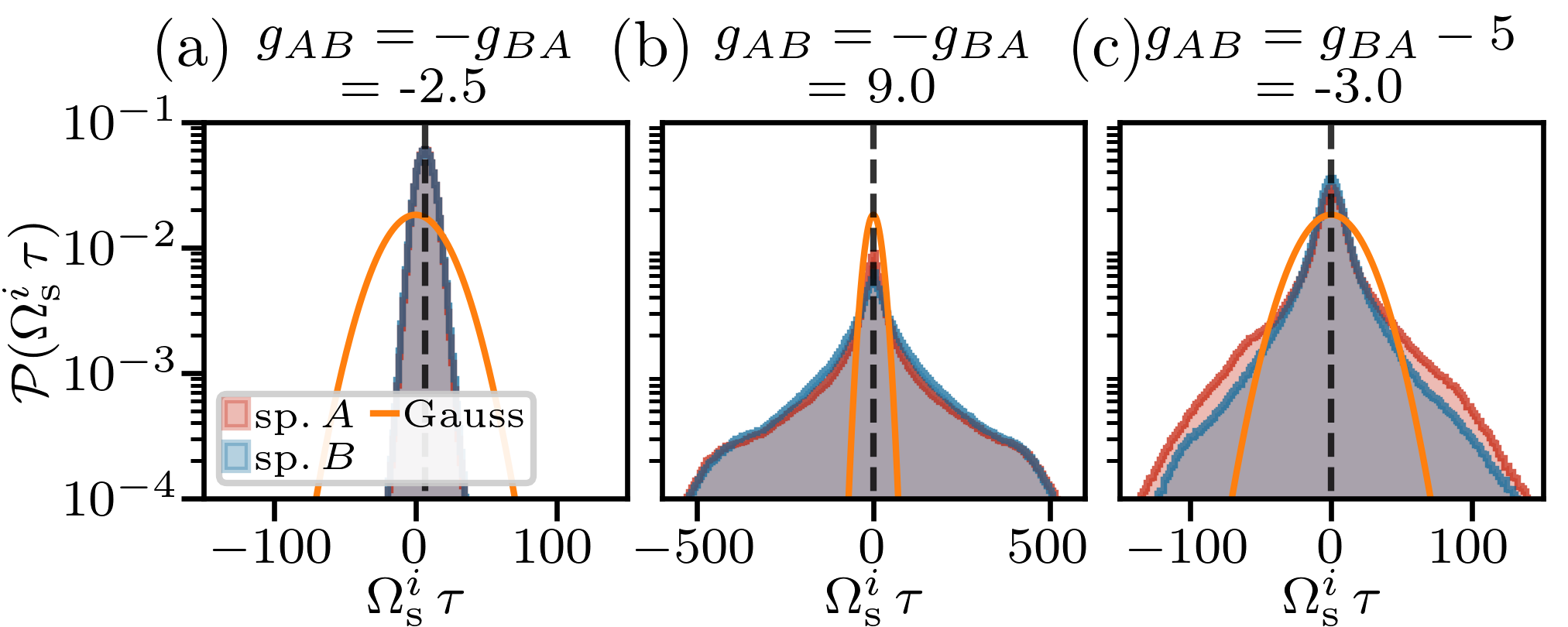}
	\caption{\label{fig:distribution_spontaneous_chirality_paper}\textbf{Distribution of phase difference rates in a single noise realization.} The distributions $\mathcal{P}(\Omega^i_{\rm s}\,\tau)$ are averaged over the phase difference rates $\Omega_{\rm s}^i\,\tau$ of all particles and times in nonreciprocal systems with (a) low and (b) high degrees of nonreciprocity, and (c) close to a critical exceptional point. The orange line indicates the Gaussian distribution induced by rotational noise alone. The black vertical dashed line indicates the average $\langle \Omega^i_{\rm s}\rangle$. The distributions correspond to the following interspecies coupling strengths ($g_{AB},g_{BA}$) and snapshots: (a) $g_{AB}=-g_{BA}=-2.5$ in Fig.~\ref{fig:stability_diagram_with_snapshots_Rtheta-10_mean-field}(e), (b) $g_{AB}=-g_{BA} = 9$ in Fig.~\ref{fig:stability_diagram_with_snapshots_Rtheta-10_mean-field}(f), and (c) $g_{AB}=g_{BA}-5=-3$ in Fig.~\ref{fig:snapshots_fixed_difference-5}(c). The characteristic time scale is given by $\tau$.}
\end{figure}

For weak nonreciprocity ($\delta = -2.5$), almost fully synchronized chiral motion emerges. The particles form two large, rotating single-species clusters as seen in the snapshot in Fig.~\ref{fig:stability_diagram_with_snapshots_Rtheta-10_mean-field}(e), and for other particle numbers in Supplementary Note 1. For $\delta < 0$, $A$-particles want to antialign with $B$-particles, whereas $B$-particles want to align with $A$-particles. As a result, all $A$-particles form part of the same big cluster. The time-evolution of the polarization, shown in Fig.~\ref{fig:synchronization_over_time_paper}(b), indicates that these $A$-particles are nearly fully synchronized (with $\langle P_{A} \rangle = 0.98$). On the other hand, particles of species $B$ are slightly less synchronized (with $\langle P_{B} \rangle = 0.8$). As seen exemplarily from the snapshot in Fig.~\ref{fig:stability_diagram_with_snapshots_Rtheta-10_mean-field}(e), full synchronization of $B$-particles is inhibited by the trapping of some of the $B$-particles within the $A$-cluster. The physical mechanism underlying the trapping is based on the strong alignment between same-species particles, in contrast to the relatively weak interspecies couplings: a single $B$-particle stays absorbed into an $A$-cluster, where it readily follows the $A$-cluster's motion. In contrast, a single $A$-particle tends to move away from $B$-clusters and does not become trapped inside them. This mechanism is schematically visualized in the ``Trapping mechanism'' subsection in the Methods. The periodicity of $\varphi_A(t)$ and $\varphi_B(t)$ reflects the continuous change of cluster polarization in \textit{either} counterclockwise or clockwise direction. Since single-species clusters would exhibit flocking behavior in the absence of the other species, the typical rotation period of the clusters (and thus the spontaneous chirality) in this regime is determined by the time it takes for clusters of different species to encounter each other again. Importantly, $\varphi_A(t)$ and $\varphi_B(t)$ are phase-shifted at all times, which indicates that particles of different species are phase-locked. As expected from continuum analyses \cite{fruchart_2021_non-reciprocal_phase_transitions}, the phase shift persists over time and depends on the interspecies coupling strengths [see Supplementary Note 2]. The overall highly periodic and synchronized rotational motion leads to the distribution $\mathcal{P}(\Omega^i_{\rm s}\,\tau)$ shown in Fig.~\ref{fig:distribution_spontaneous_chirality_paper}(a). Due to synchronization, the width of $\mathcal{P}(\Omega^i_{\rm s}\,\tau)$ is reduced compared to the purely noise-induced Gaussian case (orange line). The mean of the distribution (black dashed line) is shifted to a non-zero value, signifying a bias towards either clockwise or counterclockwise rotation in a single noise realization. Nevertheless, particles still sometimes turn in the opposite direction due to rotational noise. Averaging over several realizations reveals the spontaneous nature of this rotation, with equal probability for both directions, yielding zero mean chirality [see Supplementary Note 2].

For strong nonreciprocity ($\delta = 9$), the overall collective behavior is quite different: chimera-like, partially synchronized states with multiple smaller clusters emerge [snapshot in Fig.~\ref{fig:stability_diagram_with_snapshots_Rtheta-10_mean-field}(f)]. Additionally, a relatively large fraction of the $A$-particles forms one extended, synchronized cluster with high value of polarization. This `asymmetric clustering' of predominantly species $A$ is caused by the relatively large $\delta >0$ compared to the intraspecies alignment $g$. The phenomenon occurs already in the weak-coupling regime and is discussed in detail in \cite{kreienkamp_klapp_2024_non-reciprocal_alignment_induces_asymmetric_clustering,kreienkamp_klapp_2024_dynamical_structures_phase_separating_nonreciprocal_polar_active_mixtures}. As seen in Fig.~\ref{fig:stability_diagram_with_snapshots_Rtheta-10_mean-field}(f), the multiple clusters of the same species are not necessarily synchronized. This leads to a generally smaller degree of synchronization and periodicity, which are reflected in the polarization, with $\langle P_B \rangle = 0.28 < \langle P_A \rangle = 0.33$ and in an irregular time-dependency of $\varphi_A(t)$ and $\varphi_B(t)$ in Fig.~\ref{fig:synchronization_over_time_paper}(c). Here, not all particles in a single ensemble rotate in the same direction. Instead, particles rotate both clockwise and counterclockwise, balancing out the average chirality such that $\langle \Omega_{\rm s}^i \rangle = 0$ [Fig.~\ref{fig:distribution_spontaneous_chirality_paper}(b)].
The distribution itself is much wider than the noise-induced Gaussian, showing that strong nonreciprocity enhances the rotational motion of particles.

\begin{figure}
	\includegraphics[width=1\linewidth]{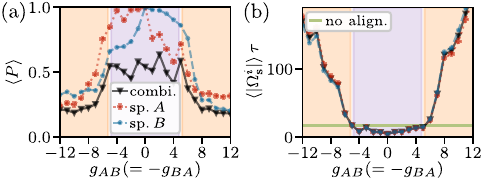}
	\caption{\label{fig:polarisation_chirality_rAlign-10_paper}\textbf{Polarization and spontaneous chirality for an antisymmetric system.} The interspecies coupling strengths are $g_{AB}=-g_{BA}=\delta$. The data points in (a) and (b) represent ensemble- and time-averaged polarization $\langle P \rangle$ and spontaneous chirality $\langle \vert \Omega_{\rm s}^i \vert \rangle$ for species $A$ (red), species $B$ (blue), and all particles combined (black) [see legend in (a)]. In (b), the additional green line shows the spontaneous chirality in a system without alignment couplings. Purple and orange backgrounds mark regimes of low and high degrees of nonreciprocity, respectively, corresponding to qualitatively different dynamical behavior. The characteristic time scale is given by $\tau$.}
\end{figure}

For a more complete picture of the $\delta$-dependency of the chiral dynamics, we now focus on the time- and ensemble-averaged \textit{absolute value} of spontaneous chirality, $\langle \vert  \Omega^i_{\rm s} \vert \rangle$ (ensuring that clockwise and counterclockwise rotations do not cancel out). Rotational diffusion alone leads to $\langle \vert \Omega_{\rm s}^{\rm no\ align.} \vert \rangle \, \tau = 2\,\sqrt{D'_{\rm r}/(\pi\,\Delta t)} > 0$. Further we take the time- and ensemble-averaged polarization $\langle P_a \rangle$ as an indicator for the size of synchronized $a$-clusters in the system (see the ``Classification of synchronized states'' subsection in the Methods).

The polarization and spontaneous chirality are shown as functions of $\delta$ in Fig.~\ref{fig:polarisation_chirality_rAlign-10_paper} for both species individually and combined. The corresponding data points are shown as cyan triangles in Fig.~\ref{fig:stability_diagram_with_snapshots_Rtheta-10_mean-field}.
For $\delta = 0$, particles of the same species align, whereas particle of different species have no orientational couplings. This results in $\langle P_A \rangle = \langle P_B \rangle = 1$, while the species-combined polarization is $\langle P \rangle \approx 0.5$ [Fig.~\ref{fig:polarisation_chirality_rAlign-10_paper}(a)]. Due to intraspecies alignment, which induces coherent motion of same-species particles, the absolute value of spontaneous chirality, $\langle \vert  \Omega^i_{\rm s} \vert \rangle$, is \textit{reduced} compared to the purely noise-induced case without any alignment [$\langle \vert  \Omega^{\rm no\ align.}_{\rm s} \vert \rangle$, green line in Fig.~\ref{fig:polarisation_chirality_rAlign-10_paper}(b)].\\

The behavior for small $\vert \delta \vert \lesssim 5$ (purple-shaded areas in Fig.~\ref{fig:polarisation_chirality_rAlign-10_paper}) is qualitatively the same we already discussed for $\delta=-2.5$. Here, nonreciprocal interspecies interactions are relatively weak compared to intraspecies alignment, resulting in the formation of two large, rotating clusters composed of a single species each. The combined polarization remains around $\langle P \rangle \approx 0.5$ and the spontaneous chirality remains lower than $\langle \vert  \Omega^{\rm no\ align.}_{\rm s} \vert \rangle$. While the spontaneous chirality is nearly the same for both species, the polarization differs. For $\delta>(<)0$, $B(A)$-particles remain synchronized [$\langle P_{B(A)} \rangle > 0.85$ in Fig.~\ref{fig:polarisation_chirality_rAlign-10_paper}(a)].
Yet, by trapping particles of the other species inside their cluster as seen in Fig.~\ref{fig:stability_diagram_with_snapshots_Rtheta-10_mean-field}(e), they inhibit full synchronization of the latter ($\langle P_{A(B)} \rangle < 0.85$).

When nonreciprocity becomes stronger ($\vert \delta \vert \gtrsim 5$, orange-shaded area in Fig.~\ref{fig:polarisation_chirality_rAlign-10_paper}), the overall polarization $\langle P \rangle$ gradually decreases and the spontaneous chirality increases. In this regime, nonreciprocity-induced spontaneous chirality dominates over alignment-induced coherent motion. Partially synchronized, chimera-like states emerge as exemplarily discussed above for $\delta = 9$ [Figs.~\ref{fig:synchronization_over_time_paper}(c),\ref{fig:distribution_spontaneous_chirality_paper}(b)]. Large variations in $P_{a}(t)$ result in large values of susceptibilities in this regime [see Supplementary Note 2].

For even stronger nonreciprocity [$\vert \delta \vert \geq 9$, Fig.~\ref{fig:stability_diagram_with_snapshots_Rtheta-10_mean-field}(f)], additional asymmetric clustering emerges. Here, the formation of highly polarized clusters that consist predominantly of only one species leads to $ \langle P_{B(A)} \rangle <\langle P_{A(B)} \rangle$ for $\delta>(<)0$ \cite{kreienkamp_klapp_2024_non-reciprocal_alignment_induces_asymmetric_clustering,kreienkamp_klapp_2024_dynamical_structures_phase_separating_nonreciprocal_polar_active_mixtures}. These effects are also reflected in the orientational correlation functions [see Supplementary Note 2]. 

Importantly, the different clustering behaviors at different strength of nonreciprocity can be attributed to the ratio of intraspecies alignment to interspecies nonreciprocity. They are \textit{not} a result of a special interplay between repulsion-induced motility-induced phase separation and orientational couplings. Rather, we find qualitatively the same behavior in the absence of repulsion [see Supplementary Note 6].

The above-mentioned threshold $\vert \delta \vert \lesssim 5$ for the formation of large, rotating clusters cannot be derived from the mean-field continuum analysis at wavenumber $k = 0$ [Fig.~\ref{fig:stability_diagram_with_snapshots_Rtheta-10_mean-field}(a),(b)]. However, including finite-wavenumber perturbations ($k > 0$) reveals that the homogeneous flocking state becomes unstable to both longitudinal and transverse modes for $\delta \lesssim 3.9$ [see Supplementary Note 4]. For $\delta \gtrsim 3.9$, the state remains stable against longitudinal perturbations, consistent with the formation of polarized, elongated clusters observed in particle simulations for $\delta \gtrsim 5$.

\subsection*{Coarse-grained density description}
So far, on the continuum level, we have concentrated on the polarization dynamics that is captured by the linear stability analysis at wavenumber $k=0$. In fact, even when considering the density dynamics by investigating $k>0$-instabilities of the disordered base state \cite{kreienkamp_klapp_2022_clustering_flocking_chiral_active_particles_non-reciprocal_couplings,kreienkamp_klapp_2024_non-reciprocal_alignment_induces_asymmetric_clustering}, we find that the polarization instabilities are dominant in the here considered coupling regime [see Supplementary Note 4].
Intriguingly, however, the density instabilities \textit{do} become apparent on an even higher level of coarse-graining. To see this, we perform an adiabatic elimination of the polarization densities $\bm{w}^a(\bm{r},t)$ \cite{kreienkamp_klapp_2024_non-reciprocal_alignment_induces_asymmetric_clustering}, thereby simplifying the full dynamics in terms of two continuity equations for the particle densities $\rho^A(\bm{r},t)$ and $\rho^B(\bm{r},t)$ alone, see ``Coarse-grained density dynamics'' subsection in the Methods. A linear stability analysis of this coarse-gained density description at $k>0$ predicts the type of clustering. In particular, it predicts symmetric demixing for the case $\delta=0$, where particles of different species have no orientational couplings. For increasing anti-symmetric couplings with $\delta > (<) 0$, the demixing gradually turns into predominant clustering of species $A (B)$. 
These predictions are in excellent agreement with the particle simulation results, which show how the nearly demixed large, rotating clusters at small $\vert \delta \vert$ transform into clusters of predominantly a single species at larger $\vert \delta \vert$ [see Figs.~\ref{fig:stability_diagram_with_snapshots_Rtheta-10_mean-field}(e),(f)].

\subsection*{Signature of exceptional points in particle description}
We now turn to the \textit{transition} to the chiral state and the signatures of EPs in particle dynamics. To this end, we consider nonreciprocal systems with $g_{AB} = g_{BA} - d$, where $d\neq 0$ is fixed. By varying $g_{AB}$, we cross non-critical EPs of the disordered base state twice and CEPs of the antiflocking and flocking base states once each.

\begin{figure}
	\includegraphics[width=1\linewidth]{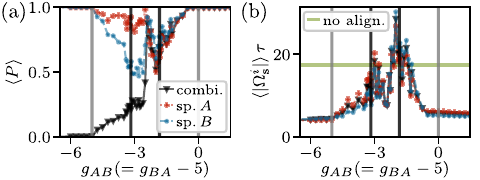}
	\caption{\label{fig:polarisation_chirality_rAlign-10_N-5000_fixed_difference-5_paper}\textbf{Polarization and spontaneous chirality when crossing exceptional points.} The data is shown for a nonreciprocal system with interspecies coupling strengths $g_{AB}=g_{BA}-5$. The vertical gray and black lines indicate non-critical and critical exceptional points, respectively. The data points in (a) and (b) represent ensemble- and time-averaged polarization $\langle P \rangle$ and spontaneous chirality $\langle \vert \Omega_{\rm s}^i \vert \rangle$ for species $A$ (red), species $B$ (blue), and all particles combined (black) [see legend in (a)]. In (b), the additional green line shows the spontaneous chirality in a system without alignment couplings. The characteristic time scale is given by $\tau$.}
\end{figure}

\begin{figure}
	\includegraphics[width=1\linewidth]{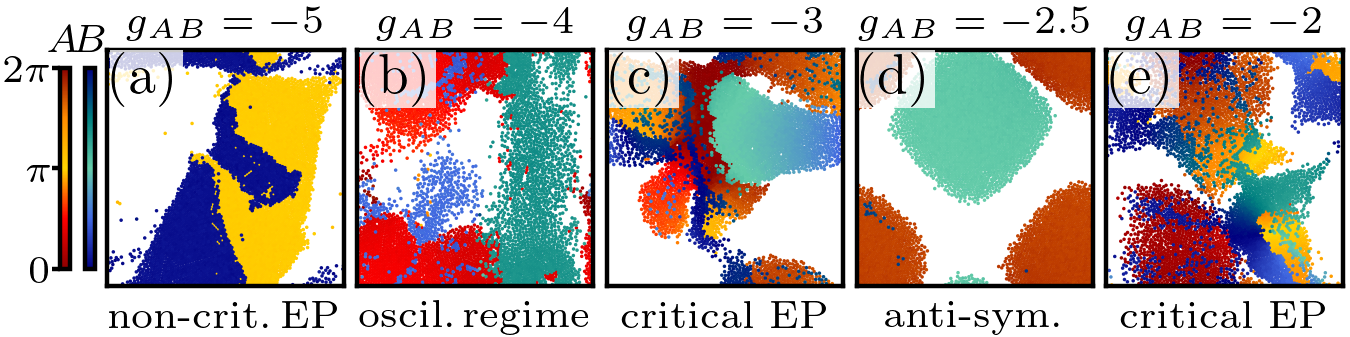}
	\caption{\label{fig:snapshots_fixed_difference-5}\textbf{Snapshot of Brownian Dynamics simulations in the vicinity of exceptional points.} The interspecies coupling strengths are $g_{AB}=g_{BA}-5$. The color code in (a) indicates the particle type and orientation in all panels (a)-(e).}
\end{figure}

In Fig.~\ref{fig:polarisation_chirality_rAlign-10_N-5000_fixed_difference-5_paper} we present the time- and ensemble-averaged polarization and spontaneous chirality along the line $g_{AB}=g_{BA}-5$ in Fig.~\ref{fig:stability_diagram_with_snapshots_Rtheta-10_mean-field}(a,b), denoted by pink triangles.
As $g_{AB}$ increases from negative to positive, the system goes from antiflocking ($\langle P \rangle = 0$) to flocking ($\langle P \rangle = 1$), both characterized by coherent motion and, consequently, small $\langle \vert \Omega^i_{\rm s} \vert \rangle$. Between these states, separated by non-critical EPs, lies the regime of oscillatory $k=0$-instabilities, where $P$ gradually increases from $0$ and $1$. At the CEPs that lie within the oscillatory regime, the single-species polarization drops, indicating reduced synchronization at the CEPs. The spontaneous chirality begins to increase once the system crosses the non-critical EPs, indicating enhanced particle rotation inside the oscillatory instability regime  [Fig.~\ref{fig:polarisation_chirality_rAlign-10_N-5000_fixed_difference-5_paper}(b)]. Strikingly, we observe peaks of this quantity at the coupling values related to CEPs. This increase is accompanied by high values of susceptibility $\chi = N\,{\rm Var}(P)$ [see Supplementary Note 2]. The enhanced susceptibility obtained in particle simulations is in line with field-theoretical predictions of diverging order parameter fluctuations close to the CEP \cite{zelle_diehl_2024_universal_phenomenology_critical_exceptional_points}.

As an illustration, we show snapshots of particle simulations close to EPs in Fig.~\ref{fig:snapshots_fixed_difference-5}. At the non-critical EP ($g_{AB}=-5$), antiflocking still persists  [Fig.~\ref{fig:snapshots_fixed_difference-5}(a)] (see Supplementary Movies 5-8). Within the oscillatory regime [$g_{AB}=-4$, Fig.~\ref{fig:snapshots_fixed_difference-5}(b)], antiflocking in constant direction does not survive anymore and the degree of synchronization among same-species particles starts to decrease. At the CEPs [$g_{AB}=-3$ and $g_{AB}=-2$ in Figs.~\ref{fig:snapshots_fixed_difference-5}(c,e)], clusters dynamically form and break up. Here, the synchronization of particles is further reduced and particles rotate in both clockwise and counterclockwise directions.
This less synchronized behavior at CEPs contrasts with the large, synchronized single-species clusters observed for the fully antisymmetric case in-between the CEPs [$g_{AB}=-g_{BA}=-2.5$, Figs.~\ref{fig:snapshots_fixed_difference-5}(d)]. The difference between these cases is also reflected in the corresponding frequency distributions of particles, shown in Figs.~\ref{fig:distribution_spontaneous_chirality_paper}(a) and (c). In the antisymmetric case, the frequency distribution is very narrow [Fig.~\ref{fig:distribution_spontaneous_chirality_paper}(a)], whereas close to the CEP, it significantly broadens [Fig.~\ref{fig:distribution_spontaneous_chirality_paper}(c)]. We observe similar CEP-related behavior along nonreciprocal paths with different values of $d$ and in the absence of repulsion [see Supplementary Notes 2 and 4]. To summarize, at parameter combinations related to CEPs, particles exhibit distinct dynamical behavior, characterized by a reduced degree of synchronization while the spontaneous chirality is increased.

Finally, we note that, although the particle behavior near both CEPs shares qualitative features, the detailed dynamics is more intricate and differs depending on which CEP is considered. For $d > 0$ (see Fig.~\ref{fig:polarisation_chirality_rAlign-10_N-5000_fixed_difference-5_paper} for $d = 5$), the CEP at smaller $g_{AB}$ exhibits $\langle P_B \rangle < \langle P_A \rangle$ and $\langle \vert \Omega_{\rm s}^i \vert \rangle_B < \langle \vert \Omega_{\rm s}^i \vert \rangle_A$, while at larger $g_{AB}$, the relation is reversed: $\langle P_A \rangle < \langle P_B \rangle$ and $\langle \vert \Omega_{\rm s}^i \vert \rangle_A < \langle \vert \Omega_{\rm s}^i \vert \rangle_B$. The distribution $\mathcal{P}(\Omega^i_{\rm s}\,\tau)$ of phase difference rates also differs between species. For $d < 0$ [see Supplementary Note 2], these trends are reversed. Moreover, irrespective of the sign of $d$, both the spontaneous chirality and the susceptibility are consistently larger at the CEP with larger $g_{AB}$.

\section*{Conclusions}
We demonstrate that nonreciprocal orientational couplings induce spontaneous chirality and synchronization on the particle level over a broad range of parameters. This behavior only emerges when the alignment between particles of the same species is strong enough to overcome reorientation due to rotational noise. While previous theoretical \cite{fruchart_2021_non-reciprocal_phase_transitions} and experimental \cite{chen_zhang_2024_emergent_chirality_hyperuniformity_active_mixture_NR} studies of similar systems -- yet, with large or even infinite interaction radii allowing simultaneous interaction between (nearly) all particles -- reported fully synchronized and homogeneously distributed phases, the system considered here, which features finite-range interactions over a maximum of 10 particle diameters, exhibits significantly different behavior.

As a first main result, we found that the strength of nonreciprocal interactions, i.e., the degree of `disagreement' between species, not only strongly affects polarization dynamics but also significantly shapes the spatial organization. For weak nonreciprocity, we observe an almost fully demixed configuration in which each species forms a large, rotating, and synchronized cluster. For stronger nonreciprocity, partially synchronized states with smaller, fragmented clusters emerge.

Secondly, we also investigate the role of exceptional points and their signatures in particle-resolved dynamics. In particular, we find that near these exceptional transitions, the nonreciprocity-induced chirality is maximized and even exceeds the values of fully antisymmetric systems. This shows that exceptional points do not only mark the transition to time-dependent states at the mean-field continuum level, but are related to actually observable, non-trivial behavior of particles.
Importantly, nonreciprocity-induced chirality is also observed in the absence of repulsion, across different system sizes, and for various coupling radii.

The chiral states in our polar active fluid persist over the timescale of the simulation. This behavior is in stark contrast to the two-dimensional nonreciprocal Ising lattice model, where nonreciprocity destroys any static or time-dependent order at the individual spin level, despite predictions from continuum models \cite{avni_vitelli_2023_non-reciprocal_Ising_model,avni_vitelli_2024_dynamical_phase_transitions_non-reciprocal_Ising_model}. Further work is needed to understand the long-time stability in different interacting nonreciprocal systems, as well as their thermodynamic consequences \cite{loos_klapp_2020_irreversibility_heat_non-reciprocal_interactions,suchanek_loos_2023_entropy_production_nonreciprocal_Cahn-Hilliard,suchanek_loos_2023_time-reversal_PT_symmetry_breaking_non-Hermitian_field_theories}, on the microscopic level. Another compelling avenue for investigations is the field-theoretical prediction concerning order parameter correlations and susceptibilities near CEPs. Our results could be tested experimentally, for instance, using robotic systems \cite{chen_zhang_2024_emergent_chirality_hyperuniformity_active_mixture_NR} or optically coupled nanoparticles \cite{reisenbauer_delic_2024_non-hermitian_dynamics_optically_coupled_particles,livska_brzobohat_2024_pt-like_phase_transition_optomechanical_oscillators}. Our findings suggest that tuning the degree of nonreciprocity, which is indeed possible in synthetic active systems like robots \cite{chen_zhang_2024_emergent_chirality_hyperuniformity_active_mixture_NR}, offers a way to harness special behavior near exceptional points -- such as enhanced chirality -- without the need for strongly competing species.

\section*{Methods}
\subsection*{Steric repulsion}
\label{ssec:methods_steric_repulsion}
On a microscopic level, the motion of particles is governed by the Langevin Eqs.~\eqref{eq:Langevin_r} and \eqref{eq:Langevin_theta}.
The translational Langevin Eq.~\eqref{eq:Langevin_r} captures the hard-sphere nature of the particles via the repulsive force $\bm{F}_{\rm rep}(\bm{r}_i^a, \bm{r}_j^b) = -\sum_{(b,j) \neq (a,i)} \nabla_{a,i} U(r^{ab}_{ij})$, where $U(r^{ab}_{ij})$ is the Weeks-Chandler-Andersen potential \cite{weeks_1971_Weeks-Chandler-Andersen_potential}
\begin{equation}
	\label{eq:WCA_potential}
	U(r^{ab}_{ij}) = \begin{cases}
		4\epsilon \left[\left(\frac{\sigma}{r^{ab}_{ij}}\right)^{12} - \left(\frac{\sigma}{r^{ab}_{ij}}\right)^6 + \frac{1}{4} \right], \ {\rm if} \ r^{ab}_{ij}<r_{\rm c}\\
		0, \ {\rm else}
	\end{cases}
\end{equation}
with $r^{ab}_{ij} = \vert \bm{r}^{ab}_{ij} \vert = \vert \bm{r}^{a}_{i} - \bm{r}^{b}_{j} \vert$. The cut-off distance is $r_{\rm c}= 2^{1/6}\,\sigma$ with particle diameter $\sigma$. The characteristic energy scale is $\epsilon = \epsilon^*\,k_{\rm B}\,T$, where $k_{\rm B}$ is the Boltzmann's constant and $T$ is the temperature. We set the thermal energy to the energy unit, i.e., $k_{\rm B}\,T = 1$.

\subsection*{Brownian Dynamics simulations}
\label{ssec:methods_BD_simulations}
We perform numerical Brownian Dynamics (BD) simulations of the Langevin Eqs.~\eqref{eq:Langevin_r} and \eqref{eq:Langevin_theta} in an $L \times L$ box with periodic boundary conditions. The dimensionless simulation parameters are chosen as following. The overall area fraction is set to $\Phi = (\rho^A_0 + \rho^B_0) \,\pi\,\ell^2 / 4 = 0.4$ with number density $\rho_0^a = N_a / L$. The number of particles of each species is equal, i.e., $\Phi_A = \Phi_B = \Phi/2$. The P\'eclet number is set to ${\rm Pe}=40$ and the repulsive strength to $\epsilon^*=100$. The diffusion constants are $D_{\rm t}' = 1\, \ell^2/\tau$ and $D_{\rm r}' = 3 \cdot 2^{-1/3} / \tau$. The dimensionless orientational coupling strengths are $g_{ab}= k_{ab}\,\mu_{\theta}\,\tau$. Focusing on strong intraspecies couplings, we set $g_{AA}=g_{BB}=9$, while $g_{AB}$, $g_{BA}$ are chosen independently. In all simulations in the main text, the cut-off radius for orientational couplings is set to $R_{\theta} = 10\,\ell$ with a total of $N=5000$ particles. The system is initialized in a random configuration. We use an Euler-Mayurama algorithm to integrate the equations of motion with a timestep of $\delta t = 10^{-5}\,\tau$. We let the simulations reach a steady state before data evaluation. Typically, we consider three independent noise realizations, start the evaluation when the systems have run for $130\,\tau$, and take the time average between $130\,\tau$ and $150\,\tau$ after initialization. The time average is thus taken over $2\cdot 10^6$ timesteps. Within this evaluation time, the systems stay in the non-equilibrium steady states discussed in the main text.

\subsection*{Trapping mechanism}
\label{ssec:methods_trapping_mechanis}

\begin{figure*}
	\includegraphics[width=0.7\linewidth]{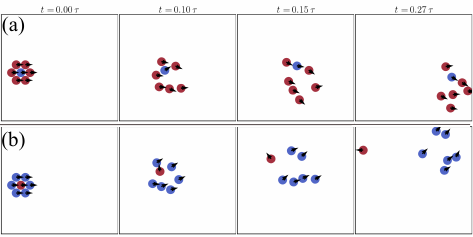}
	\caption{\label{fig:trapping_mechanism}\textbf{Schematic illustration of the trapping mechanism for weak non-reciprocity.} The mechanism is shown for interspecies coupling strengths $g_{AB}=-g_{BA}<0$. The snapshots are shown at different times $t$. (a) A single $B$-particle inside an $A$-cluster remains trapped and follows the $A$-cluster's motion. (b) A single $A$-particle inside a $B$-cluster escapes. Parameters are as in the snapshot in Fig.~\ref{fig:stability_diagram_with_snapshots_Rtheta-10_mean-field}(e). The characteristic time scale is denoted by $\tau$.}
\end{figure*}

The trapping mechanism at weak nonreciprocity is schematically illustrated in Fig.~\ref{fig:trapping_mechanism}, showing single particles within small clusters of the opposite species. For $g_{AB}=-g_{BA}<0$, $A$-particles tend to antialign with $B$-particles, while $B$-particles align with $A$-particles. At the same time, in the strong-intraspecies-coupling regime considered here, the alignment within each species is significantly stronger than the interspecies couplings.

In Fig.~\ref{fig:trapping_mechanism}(a), a single $B$-particle is initially located inside an $A$-cluster. Since $g_{BA}>0$, the $B$-particle follows the $A$-cluster's motion. Meanwhile, the $A$-cluster remains largely intact and aligned, as $g_{AA}\gg -g_{AB}$.

On the other hand, in Fig.~\ref{fig:trapping_mechanism}(b), a single $A$-particle trapped inside a $B$-cluster moves away from the $B$-cluster since $g_{AB}<0$. The $B$-cluster continues its motion without following the $A$-particle since $g_{BB}\gg g_{BA}$.

Overall, this leads to the trapping of some $B$-particles within the $A$-cluster, as seen in the snapshot in Fig.~\ref{fig:stability_diagram_with_snapshots_Rtheta-10_mean-field}(e).

At larger nonreciprocity, the dynamical behavior changes. Instead of two large clusters involving only a single species and the trapped particles, one observes the `asymmetric clustering' of predominantly species $A(B)$ for $\delta>(<)0$. Here, the nonreciprocal interspecies couplings are relatively large compared to the intraspecies alignment. The mechanism behind the asymmetric clustering is explained in detail in \cite{kreienkamp_klapp_2024_non-reciprocal_alignment_induces_asymmetric_clustering,kreienkamp_klapp_2024_dynamical_structures_phase_separating_nonreciprocal_polar_active_mixtures}.

\subsection*{Continuum model}
\label{ssec:methods_continuum_model}
The continuum equations are derived from the microscopic Langevin Eqs.~\eqref{eq:Langevin_r} and \eqref{eq:Langevin_theta} as outlined in references~\cite{kreienkamp_klapp_2022_clustering_flocking_chiral_active_particles_non-reciprocal_couplings,kreienkamp_klapp_2024_non-reciprocal_alignment_induces_asymmetric_clustering,kreienkamp_klapp_2024_dynamical_structures_phase_separating_nonreciprocal_polar_active_mixtures}. The resulting evolution equation for the density field $\rho^a = \rho^a(\bm{r},t)$ of species $a$ is
\begin{equation}
	\label{eq:continuum_eq_density_main}
	\partial_t \rho^a +  \nabla \cdot \big[v^{\rm eff}(\rho)\, \bm{w}^a -  D_{\rm t}\,\nabla\,\rho^a\big] = 0 ,
\end{equation}
where $v^{\rm eff}(\rho) = {\rm Pe} - z\,\rho$ with $\rho = \rho^A + \rho^B$ denotes the effective velocity reduction of particles in high-density regimes. The polarization densities $\bm{w}^a(\bm{r},t)$, which measure the overall orientation of particles at a certain position via $\bm{w}^a/\rho^a$, evolve according to
\begin{equation}
	\label{eq:continuum_eq_polarization_main}
	\begin{split}
		\partial_t \bm{w}^a 
		=& - \frac{1}{2} \, \nabla \,\big(v^{\rm eff}(\rho)\, \rho^a\big)  - D_{\rm r }\, \bm{w}^{a} + \sum_b g'_{ab} \, \rho^a\, \bm{w}^b \\
		&+  D_{\rm t}\,\nabla^2\,\bm{w}^a + \frac{v^{\rm eff}(\rho)}{16\,D_{\rm r}} \, \nabla^2\,\Big(v^{\rm eff}(\rho)\,\bm{w}^a \Big) \\
		&- \sum_{b,c} \frac{ g'_{ab}\,g'_{ac}}{2\,D_{\rm r}} \, \bm{w}^a \, (\bm{w}^b \cdot \bm{w}^c) \\
		& + O(\bm{w}\nabla \bm{w}) + O(\nabla \rho \nabla \bm{w}) .
	\end{split}
\end{equation}
These equations are non-dimensionalized and scaled with the average particle density $\rho_0^a$. Translational and rotational diffusion constants are $D_{\rm t}$ and $D_{\rm r}$. On the continuum level, the relative orientational coupling parameter is given by $g'_{ab} = g_{ab} \, R_{\theta}^2 \, \rho_0^b/2$. The full expressions are given in Supplementary Note 3. The continuum parameters are chosen to match corresponding particle parameters.

The mean-field polarization dynamics at wavenumber $k=0$ are obtained by neglecting all gradient terms in Eq.~\eqref{eq:continuum_eq_polarization_main}, see Supplementary Note 3. At $k=0$, the density is constant, which reflects the conservation of density.

\subsection*{Linear stability analysis}
\label{ssec:methods_linear_stability_analysis}
Linear stability analyses are analytical tools used to predict large-scale collective behavior in continuum systems. For the system at hand, the decay or growth of perturbations to the disordered base state, defined as $\rho^a_{\rm b} = 1$ and $\bm{w}_{\rm b}^a = \bm{0}$, has been previously studied in \cite{kreienkamp_klapp_2022_clustering_flocking_chiral_active_particles_non-reciprocal_couplings,kreienkamp_klapp_2024_non-reciprocal_alignment_induces_asymmetric_clustering,kreienkamp_klapp_2024_dynamical_structures_phase_separating_nonreciprocal_polar_active_mixtures}. These analyses predict the formation of polarized states, related to wavenumber $k=0$-perturbations, as well as clustering at $k>0$.

In this study, we are additionally focus on the linear stability of the anisotropic (anti)flocking state. For simplicity, we assume that the \mbox{(anti)}flocking base state is orientated along the $x$-direction. The base state is then defined as $\rho^a_{\rm b} = 1$ and $\bm{w}_{\rm b}^a = (w_0^a, 0)^{\rm T}$. The values of $w_0^A$ and $w_0^B$ may differ and are obtained as solutions to the (fixed point) continuum Eqs.~\eqref{eq:continuum_eq_density_main} and \eqref{eq:continuum_eq_polarization_main}. A flocking (antiflocking) base state is characterized by $w_0^A \, w_0^B >(<) 0$.

A linear stability analysis predicts whether perturbations to these base states grow or decay in time. A base state is considered stable if the perturbations decay over time. Growing perturbations indicate an unstable base state. Generally, perturbations are of form 
\begin{subequations} \label{eq:perturbation_form}
	\begin{align}
		\delta\rho^a(\bm{r},t) &= \int \hat{\rho}^a(\bm{k}) \, {\rm e}^{i\bm{k}\cdot\bm{r} + \sigma(\bm{k})t} \, {\rm d}\bm{k}\\
		\bm{\delta w}^a(\bm{r},t) &= \int \bm{\hat{w}}^a(\bm{k}) \, {\rm e}^{i\bm{k}\cdot\bm{r} + \sigma(\bm{k})t} \, {\rm d}\bm{k},
	\end{align}
\end{subequations}
where $\sigma(\bm{k})$ are complex growth rates and $\hat{\rho}^a(\bm{k})$, $\bm{\hat{w}}^a(\bm{k})$ denote the perturbation amplitudes.
In Supplementary Note 4, we consider the general case which involves all wave vectors $\bm{k}$. In the main text we focus on $k=0$-perturbations, which are directly related to the integrated value of perturbations via
\begin{equation}
	\int \delta\rho^a(\bm{r},t) \,{\rm d}\bm{r} = (2\,\pi)^2 \, \hat{\rho}^a(k=0) \, {\rm e}^{\sigma(0)\,t}, 
\end{equation} 
and equivalently for $\bm{\delta w}^a(\bm{r},t)$.

For the linear stability analysis, we insert $\rho^a = \rho_{\rm b} + \delta\rho^a$ and $\bm{w}^a = \bm{w}_{\rm b}^a + \bm{\delta w}^a = (w_0^a, 0)^{\rm T} + \bm{\delta w}^a $ into the time evolution Eqs.~\eqref{eq:continuum_eq_density_main} and \eqref{eq:continuum_eq_polarization_main} and neglect perturbations of order $\delta^2$. The resulting linearized time-evolution equation for perturbations to the density of species $a=A,B$ at $k=0$ is simply given by
\begin{equation}
		\partial_t \, \delta\rho^a(k=0) = 0 ,
\end{equation}
which reflects the conservation of particle densities. The polarization can be perturbed either along or transversal to the direction of the base state. For polarizations perturbations along the base state, i.e., in $x$-direction, the time-evolution equation for species $A$ at $k=0$ reads
\begin{equation}
	\begin{split}
		\partial_t\, &\delta w_x^A(k=0) \\
		= \ & \Big[ g'_{AA}\,w_0^A + g'_{AB}\,w_0^B \Big] \delta\rho^A \\
		& + \Big[ -D_r + g'_{AA}\,\rho_0^A - \frac{g_{AA}^{'2}}{2\,D_r}\,3\,(w_0^A)^2 - \frac{g_{AB}^{'2}}{2\,D_r}\,(w_0^B)^2 \\
		& \qquad - \frac{g_{AA}'\,g_{AB}'}{D_r}\,2 \, w_0^A \, w_0^B\Big] \delta w_x^A \\
		& + \Big[g'_{AB}\,\rho_0^A - \frac{g_{AA}'\,g_{AB}'}{D_r}\,(w_0^A)^2 \Big] \delta w_x^B.
	\end{split}
\end{equation}
For polarization perturbations in $y$-direction, one obtains
\begin{equation}
	\begin{split}
		\partial_t\, &\delta w_y^A(k=0) \\
		= \ & \Big[ -D_r + g'_{AA}\,\rho_0^A - \frac{g_{AA}^{'2}}{2\,D_r}\,(w_0^A)^2 - \frac{g_{AB}^{'2}}{2\,D_r}\,(w_0^B)^2 \\
		& \ \ - \frac{g_{AA}'\,g_{AB}'}{D_r} \, w_0^A \, w_0^B \Big] \delta w_y^A \\
		& + \Big[g'_{AB}\,\rho_0^A \Big] \delta w_y^B.
	\end{split}
\end{equation}
The equations for species $B$ are obtained by exchanging $A \leftrightarrow B$. To analyze the results, we consider perturbations in a different basis, that is, $\delta\rho^A \pm \delta\rho^B$ and $\delta\bm{w}^A \pm \delta\bm{w}^B$. These can be easily obtained from the equations above as shown in Supplementary Note 4.

\subsubsection*{Characterization of non-equilibrium states}
The emerging non-equilibrium states can be characterized in terms of eigenvalues and the eigenvector corresponding to the largest real eigenvalue.

For the disordered base state, the characterization follows the more detailed explanations in \cite{kreienkamp_klapp_2022_clustering_flocking_chiral_active_particles_non-reciprocal_couplings,kreienkamp_klapp_2024_non-reciprocal_alignment_induces_asymmetric_clustering,kreienkamp_klapp_2024_dynamical_structures_phase_separating_nonreciprocal_polar_active_mixtures}. The most important points are the following: Instabilities at wave number $k=0$ pertain to flocking or antiflocking instabilities. The eigenvector corresponding to the largest (real) growth rate then indicates whether flocking (in $(\bm{w}^A+\bm{w}^B)$-direction) or antiflocking (in $(\bm{w}^A-\bm{w}^B)$-direction) is predicted. At $k>0$, phase separation behavior comes into play. This case is considered in Supplementary Note 4.

\subsection*{Coarse-grained density dynamics}
\label{ssec:methods_coarse-grained_density_dynamics}
To quantify the degree of clustering predicted at the mean-field continuum level, we analyze the coarse-grained density dynamics under an adiabatic approximation of polarization fields. As shown in Ref.~ \cite{kreienkamp_klapp_2024_non-reciprocal_alignment_induces_asymmetric_clustering}, eliminating temporal and spatial derivatives, as well as higher-order moments of the polarization densities $\bm{w}^a$, allows us to describe the clustering behavior using simplified coarse-grained equations for the density of the two species alone.

The adiabatic elimination of $\bm{w}^a$ yields expressions for the polarization fields that only depend on the density fields, i.e., $\bm{w}_{\rm ad}^a = \bm{w}_{\rm ad}^a(\rho^A, \rho^B)$. One can then write down the coarse-grained density dynamics as
\begin{equation}
\label{eq:coarse-grained_density_dynamics}
	\partial_t \rho^a = - \nabla \cdot (v^{\rm eff}(\rho) \, \bm{w}_{\rm ad}^a ) + D_{\rm t}\, \nabla^2 \rho^a .
\end{equation}
As explained in Ref.~ \cite{kreienkamp_klapp_2024_non-reciprocal_alignment_induces_asymmetric_clustering} and in Supplementary Note 4, a linear stability analysis around the disordered base state yields the clustering angle $\alpha$, which indicates the type and symmetry of clustering. Specifically, the clustering angle can be computed from the eigenvector $\bm{v}_{\rho}=(\hat{\rho}_A + \hat{\rho}_B, \hat{\rho}_A - \hat{\rho}_B)^{\rm T}$ that corresponds to the largest eigenvalue in the coarse-grained density dynamics via $\alpha = {\rm arccos}(\bm{v}_{\rho} \cdot (1,0)^{\rm T})$. Symmetric clustering of both species is indicated by $\alpha=0$. Full symmetric demixing corresponds to $\alpha = \pm \pi/2$. Asymmetric clustering of species $A$ $(B)$ is indicated by $0 < \alpha < \pi/2$ $(\pi/2 < \alpha < \pi)$.

\begin{figure}
	\includegraphics[width=0.5\linewidth]{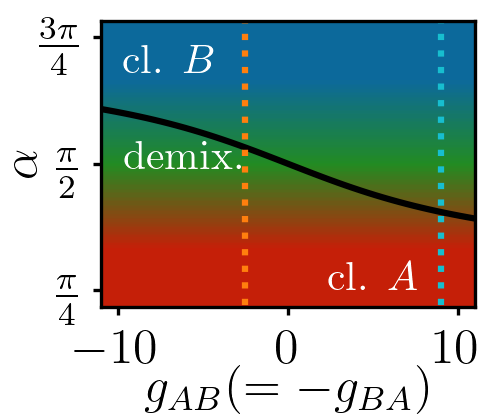}
	\caption{\label{fig:clustering_angles_gaa-9_rAlign-10}\textbf{Clustering angle from coarse-grained density dynamics.} The clustering angle $\alpha$ is obtained from the coarse-grained density dynamics given in Eq.~\eqref{eq:coarse-grained_density_dynamics}, for interspecies coupling strengths $g_{AB}=-g_{BA}$ and alignment radius $R_{\theta}=10\,\ell$. Translational diffusion is set to $D_{\rm t}=0$. Other parameters are as specified in main text. The dotted orange and purple lines mark the coupling strengths $g_{AB}= -g_{BA} = -2.5$ and $g_{AB}= -g_{BA} = 9$, corresponding to the particle simulation snapshots in Figs.~\ref{fig:stability_diagram_with_snapshots_Rtheta-10_mean-field}(e) and (f).}
\end{figure}

Following the calculations presented in Ref.~\cite{kreienkamp_klapp_2024_non-reciprocal_alignment_induces_asymmetric_clustering} and the Supplementary Note 4, we obtain the clustering angles shown in Fig.~\ref{fig:clustering_angles_gaa-9_rAlign-10} for antisymmetric couplings with $g_{AB}=-g_{BA}$. As $g_{AB}$ increases from negative to positive, the coarse-grained density dynamics predict a transition from predominant $B$-clustering to demixing to predominant $A$-clustering. These predictions qualitatively agree with particle simulations, which reveal almost demixed large, rotating clusters for small antisymmetric couplings and asymmetric clustering of predominantly one species for larger antisymmetric couplings. In Fig.~\ref{fig:clustering_angles_gaa-9_rAlign-10}, the dotted purple vertical line marks the coupling strength where predominantely $A$-clustering is observed in particle simulations [$g_{AB}= -g_{BA} = 9$, snapshot in Fig.~\ref{fig:stability_diagram_with_snapshots_Rtheta-10_mean-field}(f)], while dotted orange vertical line marks the coupling strength where large rotating clusters emerge [$g_{AB}= -g_{BA} = -2.5$, snapshot in Fig.~\ref{fig:stability_diagram_with_snapshots_Rtheta-10_mean-field}(e)].

\subsection*{Classification of synchronized states}
\label{ssec:methods_classification_synchronized_states}
The synchronization and coherence of particles is measured by the polarization order parameter $P(t)$. As mentioned in the main text, we determine the \textit{species-dependent} polarization and average phase via \cite{acebron_spigler_2005_kuramoto_model}
\begin{equation}
	P_a(t)\,{\rm e}^{i\varphi^a(t)} = \frac{1}{N_a} \sum_{j_a}^{N_a} {\rm e}^{i\theta_{j_a}(t)}
\end{equation}
for $a=A,B$.
This definition of $P_a(t)$, commonly known as the Kuramoto order parameter in synchronization theory \cite{acebron_spigler_2005_kuramoto_model}, is equivalent to another commonly used form based on heading vectors \cite{cavagna_grigera_2018_physics_of_flocking_correlations}:
\begin{equation}
	P_a(t) = \Big \vert \frac{1}{N_a} \sum_{j_a} \bm{p}_{j_a}(t) \Big \vert,
\end{equation}
where the heading vector of particle $j_a$ is $\bm{p}_{j_a} = (\cos\theta_{j_i},\sin\theta_{j_i})$.

When all particles are perfectly aligned, meaning their heading vectors point in the same direction, the polarization reaches its maximum value of $P=1$. In contrast, if particle orientations are completely uncorrelated, the polarization drops to its minimum of $P=0$.

The term \textit{synchronization} usually refers to particles exhibiting coherent \textit{rotational} motion. On the other hand, the terms \textit{polarized states} and \textit{flocking} describe states where particles move collectively in a \textit{constant} direction. In our system, both types of behavior can occur depending on the nature of interspecies alignment couplings. Therefore, we use the terms ``synchronized'' and ``polarized'' somewhat interchangeably throughout this work, depending on the context, to describe states with high polarization.

To quantify overall synchronization, we also define the \textit{combined} polarization and the average phase $\varphi(t)$ of all $N$ particles (from both species) as $P(t)\,{\rm e}^{i\varphi(t)} = N^{-1} \sum_{j}^N {\rm e}^{i\theta_j(t)}$. The time and ensemble averages of $P(t)$ and $P_{a}(t)$ are denoted by $\langle P \rangle$ and $\langle P_a \rangle$, respectively.

\subsubsection*{Thresholds}
In this study, we consider a species to be synchronized if its average polarization satisfies $\langle P_a\rangle \geq 0.85$. This threshold is motivated by polarization values obtained within the regime $\delta\lesssim 5$ (purple shaded region in Fig.~\ref{fig:polarisation_chirality_rAlign-10_paper}), where we observe large, rotating, and synchronized clusters. For $\delta \gtrsim 5$, polarization values drop rapidly to below $\langle P_a\rangle <0.45$. While the specific values of the threshold does not affect the qualitative interpretation of the results, it provides a consistent benchmark for identifying this specific synchronized behavior.

Furthermore, in the reciprocal flocking and antiflocking states, characterized in the main text by $P_A(t) = P_B(t) = 1$, the precise time and ensemble averages are $\langle P_A \rangle,\langle P_B \rangle \geq 0.998$. The species-combined polarizations are $\langle P \rangle \geq 0.998$ and $\langle P \rangle \geq 0.002$ for the flocking and antiflocking states, respectively.

\subsubsection*{\label{ssec:cluster_size}Cluster size}
In this study, we use the time- and ensemble-averaged polarization $\langle P_a \rangle$ as an indicator of the size of synchronized $a$-clusters. In particle simulations, we observe a broad variety of cluster numbers, sizes, and internal particle densities, depending on the interspecies coupling strengths. As illustrated in the snapshots in Fig.~\ref{fig:stability_diagram_with_snapshots_Rtheta-10_mean-field}(c)-(f) and Fig.~\ref{fig:snapshots_fixed_difference-5}, cluster sizes and numbers vary not only with the antisymmetric coupling strength $\delta$ but also near exceptional points.

We motivate the use of polarization as an indicator for cluster size as follows: within each cluster, particles are typically synchronized, leading to high polarization values of individual clusters (see also \cite{kreienkamp_klapp_2024_dynamical_structures_phase_separating_nonreciprocal_polar_active_mixtures}). In the present `strong-intraspecies-coupling' regime, same-species particles align strongly with neighbors, promoting local synchronization. Accordingly, we observe that for $P_{a} = 1$, all $a$-particles are synchronized and part of the same cluster. Lower values of $P_a$ reflect the presence of multiple smaller clusters, each internally synchronized but uncorrelated with others. While the polarization serves as a useful indicator, the orientational correlation function, analyzed at various parameter combinations in Supplementary Note 2, offers a more detailed view of the system's structural and orientational properties.

\section*{Data availability}
Numerical data sets are available from the corresponding author upon reasonable request.\\

\section*{Code availability}
Computer codes for numerical calculation are available from the corresponding author upon reasonable request.

\begin{acknowledgments}
This work was funded by the Deutsche Forschungsgemeinschaft (DFG, German Research
Foundation) -- Projektnummer 163436311 (SFB 910) and Projektnummer 517665044.
\end{acknowledgments}

\section*{Author contributions}
S.H.L.K.~and K.L.K.~designed the research. K.L.K.~performed and analyzed the analytical and numerical calculations. K.L.K.~and S.H.L.K.~discussed the results and wrote the manuscript.

\section*{Competing interests}
The authors declare no competing interests.


\section*{References}

%

\end{document}


\title{Supplementary Information for ``Synchronization and exceptional points in nonreciprocal active polar mixtures''}

\author{Kim L. Kreienkamp}
\email{k.kreienkamp@tu-berlin.de}
\author{Sabine H. L. Klapp}
\email{sabine.klapp@tu-berlin.de}
\affiliation{%
	Institute for Physics and Astronomy, Technische Universit\"at Berlin, Berlin, Germany
}%



\maketitle

In this supplemental (SM), we provide further details to support the analysis and findings on nonreciprocity-induced spontaneous chirality in nonreciprocal polar active matter. For the particle-level analysis, we examine finite-size effects and present supplementary results on the susceptibility and orientational correlations functions. At the continuum level, we include the complete field equations and perform a detailed linear stability analysis of polarized states, addressing arbitrary wavenumbers. To demonstrate the impact of nonreciprocal alignment for a broader range of systems, we present both particle- and continuum-level results for a smaller alignment radius than considered in the main text and, finally, discuss the influence of repulsive interactions from a particle-level perspective.

\tableofcontents

\section{\label{app:microscopic_model}Microscopic model}
On a microscopic level, the motion of particles is governed by the Langevin Eqs.~(1a) and (1b) in the main text.
The translational Langevin Eq.~(1a) captures the hard-sphere nature of the particles via the repulsive force $\bm{F}_{\rm rep}(\bm{r}_i^a, \bm{r}_j^b) = -\sum_{(b,j) \neq (a,i)} \nabla_{a,i} U(r^{ab}_{ij})$, where $U(r^{ab}_{ij})$ is the Weeks-Chandler-Andersen potential \cite{weeks_1971_Weeks-Chandler-Andersen_potential}
\begin{equation}
	\label{eq:WCA_potential}
	U(r^{ab}_{ij}) = \begin{cases}
		4\epsilon \left[\left(\frac{\sigma}{r^{ab}_{ij}}\right)^{12} - \left(\frac{\sigma}{r^{ab}_{ij}}\right)^6 + \frac{1}{4} \right], \ {\rm if} \ r^{ab}_{ij}<r_{\rm c}\\
		0, \ {\rm else}
	\end{cases},
\end{equation}
where $r^{ab}_{ij} = \vert \bm{r}^{ab}_{ij} \vert = \vert \bm{r}^{a}_{i} - \bm{r}^{b}_{j} \vert$. The cut-off distance is $r_{\rm c}= 2^{1/6}\,\sigma$ with particle diameter $\sigma$ as characteristic length scale, $\ell = \sigma = 1$. The characteristic energy scale is $\epsilon = \epsilon^*\,k_{\rm B}\,T$, where $k_{\rm B}$ is the Boltzmann's constant and $T$ is the temperature. We set the thermal energy to the energy unit, i.e., $k_{\rm B}\,T = 1$. The characteristic time scale $\tau = \sigma^2/D_{\rm t}' = 1$ is the time required for a passive particles to travel its own diameter.  

Particle positions and orientations are influenced by thermal noise. The Gaussian white noise processes $\bm{\xi}_i^a(t)$ and $\eta_i^a(t)$ have zero mean and delta-correlated variances $\langle \xi_{i,k}^a(t)\,\xi_{j,l}^b(t') \rangle = \delta_{ij}\,\delta_{ab}\,\delta_{kl}\,\delta(t-t')$ and $\langle \eta_{i}^a(t)\,\eta_{j}^b(t') \rangle = \delta_{ij}\,\delta_{ab}\,\delta(t-t')$. The mobilities $\mu_r$ and $\mu_{\theta}$ are related to thermal noise via $\mu_r = D'_{\rm t}/(k_{\rm B}\,T)$ and $\mu_{\theta} = D'_{\rm r}/(k_{\rm B}\,T)$. 

We perform numerical Brownian Dynamics (BD) simulations of the Langevin Eqs.~(1a) and (1b) in the main text in an $L \times L$ box with periodic boundary conditions. The dimensionless simulation parameters are chosen as following. The overall area fraction is set to $\Phi = (\rho^A_0 + \rho^B_0) \,\pi\,\ell^2 / 4 = 0.4$ with number density $\rho_0^a = N_a / L$. The number of particles of each species is equal, i.e., $\Phi_A = \Phi_B = \Phi/2$. The P\'eclet number is set to ${\rm Pe}=40$ and the repulsive strength to $\epsilon^*=100$. The diffusion constants are $D_{\rm t}' = 1\, \ell^2/\tau$ and $D_{\rm r}' = 3 \cdot 2^{-1/3} / \tau$. The dimensionless orientational coupling strengths are $g_{ab}= k_{ab}\,\mu_{\theta}\,\tau$. Focusing on strong intraspecies couplings, we set $g_{AA}=g_{BB}=9$, while $g_{AB}$, $g_{BA}$ are chosen independently. In all simulations in the main text, the cut-off radius for orientational couplings is set to $R_{\theta} = 10\,\ell$ with a total of $N=5000$ particles. The system is initialized in a random configuration. We use an Euler-Mayurama algorithm to integrate the equations of motion with a timestep of $\delta t = 10^{-5}\,\tau$. We let the simulations reach a steady state before data evaluation.

\subsection{\label{app:microcopic_model_finite_size}Finite-size effects}
The results presented in the main text are based on simulations with $N=5000$ particles of diameter $\ell = \sigma$ at a fixed area fraction of $\Phi = 0.4$. This corresponds to a simulation box length of $L=99\,\ell$. The orientational coupling radius considered in the main text is $R_{\theta}=10\,\ell$. In the following, we address potential finite-size effects due to an interplay between $L$ and $R_{\theta}$. (The interplay between $R_{\theta}$ and short-range repulsion is discussed separately in \ref{sec:no_repulsion}.) 

\begin{figure*}
	\includegraphics[width=\linewidth]{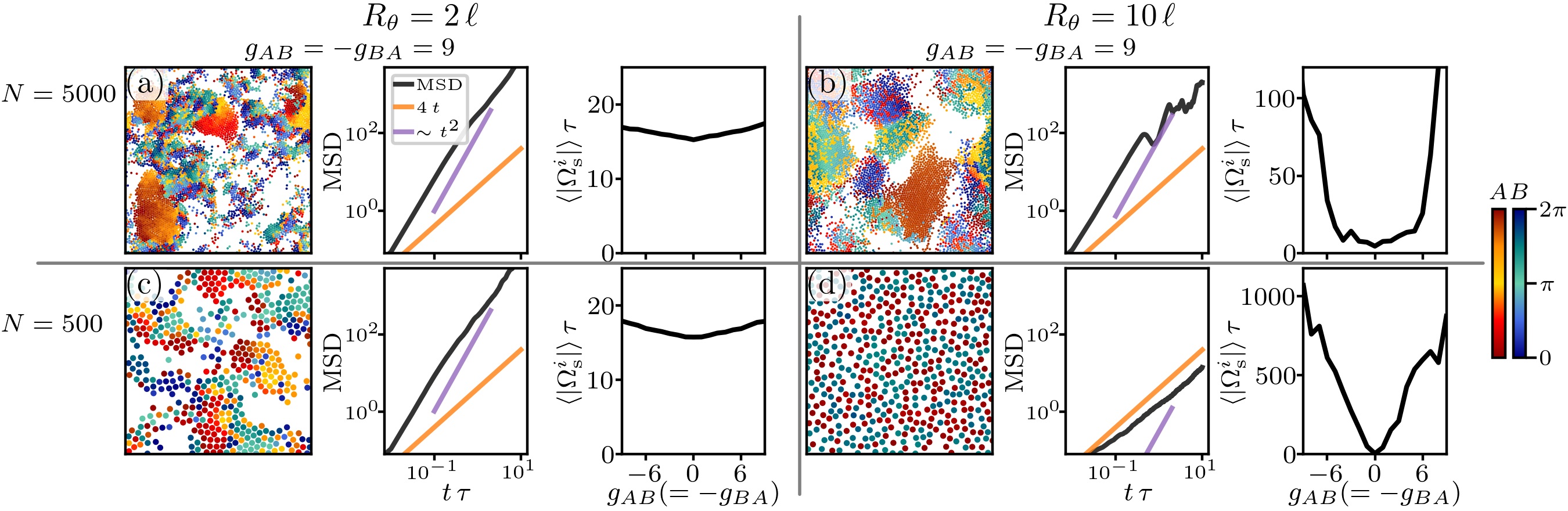}
	\caption{\label{fig:MSD}Mean-squared displacement (MSD), spontaneous chirality $\Omega^i_{\rm s}\,\tau$, and snapshots for antisymmetric systems of different size and interaction radius. The snapshots and MSDs are shown for $g_{AB}=-g_{BA}=9$. (a) $N=5000$ particles, $R_{\theta}=2\,\ell$. (b) $N=5000$ particles, $R_{\theta}=10\,\ell$. (c) $N=500$ particles, $R_{\theta}=2\,\ell$. (d) $N=500$ particles, $R_{\theta}=10\,\ell$. The simulation box length is $L=99\,\ell$ for $N=5000$ and $L_{500} = 31.3\,\ell$ for $N=500$. The MSD is defined as ${\rm MSD} = \langle (\bm{r}(t) - \bm{r}(0))^2\rangle / \ell^2$. The color code indicates the particle type and orientation.}
\end{figure*}

To study the role of finite-size effects, we first qualitatively compare simulation results for two different system sizes ($N=500$ with $L_{500}=31.3\,\ell$ and $N=5000$ with $L=99\,\ell$) and two different coupling radii ($R_{\theta}=2\,\ell$ and $R_{\theta}=10\,\ell$). To this end, we present in Fig.~\ref{fig:MSD} simulation snapshots and mean-squared displacements (MSDs) for $g_{AB}=-g_{BA}=9$, and spontaneous chiralities as functions of the antisymmetric coupling strength $g_{AB}=-g_{BA}$. The MSD, defined as $\langle (\bm{r}(t)-\bm{r}(0))^2 \rangle/\ell^2$, quantifies the translational motion by measuring particle displacements over time. This enables differentiation between dynamical and frozen states (where particle positions are essentially fixed).

When $R_{\theta} \ll L$, i.e., for $N=5000$ with (a) $R_{\theta}=2\,\ell$ or (b) $R_{\theta}=10\,\ell$, and (c) $N=500$ with $R_{\theta}=2\,\ell$, particles form clusters, with a preference for pure $A$-species clusters. This nonreciprocity-induced asymmetric clustering has been described in detail in \cite{kreienkamp_klapp_2024_non-reciprocal_alignment_induces_asymmetric_clustering,kreienkamp_klapp_2024_dynamical_structures_phase_separating_nonreciprocal_polar_active_mixtures}. As seen from the similarity of the MSDs for $R_{\theta}=2\,\ell$ and $R_{\theta}=10\,\ell$ in systems with $N=5000$ particles [Figs.~\ref{fig:MSD}(a,b)], the larger interaction radius of $R_{\theta}=10\,\ell$ does not significantly impact the asymmetric clustering of particles. The values of spontaneous chirality, on the other hand, increase with increasing alignment radius (and, thus, increasing effective coupling strength), see \ref{app:dynamics_alignment_radius}. 

However, when the alignment radius becomes comparable to the simulation box size, the qualitative behavior changes significantly. This is demonstrated for systems with $R_{\theta}=10\,\ell$ and $N=500$ particles, where $L_{500}=31.3\,\ell$, see Figs.~\ref{fig:MSD}(d). From the snapshot, we see that particles are now homogeneously distributed throughout the system without forming any clusters. The corresponding MSD clearly indicates subdiffusive behavior. The spontaneous chirality in these systems is an order of magnitude larger than for $N=5000$. The drastic change in dynamical behavior can be attributed to finite-size effects, which are absent for the results in larger systems presented in the main text. Interestingly, this fully synchronized, ``frozen'' state with essentially homogeneous particle distribution resembles the absorbing state seen in nonreciprocal robot experiments \cite{chen_zhang_2024_emergent_chirality_hyperuniformity_active_mixture_NR}.

\begin{figure}
	\includegraphics[width=0.5\linewidth]{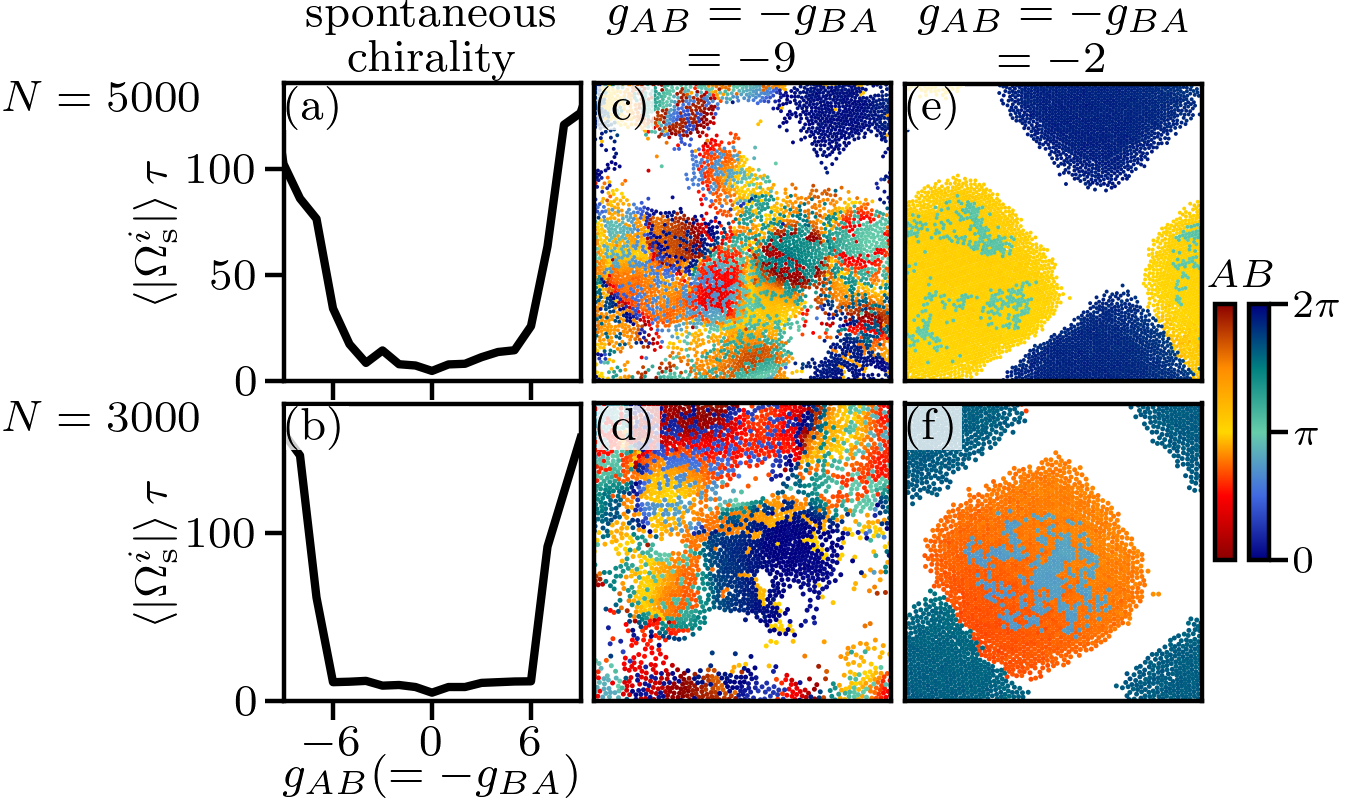}
	\caption{\label{fig:finite_size_spontaneous_chirality}Comparison of nonreciprocity-induced chirality $\Omega^i_{\rm s}\,\tau$ for systems with different numbers of particles. Spontaneous chirality as a function of antisymmetric couplings $g_{AB}(=-g_{BA})$ for (a) $N=5000$ and (b) $N=3000$ particles. Simulation snapshots for $g_{AB}=-g_{BA}=-9$ with (c) $N=5000$ and (d) $N=3000$ and for $g_{AB}=-g_{BA}=-2$ with (e) $N=5000$ and (f) $N=3000$. The alignment radius is $R_{\theta} = 10\,\ell$. The simulation box lengths are $L=99\,\ell$ for $N=5000$ and $L_{3000}=77\,\ell$ for $N=3000$ particles. The color code indicates the particle type and orientation.}
\end{figure}

To confirm that finite-size effects do not affect the qualitative dynamics of the system considered in the main text (with $N=5000$ and $R_{\theta}=10\,\ell$), we compare results for $N=5000$ and $N=3000$ particles in Fig.~\ref{fig:finite_size_spontaneous_chirality}. The spontaneous chirality as a function of antisymmetric couplings $g_{AB}(=-g_{BA})$ is shown in Figs.~\ref{fig:finite_size_spontaneous_chirality}(a,b) for $N=5000$ and $N=3000$, respectively. For both system sizes, the spontaneous chirality is small and nearly constant for moderate nonreciprocity and increases (quickly) at higher values of nonreciprocity. While the magnitude of spontaneous chirality is larger in the $N=3000$ system, the order of magnitude remains the same. Importantly, the snapshots in Figs.~\ref{fig:finite_size_spontaneous_chirality}(c-f) show that the qualitative behavior is the same for both system sizes: Asymmetric, less synchronized clusters form for strong nonreciprocity [$g_{AB}=-g_{BA}=-9$, Figs.~\ref{fig:finite_size_spontaneous_chirality}(c,d)], while large synchronized clusters of both species emerge for weaker nonreciprocity [$g_{AB}=-g_{BA}=-2$, Figs.~\ref{fig:finite_size_spontaneous_chirality}(e,f)].

These findings confirm that the results for $N=5000$ in the main text are not significantly influenced by finite-size effects.

\section{Particle-simulation results} 
In this section, we provide additional particle simulation results for the parameters given in \ref{app:microscopic_model}.

\subsection{\label{app:polarization_average_phase_species}Time evolution of polarization and average phase}
\begin{figure}
	\includegraphics[width=0.33\linewidth]{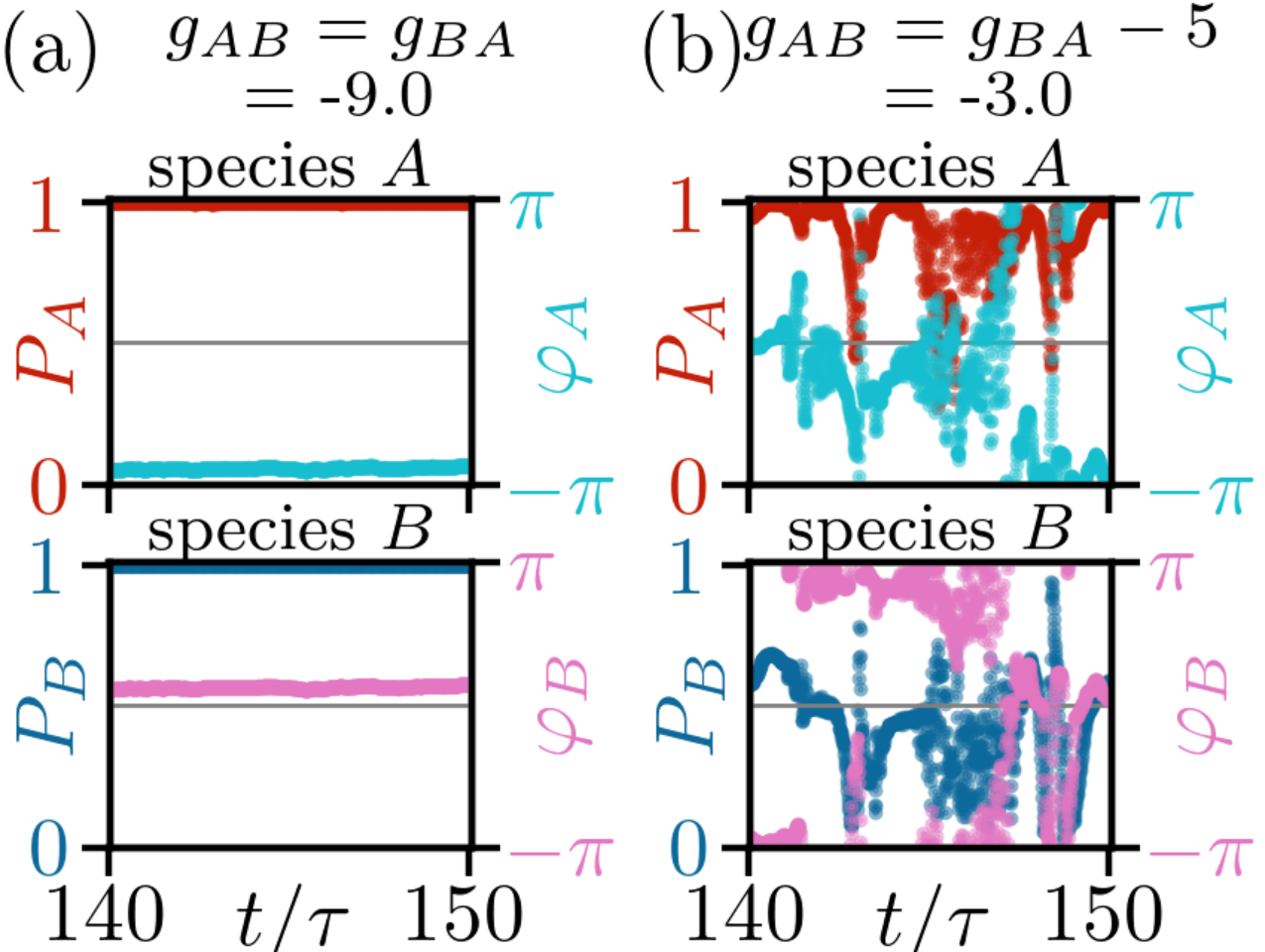}
	\caption{\label{fig:synchronization_over_time_antiflocking}Polarization $P_a$ and average phase $\varphi_a$ of particles of species $a=A,B$ in a single ensemble over time for (a) reciprocal antiflocking with $g_{AB}=g_{BA}=-9$ and (b) close to a critical exceptional point at $g_{AB} = g_{BA} -5 = -3$.}
\end{figure}

In the main text (Fig.~2), we show the characteristic time evolutions of the polarizations, $P_a(t)$, and average phases, $\varphi_a(t)$, for particles belonging to species $a=A,B$ for the cases of reciprocal flocking and non-reciprocal couplings with opposing alignment goals. Additionally, Fig.~\ref{fig:synchronization_over_time_antiflocking} shows the time evolutions of $P_a(t)$ and $\varphi_a(t)$ for the cases of reciprocal antiflocking and close to a critical exceptional point.
	
Fig.~\ref{fig:synchronization_over_time_antiflocking}(a) shows the polarizations and average phases in the antiflocking regime, with $g_{AB}=g_{BA}=-9$. Here, particles of the same species move coherently in the same direction, leading to $P_{A}(t) \approx P_{B}(t) \approx 1$. At the same time, particles of different species orient antiparallel to each other, with $\varphi_A(t) \approx \varphi_B(t) + \pi \approx \rm{const.}$. Consequently, the combined polarization of all particles vanishes, i.e., $P\approx 0$.

Fig.~\ref{fig:synchronization_over_time_antiflocking}(b) shows the time evolution close to a critical exceptional point at $g_{AB} = g_{BA} -5 = -3$ (see \ref{sec:exceptional_points}). Here, as discussed in the main text, $B$-particles get trapped inside $A$-clusters. Consequently, the polarization of $A$-particles is higher than that of $B$-particles. However, compared to the time evolutions in cases of weak and strong anti-symmetric couplings, shown in Figs.~2(b) and (c) in the main text, the polarizations and average phases of both species fluctuate heavily.

\subsection{\label{app:phase_shift}Phase shift}
In the main text, we discuss the emergence of large, rotating clusters for weak antisymmetric couplings, as illustrated in the BD simulation snapshot in Fig.~1(e). These clusters, which (almost only) consist either of particles of species $A$ or species $B$, rotate with a phase shift. Here, we take a closer look at this phase shift, defined as $\Delta \varphi = \varphi_A - \varphi_B$.

\begin{figure}
	\includegraphics[width=0.67\linewidth]{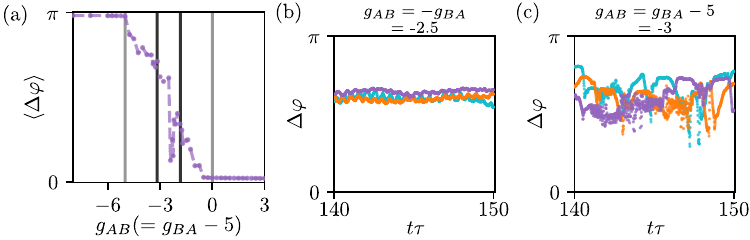}
	\caption{\label{fig:phase_shift}Phase shift $\Delta \varphi = \varphi_A - \varphi_B$ between $A$- and $B$-particles. (a) The circular mean of time- and ensemble-averaged phase shifts, $\langle \Delta \varphi \rangle$, as a function of $g_{AB} (= g_{BA} - 5)$. Phase shift $\Delta \varphi$ over time for (b) weak non-reciprocity and (c) close to critical exceptional points. Different colors indicate different noise realizations. In (a), the vertical gray and black lines indicate non-critical and critical exceptional points, respectively.}
\end{figure}

Fig.~\ref{fig:phase_shift}(a) shows the circular mean of the phase shift, $\langle \Delta \varphi \rangle$, as a function of $g_{AB} (= g_{BA} - 5)$, crossing exceptional points. In line with expectations based on the system's the polarization (see Fig.~5 in the main text), we find $\langle \Delta \varphi \rangle \approx \pi$ in the antiflocking phase and $\langle \Delta \varphi \rangle \approx 0$ in the flocking phase. Between the non-critical exceptional points, $\langle \Delta \varphi \rangle$ decreases from $\pi$ to $0$. However, this decrease is not monotonic, but exhibits a peak near the critical exceptional point close to the flocking regime.

In Fig.~\ref{fig:phase_shift}(b), we show that the phase shift $\Delta \varphi$ remains approximately constant over time in a single noise realization within the regime of weak antisymmetric couplings between species ($g_{AB} = -g_{BA} = -2.5$). In contrast, as shown in Fig.~\ref{fig:phase_shift}(c), near the critical exceptional point ($g_{AB} = g_{BA} - 5 = -3$), the phase shift exhibits strong temporal fluctuations.

\subsection{\label{app:distribution_of_phase_difference_rates}Distributions of phase difference rates}
\begin{figure}
	\includegraphics[width=0.5\linewidth]{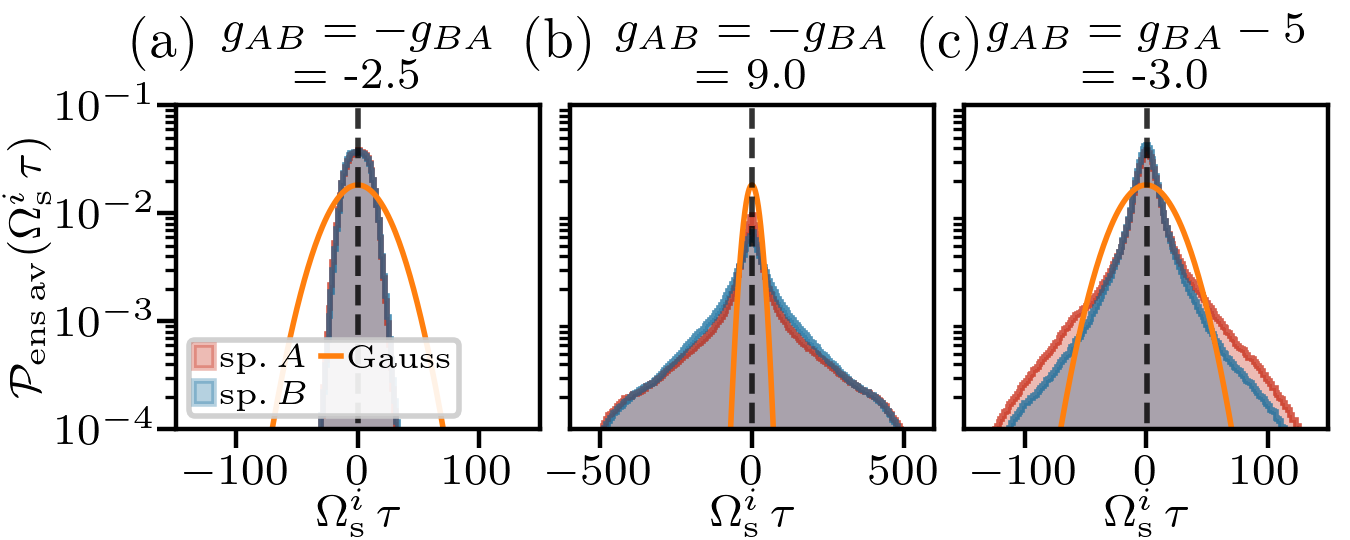}
	\caption{\label{fig:distribution_spontaneous_chirality_ensemble_average}Distribution $\mathcal{P}_{\rm ens\, av}(\Omega^i_{\rm s}\,\tau)$ of phase difference rates $\Omega^i_{\rm s}\,\tau$ averaged over two noise realizations. The orange line represents the Gaussian distribution induced by rotational noise alone. The black vertical dashed line marks the average $\langle \Omega^i_{\rm s}\rangle$.}
\end{figure}

In the main text (Fig.~3), we show the distribution $\mathcal{P}(\Omega^i_{\rm s}\,\tau)$ of phase difference rates $\Omega^i_{\rm s}\,\tau$ for a single noise realization in cases of weak and strong anti-symmetric couplings, as well as close to a critical exceptional point at $g_{AB} = g_{BA} -5 = -3$. In Fig.~\ref{fig:distribution_spontaneous_chirality_ensemble_average}, we additionally provide the \textit{ensemble-averaged} distributions.
	
The distributions for strong anti-symmetric couplings [Fig.~3(b) in main text and Fig.~\ref{fig:distribution_spontaneous_chirality_ensemble_average}(b) in SM] and near a critical exceptional point [Fig.~3(c) in main text and Fig.~\ref{fig:distribution_spontaneous_chirality_ensemble_average}(c) in SM] show little variation between single-noise realizations and ensemble averages. However, for weak anti-symmetric couplings, particles in a single-noise realization tend to rotate \textit{either} clockwise or counterclockwise, leading to a nonzero average phase difference rate, $\langle \Omega^i_{\rm s} \rangle \neq 0$ [Fig.~3(a) in main text]. When averaging over multiple noise ensembles, clockwise and counterclockwise rotations occur with equal probability, resulting in a mean of zero [Fig.~\ref{fig:distribution_spontaneous_chirality_ensemble_average}(a)].

\begin{figure}
	\includegraphics[width=0.5\linewidth]{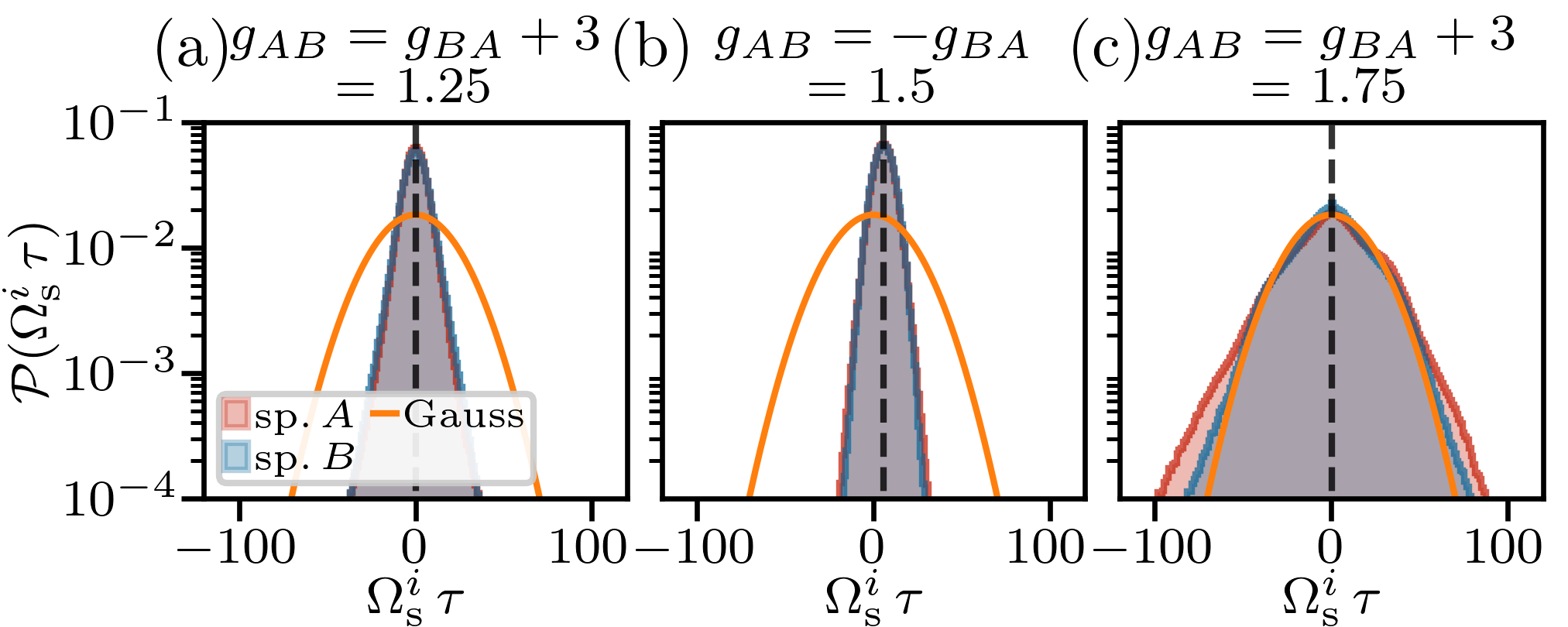}
	\caption{\label{fig:distribution_spontaneous_chirality_fd--3}Distribution $\mathcal{P}(\Omega^i_{\rm s}\,\tau)$ of phase difference rates $\Omega^i_{\rm s}\,\tau$ in a single noise realization for $g_{AB} = g_{BA} + 3$. The distribution in (b) corresponds to anti-symmetric couplings between the two species, while (a) and (c) show distributions close to critical exceptional points. The orange line indicates the Gaussian distribution induced by rotational noise alone. The black vertical dashed line indicates the average $\langle \Omega^i_{\rm s}\rangle$.}
\end{figure}

Furthermore, Fig.~\ref{fig:distribution_spontaneous_chirality_fd--3} shows the distribution $\mathcal{P}(\Omega^i_{\rm s}\,\tau)$ for a single noise realization along the path $g_{AB} = g_{BA} + 3$ in the phase diagram. As observed in the main text for $g_{AB} = g_{BA} - 5$ [Fig.~3(a) in main text], the distribution for weak anti-symmetric couplings is narrow and has a nonzero mean [Fig.~\ref{fig:distribution_spontaneous_chirality_fd--3}(b) in SM]. Close to critical exceptional points [Fig.~\ref{fig:distribution_spontaneous_chirality_fd--3}(a) and (c)], the distribution broadens significantly.

\subsection{\label{app:polarization_rotation_different_path}Polarization and spontaneous chirality along another path}
In the main text (Fig.~5), we present the time- and ensemble-averaged polarization and spontaneous chirality along the path $g_{AB}=g_{BA}-5$. Here, we provide results obtained along a different path in the stability diagram, defined by $g_{AB}=g_{BA}+3$.

\begin{figure}
	\includegraphics[width=0.5\linewidth]{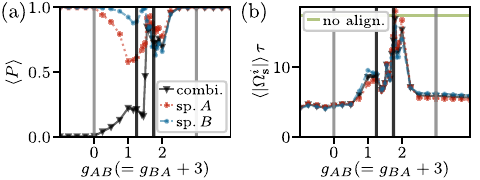}
	\caption{\label{fig:polarisation_chirality_rAlign-10_N-5000_fixed_difference--3}Polarization and spontaneous chirality as the system crosses exceptional points in non-reciprocal system with $g_{AB}=g_{BA}+3$. The vertical gray and black lines indicate non-critical and critical exceptional points, respectively.}
\end{figure}

Fig.~\ref{fig:polarisation_chirality_rAlign-10_N-5000_fixed_difference--3} shows the time- and ensemble-averaged polarization and spontaneous chirality along the path $g_{AB}=g_{BA}+3$. As discussed in the main text, the system transitions from antiflocking ($P\approx 0$) to flocking ($P\approx 1$) as $g_{AB}$ increases from negative to positive. Both of these non-equilibrium states exhibit coherent motion and, consequently, small values of $\langle \vert \Omega^i_{\rm s} \vert \rangle$. Between these states lies the regime of oscillatory $k=0$-instabilities, which is bounded by non-critical exceptional points. Within this regime, $P$ gradually increases from $0$ and $1$. Near the critical exceptional points, the polarization is reduced. The spontaneous chirality begins to increase once the system crosses the non-critical exceptional point, indicating enhanced particle rotation within the oscillatory instability regime. At the parameters associated with critical exceptional points, the spontaneous chirality exhibits peaks. This behavior qualitatively matches the results presented in the main text for $g_{AB}=g_{BA}-5$.

\subsection{\label{app:susceptibilities}Susceptibilities}
The susceptibility $\chi = N \, {\rm Var}(P)$ measures polarization fluctuations in the system. It can be determined as
\begin{equation}
	\chi(P) = \frac{1}{N} \left(\left\langle \Bigg\vert \sum_{i=1}^N \bm{p}_i \Bigg\vert \, \Bigg\vert \sum_{j=1}^N \bm{p}_j \Bigg\vert \right\rangle - \left\langle \Bigg\vert \sum_{i=1}^N \bm{p}_i \Bigg\vert \right\rangle^2\right)
\end{equation}
from individual particle orientation vectors $\bm{p}_i = (\cos(\theta_i), \sin(\theta_i))^{\rm T}$. Near flocking transitions, the susceptibility typically peaks \cite{Baglietto_Albano_2008_Finite-size_scaling_analysis_self-driven_indiviudals,kreienkamp_klapp_2024_dynamical_structures_phase_separating_nonreciprocal_polar_active_mixtures}. In the strong-coupling regime considered here, stationary \mbox{(anti)}flocking or oscillatory $k=0$-instabilities occur for all interspecies couplings. This means there is no ``standard'' flocking transition from a disordered to an ordered state.

\begin{figure}
	\includegraphics[width=0.5\linewidth]{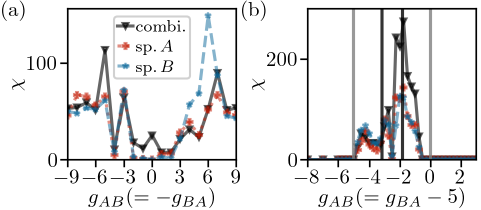}
	\caption{\label{fig:susceptibility}Susceptibility for (a) $g_{AB} = -g_{BA}$ and (b) $g_{AB} = g_{BA} - 5$. Exceptional points of the disordered and \mbox{(anti)}flocking base states are marked as gray and black vertical lines, respectively. The alignment interaction radius is $R_{\theta} = 10\,\ell$. Other parameters are specified in \ref{app:microscopic_model}.}
\end{figure}

In antisymmetric systems with $g_{AB}=-g_{BA} = \delta \neq 0$, the continuum theory predicts oscillatory $k=0$-instabilities, without crossing any exceptional points. In this case, the susceptibility calculated from BD simulations increases with the strength of nonreciprocity, $\vert \delta\vert$, as shown in Fig.~\ref{fig:susceptibility}(a).

For $g_{AB} = g_{BA} - d$ (with fixed $d \neq 0$), exceptional points of both the disordered and (anti)flocking base states are crossed upon varying $g_{AB}$. As shown in Fig.~\ref{fig:susceptibility}(b), the susceptibility is zero outside and non-zero within the regime of oscillatory instabilities, marked by the gray vertical lines. Interestingly, the susceptibility is largest between the critical exceptional points of the flocking and antiflocking states, indicated by the black vertical lines.

\subsection{\label{app:orientational_correlations}Orientational correlation functions}
\begin{figure}
	\includegraphics[width=0.77\linewidth]{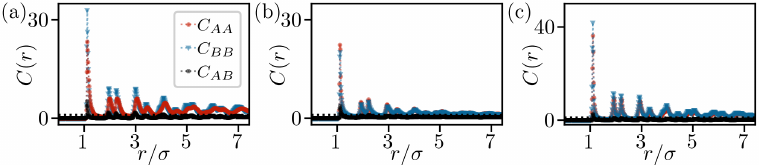}
	\caption{\label{fig:orientational_correlation}Orientational correlation function for the antisymmetric cases (a) $g_{AB} = -g_{BA} = 2$ and (b) $g_{AB} = -g_{BA} = 9$, and close to the exceptional point (c) $g_{AB}=g_{BA}-5= -3$.}
\end{figure}

Here we comment on the orientational spatial structure in systems with fully antisymmetric couplings $g_{AB}=-g_{BA}$ and close to exceptional points. These can be analyzed in terms of the orientational correlation function \cite{cavagna_grigera_2018_physics_of_flocking_correlations,kreienkamp_klapp_2024_dynamical_structures_phase_separating_nonreciprocal_polar_active_mixtures}
\begin{equation}
	C_{ab}(r) = \frac{1}{\mathcal{N}} \sum_{i}^{N_a} \sum_{j}^{N_b} \bm{p}_i^a \cdot \bm{p}_j^b \, \delta(r - \vert \bm{r}_i^a - \bm{r}_j^b \vert), 
\end{equation}
where $\mathcal{N} = N_{a}\,N_{b}\,4\,\pi\,r^2/V$ and $\bm{p}_i^a$, $\bm{p}_j^b$ denote the orientation vectors of particles $i$ and $j$ of species $a$ and $b$, respectively. In nonreciprocal mixtures, the single-species correlations generally differ, i.e., $C_{AA} \neq C_{BB}$ \cite{Bartnick_Loewen_2015_structural_correlations_in_nonreciprocal_systems,kreienkamp_klapp_2024_dynamical_structures_phase_separating_nonreciprocal_polar_active_mixtures}. The orientational correlation functions $C_{ab}(r)$ are shown in Fig.~\ref{fig:orientational_correlation} for different strengths of nonreciprocity.

For moderate nonreciprocity ($0 < \delta < 9$), the polarization of the species $B$ is generally larger compared to those of the species $A$ [Fig.~4(a) in main text]. Then, for $g_{AB} = -g_{BA} = \delta = 2$, the stronger synchronization of $B$ particles is reflected by $C_{BB} \geq C_{AA}$ [Fig.~\ref{fig:orientational_correlation}(a)]. As nonreciprocity increases, the correlation range of particle orientations decreases, see Fig.~\ref{fig:orientational_correlation}(b). For $\delta = 9$, asymmetric $A$-clustering leads to $C_{AA} \geq C_{BB}$ \cite{kreienkamp_klapp_2024_non-reciprocal_alignment_induces_asymmetric_clustering,kreienkamp_klapp_2024_dynamical_structures_phase_separating_nonreciprocal_polar_active_mixtures}. The orientational correlations $C_{AB}$ between different particle species are generally weak.

Finally, the orientational correlation function close to a critical exceptional point, $g_{AB}=g_{BA}-5=-3$, is shown in Fig.~\ref{fig:orientational_correlation}(c). For small distances $r\leq 3\,\sigma$, the magnitude of $C(r)$ is enhanced compared to the case of moderate nonreciprocity [Fig.~\ref{fig:orientational_correlation}(a)]. 

\section{\label{app:continuum_model}Continuum model}

\subsection{\label{sapp:full_continuum_equations}Continuum equations}
The continuum equations are derived from the microscopic Langevin Eqs.~(1a) and (1b) given the main text as outlined in \cite{kreienkamp_klapp_2022_clustering_flocking_chiral_active_particles_non-reciprocal_couplings,kreienkamp_klapp_2024_non-reciprocal_alignment_induces_asymmetric_clustering,kreienkamp_klapp_2024_dynamical_structures_phase_separating_nonreciprocal_polar_active_mixtures}. The evolution equation for the density field $\rho^a = \rho^a(\bm{r},t)$ of species $a$ is given as
\begin{equation}
	\label{eq:continuum_eq_density}
	\partial_t \rho^a +  \nabla \cdot \bm{j}_{a} = 0
\end{equation}
with flux
\begin{equation}
	\label{eq:continuity_flux}
	\bm{j}_a = v^{\rm eff}(\rho)\, \bm{w}^a -  D_{\rm t}\,\nabla\,\rho^a .
\end{equation} 
The polarization density $\bm{w}^a = \bm{w}^a(\bm{r},t)$ evolves like
{\small
	\begin{equation}
		\label{eq:continuum_eq_polarization}
		\begin{split}
			\partial_t \bm{w}^a  
			=& - \frac{1}{2} \, \nabla \,\big(v^{\rm eff}(\rho)\, \rho^a\big)  - D_{\rm r }\, \bm{w}^{a} + \sum_b g'_{ab} \, \rho^a\, \bm{w}^b 
			+  D_{\rm t}\,\nabla^2\,\bm{w}^a + \frac{v^{\rm eff}(\rho)}{16\,D_{\rm r}} \, \nabla^2\,\Big(v^{\rm eff}(\rho)\,\bm{w}^a \Big)
			- \sum_{b,c} \frac{ g'_{ab}\,g'_{ac}}{2\,D_{\rm r}} \, \bm{w}^a \, (\bm{w}^b \cdot \bm{w}^c) \\
			&- \frac{z}{16\,D_{\rm r}} \nabla\rho \cdot \big[\nabla \big(v^{\rm eff}(\rho)\,\bm{w}^a\big) - \nabla^* \big(v^{\rm eff}(\rho)\,\bm{w}^{a*} \big) \big]
			+ \sum_b \frac{g'_{ab}}{8 \,D_{\rm r}} \Big[ \bm{w}^b \cdot \nabla \big(v^{\rm eff}(\rho)\, \bm{w}^a\big) 
			+ \bm{w}^{b*} \cdot \nabla \big(v^{\rm eff}(\rho)\, \bm{w}^{a*}\big) \\
			& \qquad -2 \, \Big\{ v^{\rm eff}(\rho)\,\bm{w}^a \cdot \nabla \bm{w}^b 
			+\bm{w}^b \, \nabla \cdot \big(v^{\rm eff}(\rho)\,\bm{w}^a\big) - v^{\rm eff}(\rho)\, \bm{w}^{a*} \cdot \nabla \bm{w}^{b*} 
			- \bm{w}^{b*}\,  \nabla \cdot \big( v^{\rm eff}(\rho)\, \bm{w}^{a*}\big) \Big \}  \Big] ,
		\end{split}
\end{equation} }%
where $\bm{w}^*=(w_y, -w_x)^{\rm T}$ and $\nabla^*=(\partial_y, -\partial_x)^{\rm T}$. As explained in \cite{kreienkamp_klapp_2022_clustering_flocking_chiral_active_particles_non-reciprocal_couplings,kreienkamp_klapp_2024_non-reciprocal_alignment_induces_asymmetric_clustering,kreienkamp_klapp_2024_dynamical_structures_phase_separating_nonreciprocal_polar_active_mixtures}, the continuity Eq.~\eqref{eq:continuum_eq_density} for the particle densities describe that the motion of particles results from self-propulsion in direction $\bm{w}^a$ and translational diffusion of strength $D_{\rm t}$. Importantly, the self-propulsion velocity is not constant but depends on the density like $v^{\rm eff}(\rho)={\rm Pe}- z\,\rho$ with $\rho=\sum_b\rho^b$. The density-dependent velocity models the effect of steric repulsion, as it reflects the slowing down of particles in crowded situations, where free motion is hindered by others.
The main contributions to the evolution of polarization densities [Eq.~\eqref{eq:continuum_eq_polarization}] are the drift of particles towards low-density regions (first term on right-hand side), the decay of polarization induced by rotational diffusion (second term), and the orientational couplings between particles of all species (third term). The other contributions are of diffusional origin or non-linear, smoothing out low- and high-polarization regimes.  

The continuum equations are non-dimensionalized with characteristic time ($\tau$) and length ($\ell$) scales. Furthermore, $\rho^a$ and $\bm{w}^a$ are scaled with the average particle density $\rho_0^a$. The control parameters are the P\'eclet number ${\rm Pe}=v_0\,\tau/\ell$, the velocity-reduction parameter $z = \zeta\,\rho^a_0\,\tau/\ell$, the translational diffusion coefficient $D_{\rm t}=\xi\,\tau/\ell^2$, the rotational diffusion coefficient $D_{\rm r}=\eta\,\tau$, and the relative orientational coupling parameter $g'_{ab} = k_{ab}\,\mu_{\theta}\,R_{\theta}^2\,\rho_0^b\,\tau/2$. Note that $g'_{ab}$ scales with the average density $\rho^b$ and the alignment radius $R_{\theta}^2$. 

\subsection{\label{sapp:continuum_mean-field}Infinite-wavelength limit}
We are mainly interested in $k=0$-fluctuations of the polarization fields. These can be studied in a mean-field approximation of the full model \eqref{eq:continuum_eq_density}-\eqref{eq:continuum_eq_polarization} by neglecting all gradient terms. The mean-field ($k=0$-)polarization dynamics is given by \cite{fruchart_2021_non-reciprocal_phase_transitions,kreienkamp_klapp_2024_non-reciprocal_alignment_induces_asymmetric_clustering}
\begin{equation}
	\label{eq:mean-field_polarization_matrix_equation}
	\partial_t \begin{pmatrix}
		\bm{w}^A \\
		\bm{w}^B
	\end{pmatrix} = \begin{pmatrix}
		g'-D_{\rm r} - \frac{\bm{\mathcal{Q}}_A^2}{2 D_{\rm r}} & g'_{AB} \\
		g'_{BA} & g'-D_{\rm r} - \frac{\bm{\mathcal{Q}}_B^2}{2 D_{\rm r}}
	\end{pmatrix} \cdot \begin{pmatrix}
		\bm{w}^A \\
		\bm{w}^B
	\end{pmatrix}
\end{equation}
with
\begin{equation}
	\bm{\mathcal{Q}}_a = g'\,\bm{w}^a + g'_{ab}\,\bm{w}^b \ \text{with} \ b \neq a.
\end{equation}
Note that density fields are constant because they are conserved.

\subsection{\label{sapp:parameter_choice_continuum}Parameter choice}
Most parameters of the continuum model can be directly adopted from the considered particle simulation parameters. We use the same P\'eclet number, ${\rm Pe}=40$, and the same rotational diffusion constant, $D_r = \eta\,\tau = {3 \cdot 2^{-1/3}}$. The number density $\rho_0 = 2\, \rho_0^a = 4/\pi\,\Phi$, where $\rho^a_0 = 2/\pi\,\Phi$, is obtained from the area fraction $\Phi = 0.4$ in particle simulations. For the alignment radius $R_{\theta}=10\,\ell$ considered in the main text, the orientational coupling strengths on the continuum level ($g'_{ab}$) are related to those in the particle simulations ($g_{ab}$) via $g_{ab}'=12.73\,g_{ab}$. When we compare our results to the case with smaller $R_{\theta}=2\,\ell$, $g_{ab}'=0.51\,g_{ab}$. In this study, we focus on the strong-intraspecies-coupling regime with fixed $g_{AA}=g_{BB}=9$, while interspecies couplings, $g_{AB}$ and $g_{BA}$, are chosen independently.

The last two continuum parameters cannot be directly obtained from particle parameters. Yet, as explained in \cite{kreienkamp_klapp_2024_dynamical_structures_phase_separating_nonreciprocal_polar_active_mixtures}, the velocity reduction parameter, $\zeta$, can be determined from pair correlation functions. For the chosen P\'eclet number and area fraction, one obtains $z = \zeta \, \rho^a_0 \, \tau/\ell = 57.63 \, \rho^a_0 \, \tau/\ell = 0.37 \, {\rm Pe}/\rho_0^{\rm con}$ with $\rho_0^{\rm con} = 1$.
Further, we choose $D_{\rm t}=9$ in our continuum description \cite{kreienkamp_klapp_2024_dynamical_structures_phase_separating_nonreciprocal_polar_active_mixtures}.

\section{Field-theoretical results}
In this section, we provide details on the results obtained from the continuum model introduced in \ref{app:continuum_model}.

\subsection{\label{app:linear_stability_analysis_flocking_base_states}Mean-field linear stability analysis around (anti)flocking base states}
Linear stability analyses are analytical tools used to predict large-scale collective behavior in continuum systems. For the system at hand, the decay or growth of perturbations to the disordered base state, defined as $\rho^a_{\rm b} = 1$ and $\bm{w}_{\rm b}^a = \bm{0}$, has been previously studied in \cite{kreienkamp_klapp_2022_clustering_flocking_chiral_active_particles_non-reciprocal_couplings,kreienkamp_klapp_2024_non-reciprocal_alignment_induces_asymmetric_clustering,kreienkamp_klapp_2024_dynamical_structures_phase_separating_nonreciprocal_polar_active_mixtures}. These analyses predict the formation of polarized states, related to wavenumber $k=0$-perturbations, as well as clustering at $k>0$.

In this study, we are additionally focus on the linear stability of the (anti)flocking state. For simplicity, we assume that the \mbox{(anti)}flocking base state is orientated along the $x$-direction. The base state is then defined as $\rho^a_{\rm b} = 1$ and $\bm{w}_{\rm b}^a = (w_0^a, 0)^{\rm T}$, where $w_0^A$ and $w_0^B$ may differ. A flocking (antiflocking) base state is characterized by $w_0^A \, w_0^B >(<) 0$. Importantly, the constant base state needs to fulfill the (fixed point) continuum Eqs.~\eqref{eq:continuum_eq_density}-\eqref{eq:continuum_eq_polarization}.

We are interested in the time evolution of perturbations to these base states. If the perturbations decay over time, the considered base state is stable. On the other hand, growing perturbations indicate the instability of the base state. Unlike the isotropic disordered base state considered in previous works \cite{kreienkamp_klapp_2022_clustering_flocking_chiral_active_particles_non-reciprocal_couplings,kreienkamp_klapp_2024_non-reciprocal_alignment_induces_asymmetric_clustering,kreienkamp_klapp_2024_dynamical_structures_phase_separating_nonreciprocal_polar_active_mixtures}, the (anti)flocking base states are anisotropic. Thus, we need to distinguish between perturbations along the direction of the base state (longitudinal) and transversal to it. In the main text, we focus on $k=0$-perturbations. In \ref{app:linear_stability_analysis_flocking_base_states_larger_wavenumber}, we consider also the case of arbitrary $\bm{k}$.

We consider perturbations
\begin{subequations} \label{eq:perturbation_form}
	\begin{align}
		\delta\rho^a(\bm{r},t) &= \int \hat{\rho}^a(\bm{k}) \, {\rm e}^{i\bm{k}\cdot\bm{r} + \sigma(\bm{k})t} \, {\rm d}\bm{k}\\
		\bm{\delta w}^a(\bm{r},t) &= \int \bm{\hat{w}}^a(\bm{k}) \, {\rm e}^{i\bm{k}\cdot\bm{r} + \sigma(\bm{k})t} \, {\rm d}\bm{k},
	\end{align}
\end{subequations}
expressed as plane waves with complex growth rates $\sigma(\bm{k})$ and amplitudes $\hat{\rho}^a(\bm{k})$ and $\bm{\hat{w}}^a(\bm{k})$. The perturbations \eqref{eq:perturbation_form} involve all wave vectors $\bm{k}$. The growth rate $\sigma$ depends on the magnitude and direction of $\bm{k}$.

We then insert $\rho^a = \rho_{\rm b} + \delta\rho^a$ and $\bm{w}^a = \bm{w}_{\rm b}^a + \bm{\delta w}^a = (w_0^a, 0)^{\rm T} + \bm{\delta w}^a $ into the time evolution Eqs.~\eqref{eq:continuum_eq_density}-\eqref{eq:continuum_eq_polarization} and neglect perturbations of order $\delta^2$. Note that the (constant) terms cancel since the base state is a fixed point.

The resulting linearized time-evolution equation for perturbations to the density of species $A$ (for arbitrary $\bm{k}$) is given by
\begin{equation}
	\begin{split}
		\partial_t \, \delta\rho^A & = \delta\rho^A \Big[-D_{\rm t}\,k^2 + z \, w_0^A \,i\,k_x \Big] + \delta\rho^B \Big[z\,w_0^A\,i\,k_x \Big] 
		+ \delta w_x^A \Big[-i\,k_x\,v^{\rm eff}(\rho_0) \Big] + \delta w_y^A \Big[-i\,k_y\,v^{\rm eff}(\rho_0) \Big] \\
		& =: A_{\rho^A}^{\rho^A} \delta\rho^A + A_{\rho^A}^{\rho^B} \delta\rho^B + A_{\rho^A}^{w_x^A} \delta w_x^A 
		+ A_{\rho^A}^{w_y^A} \delta w_y^A + A_{\rho^A}^{w_x^B} \delta w_x^B + A_{\rho^A}^{w_y^B} \delta w_y^B .
	\end{split}
\end{equation}
The polarization can be perturbed either along or transversal to the direction of the base state. For polarizations perturbations along the base state, i.e., in $x$-direction, the time-evolution equation for species $A$ reads
\begin{equation}
	\begin{split}
		\partial_t\, \delta w_x^A & = A_{w_x^A}^{\rho^A} \delta\rho^A + A_{w_x^A}^{\rho^B} \delta\rho^B + A_{w_x^A}^{w_x^A} \delta w_x^A 
		+ A_{w_x^A}^{w_y^A} \delta w_y^A + A_{w_x^A}^{w_x^B} \delta w_x^B + A_{w_x^A}^{w_y^B} \delta w_y^B .
	\end{split}
\end{equation}
with 
\begin{equation}
	\begin{split}
		A_{w_x^A}^{\rho^A} &= -\tfrac{1}{2}\,i\,k_x\,\big(v^{\rm eff}(\rho_0) - z \, \rho_0^a \big) + g'_{AA}\,w_0^A   
		+ g'_{AB}\,w_0^B + z\,w_0^A\,k^2 + \frac{g'_{AA}}{8\,D_r} \, (w_0^A)^2 \, i\, k_x \, z 
		+ \frac{g'_{AB}}{8\,D_r} \, w_0^A \, w_0^B \, i\, k_x \, z ,
	\end{split}
\end{equation}
\begin{equation}
	\begin{split}
		A_{w_x^A}^{\rho^B} &= \tfrac{1}{2}\,i\,k_x\,z\,\rho_0^a + z \, w_0^A \, k^2 
		+ \frac{g'_{AA}}{8\,D_r} \, (w_0^A)^2 \, i\, k_x \, z + \frac{g'_{AB}}{8\,D_r} \, w_0^A \, w_0^B \, i\, k_x \, z ,
	\end{split}
\end{equation}
\begin{equation}
	\begin{split}
		A_{w_x^A}^{w_x^A} &= -D_r + g'_{AA}\,\rho_0^A - D_{\rm t}\,k^2 - \frac{(v^{\rm eff}(\rho_0))^2}{16\,D_r}\,k^2 
		- \frac{g_{AA}^{'2}}{2\,D_r}\,3\,(w_0^A)^2 - \frac{g_{AB}^{'2}}{2\,D_r}\,(w_0^B)^2 
		- \frac{g_{AA}'\,g_{AB}'}{D_r}\,2 \, w_0^A \, w_0^B \\
		& \quad  + \frac{g'_{AA}}{8\,D_r}\big\{-3\,w_0^A\,i\,k_x\,v^{\rm eff}(\rho_0)\big\} 
		+ \frac{g'_{AB}}{8\,D_r}\big\{-\,w_0^B\,i\,k_x\,v^{\rm eff}(\rho_0) \big\} ,
	\end{split}
\end{equation}
\begin{equation}
	\begin{split}
		A_{w_x^A}^{w_y^A} &= \frac{g'_{AA}}{8\,D_r}\big\{-5\,w_0^A\,i\,k_y\,v^{\rm eff}(\rho_0)\big\} 
		+ \frac{g'_{AB}}{8\,D_r}\big\{-3\,w_0^B\,i\,k_y\,v^{\rm eff}(\rho_0) \big\} ,
	\end{split}
\end{equation}
\begin{equation}
	\begin{split}
		A_{w_x^A}^{w_x^B} &= g'_{AB}\,\rho_0^A - \frac{g_{AA}'\,g_{AB}'}{D_r}\,(w_0^A)^2 - \frac{g_{AB}^{'2}}{2\,D_r} \, 2\, w_0^A \, w_0^B 
		+ \frac{g'_{AB}}{8\,D_r}\big\{-2\,w_0^A\,i\,k_x\,v^{\rm eff}(\rho_0) \big\} ,
	\end{split}
\end{equation}
and
\begin{equation}
	\begin{split}
		A_{w_x^A}^{w_y^B} &= \frac{g'_{AB}}{8\,D_r}\big\{-2\,w_0^A\,i\,k_y\,v^{\rm eff}(\rho_0) \big\} .
	\end{split}
\end{equation}
For polarization perturbations in $y$-direction, one obtains
\begin{equation}
	\begin{split}
		\partial_t\, \delta w_y^A 
		& = A_{w_y^A}^{\rho^A} \delta\rho^A + A_{w_y^A}^{\rho^B} \delta\rho^B + A_{w_y^A}^{w_x^A} \delta w_x^A 
		+ A_{w_y^A}^{w_y^A} \delta w_y^A + A_{w_y^A}^{w_x^B} \delta w_x^B + A_{w_y^A}^{w_y^B} \delta w_y^B .
	\end{split}
\end{equation}
with 
\begin{equation}
	\begin{split}
		A_{w_y^A}^{\rho^A} &= -\tfrac{1}{2}\,i\,k_y\,\big(v^{\rm eff}(\rho_0) - z \, \rho_0^a \big)
		+ \frac{g'_{AA}}{8\,D_r} \, (-3\,(w_0^A)^2) \, i\, k_y \, z 
		+ \frac{g'_{AB}}{8\,D_r} \, (-3\,w_0^A \, w_0^B) \, i\, k_y \, z ,
	\end{split}
\end{equation}
\begin{equation}
	\begin{split}
		A_{w_y^A}^{\rho^B} &= \tfrac{1}{2}\,i\,k_y\,z\,\rho_0^a + \frac{g'_{AA}}{8\,D_r} \, (-3\,(w_0^A)^2) \, i\, k_y \, z 
		+ \frac{g'_{AB}}{8\,D_r} \, (-3\,w_0^A \, w_0^B) \, i\, k_y \, z ,
	\end{split}
\end{equation}
\begin{equation}
	\begin{split}
		A_{w_y^A}^{w_x^A} &= \frac{g'_{AA}}{8\,D_r}\big\{5\,w_0^A\,i\,k_y\,v^{\rm eff}(\rho_0)\big\} 
		+ \frac{g'_{AB}}{8\,D_r}\big\{3\,w_0^B\,i\,k_y\,v^{\rm eff}(\rho_0) \big\} ,
	\end{split}
\end{equation}
\begin{equation}
	\begin{split}
		A_{w_y^A}^{w_y^A} &= -D_r + g'_{AA}\,\rho_0^A - D_{\rm t}\,k^2 - \frac{(v^{\rm eff}(\rho_0))^2}{16\,D_r}\,k^2 
		- \frac{g_{AA}^{'2}}{2\,D_r}\,(w_0^A)^2 - \frac{g_{AB}^{'2}}{2\,D_r}\,(w_0^B)^2 
		- \frac{g_{AA}'\,g_{AB}'}{D_r} \, w_0^A \, w_0^B\\
		& \quad  + \frac{g'_{AA}}{8\,D_r}\big\{-3\,w_0^A\,i\,k_x\,v^{\rm eff}(\rho_0)\big\} 
		+ \frac{g'_{AB}}{8\,D_r}\big\{-\,w_0^B\,i\,k_x\,v^{\rm eff}(\rho_0) \big\} ,
	\end{split}
\end{equation}
\begin{equation}
	\begin{split}
		A_{w_y^A}^{w_x^B} &= \frac{g'_{AB}}{8\,D_r}\big\{2\,w_0^A\,i\,k_y\,v^{\rm eff}(\rho_0) \big\} ,
	\end{split}
\end{equation}
and
\begin{equation}
	\begin{split}
		A_{w_y^A}^{w_y^B} &= g'_{AB}\,\rho_0^A + \frac{g'_{AB}}{8\,D_r}\big\{-2\,w_0^A\,i\,k_x\,v^{\rm eff}(\rho_0) \big\} .
	\end{split}
\end{equation}
The equations for species $B$ are obtained by exchanging $A \leftrightarrow B$.

To analyze the results, we consider perturbations in a different basis, that is, $\delta\rho^A \pm \delta\rho^B$ and $\delta\bm{w}^A \pm \delta\bm{w}^B$. These can be easily obtained by rewriting
\begin{equation}
	\delta\rho^A = \tfrac{1}{2} \big((\delta\rho^A + \delta\rho^B) + (\delta\rho^A - \delta\rho^B) \big)
\end{equation} 
and
\begin{equation}
	\delta\rho^B = \tfrac{1}{2} \big((\delta\rho^A + \delta\rho^B) - (\delta\rho^A - \delta\rho^B) \big).
\end{equation}
The time evolution of $\delta\rho^A \pm \delta\rho^B$ is then given by
\begin{equation}
	\begin{split}
		2\,\partial_t \, (\delta\rho^A \pm \delta\rho^B)
		=& \ (\delta\rho^A + \delta\rho^B)  \big(\big(A_{\rho^A}^{\rho^A} \pm A_{\rho^B}^{\rho^A} \big) + \big(A_{\rho^A}^{\rho^B} \pm A_{\rho^B}^{\rho^B} \big) \big) \\
		& + (\delta\rho^A - \delta\rho^B)  \big(\big(A_{\rho^A}^{\rho^A} \pm A_{\rho^B}^{\rho^A} \big) - \big(A_{\rho^A}^{\rho^B} \pm A_{\rho^B}^{\rho^B} \big) \big) \\
		& + (\delta w^A_x + \delta w^B_x) \big(\big(A_{\rho^A}^{w^A_x} \pm A_{\rho^B}^{w_x^A} \big) + \big(A_{\rho^A}^{w_x^B} \pm A_{\rho^B}^{w_x^B} \big) \big) \\
		& + {\rm other \ contributions \ from \ } (\delta\bm{w}^A \pm \delta\bm{w}^B). 
	\end{split}
\end{equation}
For $\delta\bm{w}^A \pm \delta\bm{w}^B$, the expressions can be obtained equivalently.

\subsection{Characterization of emerging states}
The emerging non-equilibrium states can be characterized in terms of eigenvalues and the eigenvector corresponding to the largest real eigenvalue.

For the disordered base state, the characterization follows the more detailed explanations in \cite{kreienkamp_klapp_2022_clustering_flocking_chiral_active_particles_non-reciprocal_couplings,kreienkamp_klapp_2024_non-reciprocal_alignment_induces_asymmetric_clustering,kreienkamp_klapp_2024_dynamical_structures_phase_separating_nonreciprocal_polar_active_mixtures}. The most important points are the following: Instabilities at wave number $k=0$ pertain to flocking or antiflocking instabilities. The eigenvector corresponding to the largest (real) growth rate then indicates whether flocking (in $(\bm{w}^A+\bm{w}^B)$-direction) or antiflocking (in $(\bm{w}^A-\bm{w}^B)$-direction) is predicted. At $k>0$, phase separation behavior comes into play. If the maximum growth rate is found at a finite $k>0$, phase separation is predicted. Phase separation can occur alone or in combination with flocking. We use the angle between perturbations of total density ($\rho^A+\rho^B$) and composition ($\rho^A-\rho^B$) at small $k$ to predict the type of phase separation. Both species can cluster symmetrically, demix or partially demix in form of asymmetric clustering of predominately one species. Imaginary growth rates with positive real parts indicate oscillatory instabilities, which are generally related to non-stationary behavior.

For the flocking and antiflocking base states, we reduce the complexity of our analysis. This is motivated by the mere fact that the analysis of the six-dimensional problem quickly becomes very complex. For the polarized base states, we restrict ourselves to statements concerning their general stability or instability regardless of the wavenumber at which the maximum of the growth rate is reached, and regardless of the direction of the eigenvector. Thus, we say that an antiflocking or flocking state is stable at $k=0(\geq 0)$ if the real parts of all six growth rates are smaller than zero at $k=0(\geq 0)$. Otherwise, the state is unstable.

In the main text, we focus on the stability of base states against $k=0$-perturbations and the appearance exceptional points (which appear only at $k=0$, see \ref{sec:exceptional_points}). The case of arbitrary $\bm{k}$ is considered in \ref{app:linear_stability_analysis_flocking_base_states_larger_wavenumber}.

\subsection{\label{sec:exceptional_points}Exceptional points}
Exceptional points have been related to parity-time symmetry breaking transitions in nonreciprocal scalar \cite{You_Baskaran_Marchetti_2020_pnas,suchanek_loos_2023_irreversible_fluctuations_emergence_dynamical_phases} and polar active systems \cite{fruchart_2021_non-reciprocal_phase_transitions}. They are defined as points where eigenvalues of a linear stability matrix coalesce and eigenvectors become parallel \cite{el-ganainy_demetrios_2018_non-Hermitian_physics_PT-symmetry_breaking}. When exceptional points coincide with a bifurcation, a point where a qualitative change in the system's behavior takes place, the exceptional point is referred to as ``critical exceptional point'' \cite{suchanek_loos_2023_entropy_production_nonreciprocal_Cahn-Hilliard}.

\begin{figure}
	\includegraphics[width=0.5\linewidth]{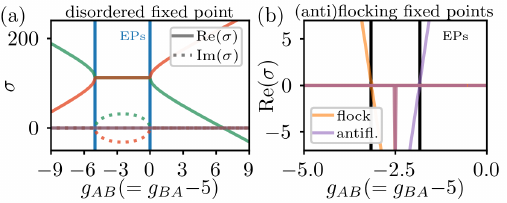}
	\caption{\label{fig:growth_rates_exceptional_points_rAlign-10}Growth rates at $k=0$ and exceptional points of the (a) disordered and (b) flocking and antiflocking base states for $g_{AB}=g_{BA}-5$ and $R_{\theta}=10\,\ell$. Exceptional points (EPs) are indicated as blue and black vertical lines in (a) and (b), respectively. The six eigenvalues of the disordered base state are all colored differently. Real parts are indicated as solid line, imaginary parts as dotted lines. The six eigenvalues of the flocking (antiflocking) base state are all colored in orange (purple). Other parameters are as specified in \ref{sapp:parameter_choice_continuum}.}
\end{figure}

Our system features three homogeneous base states. These are the disordered, flocking, and antiflocking base states. The growth rates of $k=0$-perturbations to these base states and corresponding exceptional points are shown for nonreciprocal systems with $g_{AB}=g_{BA}-d$ in Fig.~\ref{fig:growth_rates_exceptional_points_rAlign-10}.

We can make the following observations. Exceptional points only occur at wave number $k=0$. 
Starting from the disordered base state [Fig.~\ref{fig:growth_rates_exceptional_points_rAlign-10}(a)], exceptional points have a positive real growth rate, ${\rm Re}(\sigma^{\rm dis})>0$. Two pairs of eigenvalues meet at $g_{AB}=0$ and $g_{AB}=-d$, where we obtain two sets of two parallel eigenvectors. Between the exceptional points, the growth rates becomes imaginary.

For the flocking and antiflocking base states [Figs.~\ref{fig:growth_rates_exceptional_points_rAlign-10}(b)], exceptional points occur in combination with bifurcations, i.e., the largest real growth rates are ${\rm Re}(\sigma^{\rm fl})={\rm Re}(\sigma^{\rm antifl})=0$. Thus, these exceptional points are critical exceptional points. The imaginary part of the growth rate is zero at all $g_{AB}$. Both, flocking and antiflocking base states are stable between the exceptional points (${\rm Re}(\sigma^{\rm fl}),{\rm Re}(\sigma^{\rm antifl})\leq 0$). Then, at the exceptional point, either flocking or antiflocking becomes unstable with at least one ${\rm Re}(\sigma^{\rm fl/antifl})>0$.

\subsection{\label{app:mapping_coarse-graining}Coarse-grained density dynamics}
The orientational couplings between particles significantly influence their spatial distribution. For weak intraspecies alignment, nonreciprocal orientational couplings have been shown to induce asymmetric density behavior, characterized by the formation of single-species clusters \cite{kreienkamp_klapp_2024_non-reciprocal_alignment_induces_asymmetric_clustering,kreienkamp_klapp_2024_dynamical_structures_phase_separating_nonreciprocal_polar_active_mixtures}.

Also in the present case of strong intraspecies alignment, the particle-level results discussed in the main text reveal distinct clustering behaviors depending on the strength nonreciprocal orientational couplings (see Figs.~1). Specifically, weak antisymmetric couplings lead to the formation of two large, synchronized, rotating clusters, each composed predominantly of a single species. This behavior corresponds to near-complete demixing of particles based on their species. In contrast, strong antisymmetric couplings result in the formation of numerous smaller, less synchronized clusters, along with a single, larger, polarized cluster composed of either species $A$ (for $g_{AB}=-g_{BA}\gg 0$) or species $B$ (for $g_{AB}=-g_{BA}\ll 0$). This behavior closely resembles the asymmetric cluster formation observed at lower intraspecies coupling strengths \cite{kreienkamp_klapp_2024_non-reciprocal_alignment_induces_asymmetric_clustering}.

To quantify the degree of clustering at the mean-field continuum level, we introduce a further step of coarse-graining. To this end, we analyze the coarse-grained density dynamics under an adiabatic approximation of polarization fields. As shown in \cite{kreienkamp_klapp_2024_non-reciprocal_alignment_induces_asymmetric_clustering}, eliminating temporal and spatial derivatives, as well as higher-order moments of the polarization densities $\bm{w}^a$, allows us to describe the clustering behavior using simplified coarse-grained equations for the density of the two species alone.

In particular, the adiabatic elimination of $\bm{w}^a$ yields expressions for the polarization fields that depend only on the density fields, i.e., $\bm{w}_{\rm ad}^a = \bm{w}_{\rm ad}^a(\rho^A, \rho^B)$. This allows us to express the coarse-grained density dynamics as
	\begin{equation}
		\label{eq:coarse-grained_density_dynamics}
		\partial_t \rho^a = - \nabla \cdot (v^{\rm eff}(\rho) \, \bm{w}_{\rm ad}^a ) + D_{\rm t}\, \nabla^2 \rho^a .
\end{equation}
Linearizing this equation around the disordered base state yields the eigenvalue equation \cite{kreienkamp_klapp_2024_non-reciprocal_alignment_induces_asymmetric_clustering}
\begin{equation}
	\sigma_{\rho} \, \begin{pmatrix}
		\hat{\rho}_A + \hat{\rho}_B \\
		\hat{\rho}_A - \hat{\rho}_B
	\end{pmatrix} = \bm{\mathcal{M}}_{\rm ad} \cdot \begin{pmatrix}
	\hat{\rho}_A + \hat{\rho}_B \\
	\hat{\rho}_A - \hat{\rho}_B
\end{pmatrix} .
\end{equation}
For reciprocal couplings, the stability matrix $\bm{\mathcal{M}}_{\rm ad}$ is diagonal. However, for nonreciprocal couplings, with antisymmetric $g'_{AB} = g'_{BA} = \delta'$, it takes the form \cite{kreienkamp_klapp_2024_non-reciprocal_alignment_induces_asymmetric_clustering}
\begin{equation}
	\bm{\mathcal{M}}^{\rm nr}_{\rm ad} = \tfrac{V\, k^2}{2(\delta^2 + (D_{\rm r}-g')^2)}  \begin{pmatrix}
		(g'- D_{\rm r}) (V-2\,z) & V\,\delta \\
		-(V-2\,z)\,\delta & (g'- D_{\rm r})V
	\end{pmatrix}.
\end{equation}
By computing the eigenvalues and eigenvectors of $\bm{\mathcal{M}}_{\rm ad}$, we can determine the existence, type and symmetry of the density instability. Note that the matrix elements exhibit a simple $k^2$-dependence on the wavenumber. The following remarks therefore hold for any $k>0$.

A positive real part of the eigenvalue, i.e., ${\rm Re}(\sigma_{\rho})>0$, indicates a density instability associated with phase separation. The type and symmetry of the density instability are encoded in the eigenvector $\bm{v}_{\rho}=(\hat{\rho}_A + \hat{\rho}_B, \hat{\rho}_A - \hat{\rho}_B)^{\rm T}$ that corresponds to the largest eigenvalue in the coarse-grained density dynamics \cite{kreienkamp_klapp_2024_non-reciprocal_alignment_induces_asymmetric_clustering}. This eigenvector defines the clustering angle $\alpha = {\rm arccos}(\bm{v}_{\rho} \cdot (1,0)^{\rm T})$, which provides a quantitative measure of the type and symmetry of emerging clusters. Specifically, symmetric clustering of both species is indicated by $\alpha=0$. Full symmetric demixing corresponds to $\alpha = \pm \pi/2$. Asymmetric clustering of species $A$ $(B)$ is indicated by $0 < \alpha < \pi/2$ $(\pi/2 < \alpha < \pi)$.

We obtain the clustering angles shown in Fig.~7 of the main text for antisymmetric couplings with $g_{AB}=-g_{BA}$. As $g_{AB}$ increases from negative to positive, the coarse-grained density dynamics predict a transition from asymmetric $B$-clustering to demixing to asymmetric $A$-clustering. These predictions qualitatively agree with particle simulations, which reveal demixing for small antisymmetric couplings and asymmetric clustering for larger antisymmetric couplings.

This agreement is particularly notable given that the derivation of the coarse-grained density dynamics assumes the absence of orientational order. However, as explained in \ref{app:linear_stability_analysis_flocking_base_states_larger_wavenumber}, in the strong-intraspecies-alignment regime considered here, polarization dynamics dominate over finite-wavelength perturbations in the full hydrodynamic description. Therefore, asymmetric clustering can only be captured at the level of coarse-grained density dynamics and remains ``hidden'' in the full dynamical description that includes the polarization dynamics.

\subsection{\label{app:linear_stability_analysis_flocking_base_states_larger_wavenumber}Results for finite-wavelength perturbations}
We now turn back to the collective dynamics of the full, six-dimensional problem. In the main text, we ignore perturbations of wavenumber $k>0$ and focus on infinite-wavelength instabilities with $k=0$. In this limit, density and polarization dynamics decouple. This is not anymore the case at $k>0$, where we have to consider the full six-dimensional problem. The consequence is that the stability analysis against perturbations of arbitrary wavenumbers in arbitrary directions (that is, longitudinal or transversal to the direction of polarization) quickly becomes very complex. In the following, we therefore only make statements about the general stability or instability of the flocking and antiflocking base state, without going into details of their more specific characteristics.

\begin{figure}
	\includegraphics[width=0.5\linewidth]{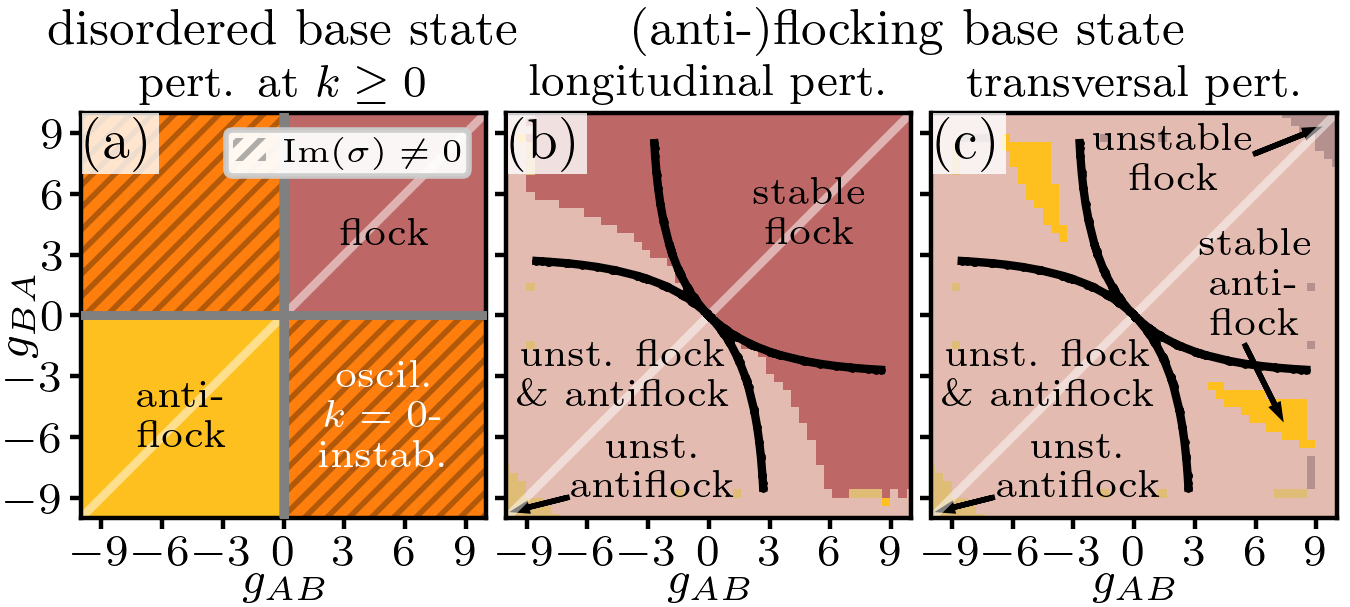}
	\caption{\label{fig:diagrams_arbitrary_k_Rtheta-10}Stability diagrams for arbitrary $k\geq 0$-instabilities. The diagrams are obtained from linear stability analyses of the (a) isotropic disordered and (b,c) homogeneous (anti)flocking base states of the continuum Eqs.~\eqref{eq:continuum_eq_density}-\eqref{eq:continuum_eq_polarization}. Perturbations to the (anti)flocking base states can be (b) longitudinal or (c) transversal to the base state. Exceptional points of the disordered and (anti)flocking base states are indicated as gray and black lines, respectively. The alignment radius is $R_{\theta}=10\,\ell$. Other parameters are specified in \ref{sapp:parameter_choice_continuum}.}
\end{figure} 

The non-equilibrium phase diagram for perturbations of arbitrary $k\geq 0$ is shown in Fig.~\ref{fig:diagrams_arbitrary_k_Rtheta-10}. The stability of the disordered base state [Fig.~\ref{fig:diagrams_arbitrary_k_Rtheta-10}(a)] is identical for $k\geq 0$ and $k=0$. This indicates that the instabilities of the disordered state are dominated by $k=0$-instabilities, which are related to polarization dynamics. We note that, for smaller alignment radii and thus reduced effective interactions, additional (a)symmetric clustering instabilities arise, see \ref{app:dynamics_alignment_radius}. 

The stability of the flocking and antiflocking states against longitudinal and transversal $k\geq 0$-perturbations is shown in Figs.~\ref{fig:diagrams_arbitrary_k_Rtheta-10}(b,c). The phase diagrams incorporate the $k=0$-case of the main text, but contain a more detailed picture with additional instabilities. In particular, different to the case with just $k=0$-perturbations, the phase diagrams indicate that neither homogeneous flocking nor homogeneous antiflocking are stable against perturbations. This holds for arbitrary wavenumbers, arbitrary direction, and all coupling strengths. Although homogeneous flocking is stable against longitudinal perturbations in the triangular region in the upper right of phase diagram, it is unstable against transversal perturbations throughout the phase diagram. Homogeneous antiflocking, on the other hand, is unstable against longitudinal and transversal perturbations at almost all coupling strengths, including, in particular, $g_{AB} \approx g_{BA} < 0$.

These continuum predictions are consistent with the particle simulation results shown in the snapshots in Fig.~1(c,d) in the main text. The reciprocal flocking state [Fig.~1(c)] exhibits a more homogeneous density distribution, while the reciprocal antiflocking state [Fig.~1(d)] is characterized by unstable, bending bands.

Finally, we consider fully antisymmetric couplings ($g_{AB}=-g_{BA}$). For weak antisymmetric couplings, large, synchronized, rotating clusters emerge, consisting almost entirely of a single species. In this regime, where nonreciprocity is weak, both homogeneous flocking and antiflocking are unstable against longitudinal and transversal perturbations, as shown in Figs.~\ref{fig:diagrams_arbitrary_k_Rtheta-10}(b) and (c). At the particle level, the resulting clusters are nearly circular.

In contrast, for strong antisymmetric couplings, less synchronized behavior emerges. Additionally, asymmetric clusters of either species $A$ (for $g_{AB}=-g_{BA}\gg 0$) or species $B$ (for $g_{AB}=-g_{BA}\ll 0$) form. This asymmetric density behavior is discussed in more detail in \ref{app:mapping_coarse-graining} and in \cite{kreienkamp_klapp_2024_non-reciprocal_alignment_induces_asymmetric_clustering,kreienkamp_klapp_2024_dynamical_structures_phase_separating_nonreciprocal_polar_active_mixtures}. Importantly, in this strong-nonreciprocity regime, the homogeneous flocking state is stable against longitudinal perturbations but remains unstable against transversal perturbations. This is in accordance with particle simulations, which show that the asymmetric single-species clusters are highly polarized. They are elongated along the direction of motion and compressed perpendicular to it.

\section{\label{app:dynamics_alignment_radius}Smaller alignment radius: continuum versus particle scale}
In the main text, we focus on a system with relatively large radius of alignment interactions, $R_{\theta}=10\,\ell$. Here, we show that also at smaller alignment radii, spontaneous chirality emerges due to nonreciprocal couplings. However, reducing the interaction radius decreases the ``effective'' alignment strength, which scales as $g'_{ab} \sim R_{\theta}^2$ on the continuum level. This reduction has consequences on both the continuum and particle levels, which we address in the following.

\begin{figure}
	\includegraphics[width=0.5\linewidth]{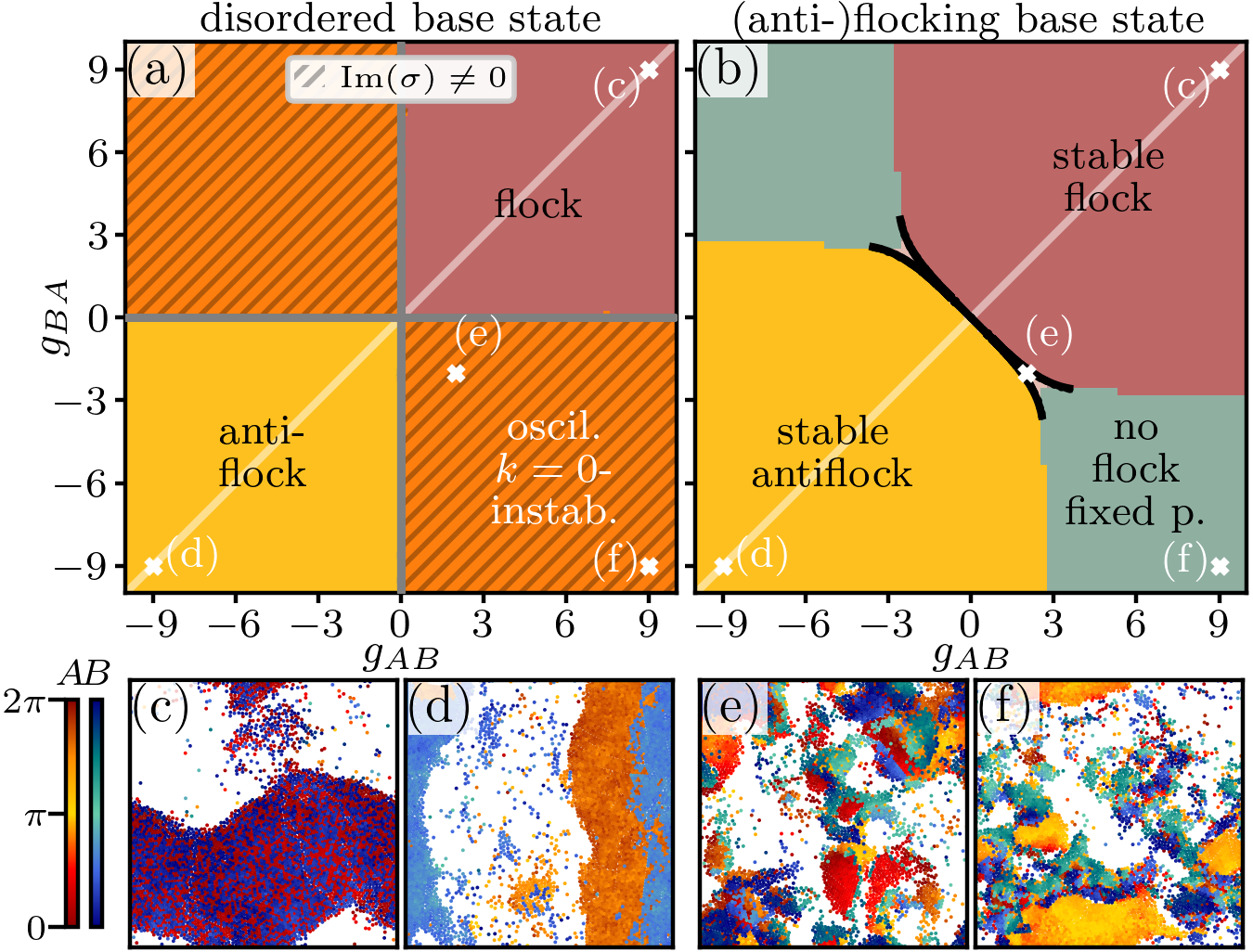}
	\caption{\label{fig:stability_diagram_with_snapshots_Rtheta-2_mean-field}Stability diagrams at $k=0$ and particle simulation snapshots for alignment radius $R_{\theta}=2\,\ell$. The stability diagrams are obtained from linear stability analyses of the (a) uniform disordered and (b) homogeneous (anti)flocking base states of the $k=0$-continuum Eq.~\eqref{eq:mean-field_polarization_matrix_equation}. Exceptional points of the disordered and (anti)flocking base states are indicated as gray and black lines, respectively. BD simulation snapshots at (c) $g_{AB}= g_{BA}=9$, (d) $g_{AB}= g_{BA}=-9$, (e) $g_{AB}=-g_{BA}=2$, and (f) $g_{AB}=-g_{BA}=9$. Other parameters are specified in \ref{sapp:parameter_choice_continuum}. The color code in (c) indicates the particle type and orientation.}
\end{figure}

\begin{figure}
	\includegraphics[width=0.5\linewidth]{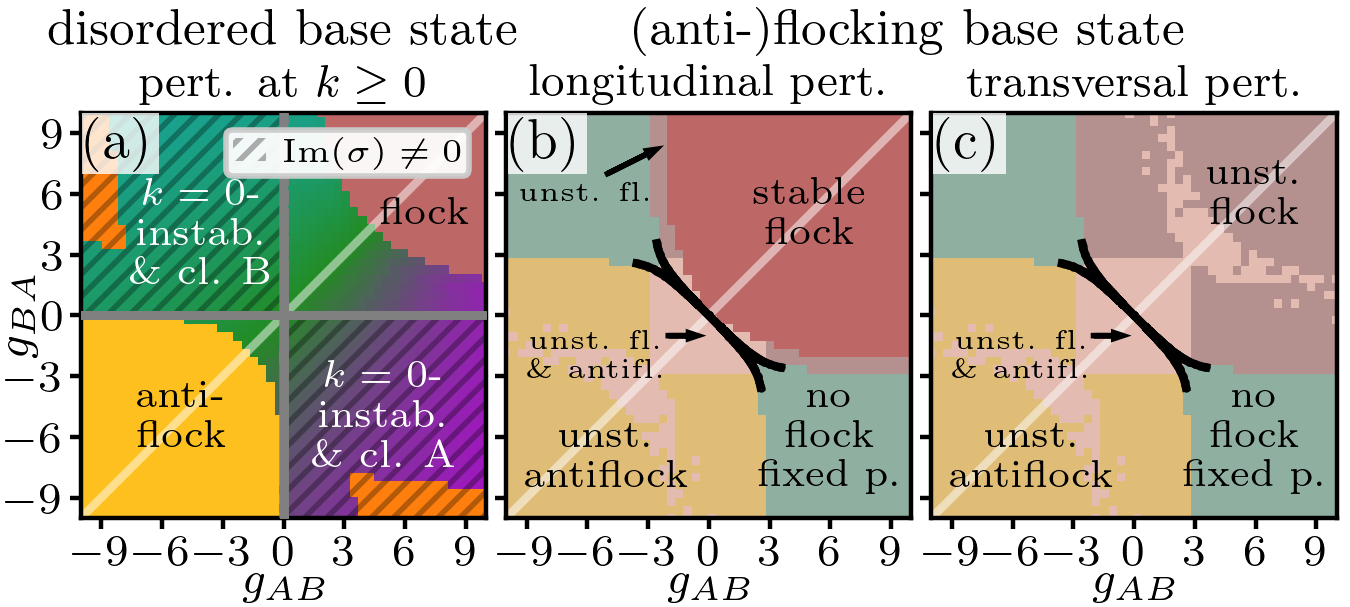}
	\caption{\label{fig:diagrams_arbitrary_k_Rtheta-2}Stability diagrams for arbitrary $k\geq 0$-instabilities for alignment radius $R_{\theta}=2\,\ell$. The stability diagrams are obtained from linear stability analyses of the (a) isotropic disordered and (b,c) homogeneous (anti)flocking base states of the continuum Eqs.~\eqref{eq:continuum_eq_density}-\eqref{eq:continuum_eq_polarization}. Perturbations to the (anti)flocking base states can be (b) longitudinal and (c) transversal to the direction of the base state. Exceptional points of the disordered and (anti)flocking base states are indicated as gray and black lines, respectively. Other parameters are specified in \ref{sapp:parameter_choice_continuum}.}
\end{figure} 

On the continuum level, there are indeed distinct differences in the predicted non-equilibrium behavior when we consider smaller $R_{\theta}$, e.g., $R_{\theta}=2\,\ell$. The corresponding non-equilibrium phase diagrams for $k=0$ and $k\geq 0$ are shown in Figs.~\ref{fig:stability_diagram_with_snapshots_Rtheta-2_mean-field} and \ref{fig:diagrams_arbitrary_k_Rtheta-2}, respectively. The disordered base state [Fig.~\ref{fig:stability_diagram_with_snapshots_Rtheta-2_mean-field}(a)] exhibits stationary flocking and antiflocking \mbox{($k=0$-)}instabilities, as well as oscillatory $k=0$-instabilities when species have opposite alignment goals ($g_{AB}\,g_{BA}<0$), exactly like for larger $R_{\theta}$. However, finite-wavelength perturbations with $k>0$ [Fig.~\ref{fig:diagrams_arbitrary_k_Rtheta-2}(a)] predict additional clustering instabilities, which are suppressed for larger $R_{\theta}$ [Fig.~\ref{fig:diagrams_arbitrary_k_Rtheta-10}(a)]. The phenomenon of nonreciprocity-induced asymmetric clustering and its origin is discussed in \cite{kreienkamp_klapp_2024_non-reciprocal_alignment_induces_asymmetric_clustering,kreienkamp_klapp_2024_dynamical_structures_phase_separating_nonreciprocal_polar_active_mixtures}. The predicted clustering instabilities qualitatively agree with particle simulation results, where snapshots [Figs.~\ref{fig:stability_diagram_with_snapshots_Rtheta-2_mean-field}(e,f)] show the formation of single-species clusters of species $A$ for $g_{AB}=-g_{BA}=\delta >0$. 

Perturbations to the (anti)flocking base states reveal differences compared to larger $R_{\theta}$ already at $k=0$, see Fig.~\ref{fig:stability_diagram_with_snapshots_Rtheta-2_mean-field}(b). When species have strongly opposing alignment goals, neither flocking nor antiflocking states are fixed point solutions to the continuum Eq.~\eqref{eq:mean-field_polarization_matrix_equation}. The regimes of stable flocking and antiflocking are separated from the regimes without (anti)flocking fixed points by exceptional points [black dots in Figs.~\ref{fig:stability_diagram_with_snapshots_Rtheta-2_mean-field}(a,b)].
For $k\geq 0$, the non-equilibrium phase diagrams at $R_{\theta}=2\,\ell$ for longitudinal and transversal perturbations are shown in Figs.~\ref{fig:diagrams_arbitrary_k_Rtheta-2}(b,c). The antiflocking base state is unstable to both types of perturbations, whereas the flocking base state is stable against longitudinal but unstable to transversal ones.

These continuum theory results conform with particle simulations, as seen in the snapshots in Figs.~\ref{fig:stability_diagram_with_snapshots_Rtheta-2_mean-field}(c,d). At large $g_{AB},g_{BA}>0$, flocking is predicted and highly polarized bands are observed at the particle level. The formation of these bands confirms the flocking state's stability against longitudinal and instability against transversal perturbations. In contrast, the bands in the antiflocking state at large $g_{AB},g_{BA}<0$ are more dynamic, reflecting the instability against both longitudinal and transversal perturbations.

\begin{figure}
	\includegraphics[width=0.5\linewidth]{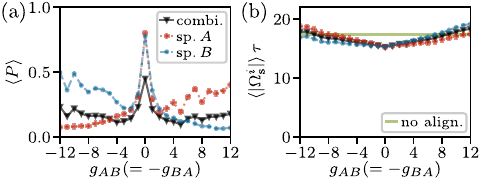}
	\caption{\label{fig:polarisation_chirality_anti-symmetric_rAlign-2}Polarization and spontaneous chirality for antisymmetric system with $g_{AB}=-g_{BA}$ and $R_{\theta}=2\,\ell$. The data points represent ensemble and time averages of the polarization and spontaneous chirality for species $A$ (red), species $B$ (blue), and all particles combined (black). In (b), the green line shows the spontaneous chirality in a system without alignment couplings.}
\end{figure}

\begin{figure}
	\includegraphics[width=0.5\linewidth]{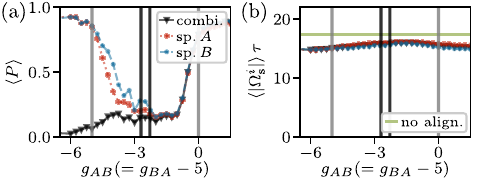}
	\caption{\label{fig:polarisation_chirality_fixed_difference_rAlign-2}Polarization and spontaneous chirality for antisymmetric system with $g_{AB}=g_{BA}-5$ and $R_{\theta} = 2\,\ell$. Exceptional points of the disordered and \mbox{(anti)}flocking base states are indicated as gray and black vertical lines, respectively. The data points represent ensemble and time averages of the polarization and spontaneous chirality for species $A$ (red), species $B$ (blue), and all particles combined (black). In (b), the green line shows the spontaneous chirality in a system without alignment couplings.}
\end{figure}

To quantitatively analyze the effect of nonreciprocity at smaller alignment radii, we examine the time- and noise-averaged polarization and spontaneous chirality for antisymmetric couplings in Fig.~\ref{fig:polarisation_chirality_anti-symmetric_rAlign-2} and for cases crossing exceptional points in Fig.~\ref{fig:polarisation_chirality_fixed_difference_rAlign-2}. 

In systems with antisymmetric couplings, $g_{AB} = -g_{BA} = \delta$, the combined polarization remains small ($P_{\rm combi} < 0.3$) for all non-zero $\delta$ [Fig.~\ref{fig:polarisation_chirality_anti-symmetric_rAlign-2}(a)]. As seen in the snapshot in Fig.~\ref{fig:stability_diagram_with_snapshots_Rtheta-2_mean-field}(e), weak nonreciprocity ($\delta = 2$) leads to the formation of many small single-species clusters, predominantly of the more aligning species $A$. Thus, for small $\delta > 0$, $P_{A}> P_{B}$. This behavior contrasts with systems with $R_{\theta}=10\,\ell$. There, weak nonreciprocity produces one large polarized cluster for both species each. The antialigning particles trap some aligning particles inside their cluster, leading to $P_{B}> P_{A}$ for small $\delta > 0$ [Fig.~1(e) in the main text]. At larger nonreciprocity [$\delta = 9$, Figs.~1(f) in the main text and \ref{fig:stability_diagram_with_snapshots_Rtheta-2_mean-field}(f) in this SM], large polarized single-species clusters form regardless of the alignment radius, leading to $P_{A}> P_{B}$ for $\delta > 0$. 
The nonreciprocity-induced spontaneous chirality at $R_{\theta}=2\,\ell$ [Fig.~\ref{fig:polarisation_chirality_anti-symmetric_rAlign-2}(b)], increases with increasing strength $\vert \delta \vert$ of nonreciprocity, but remains below the noise-induced chirality for $\vert \delta \vert < 9$. Only at higher $\delta \geq 9$, it exceeds the noise-induced chirality.

To summarize, the nonreciprocity-induced chirality is weaker for $R_{\theta}=2\,\ell$ compared to $R_{\theta}= 10\,\ell$, but it remains measurable. However, the signatures of exceptional points become less prominent for $R_{\theta}=2\,\ell$. In oscillatory instability regime, the polarization increases with increasing $g_{AB} = (g_{BA} - 5)$ [Fig.~\ref{fig:polarisation_chirality_fixed_difference_rAlign-2}(a)], but clear peaks in the spontaneous chirality near the critical exceptional points are not observed [Fig.~\ref{fig:polarisation_chirality_fixed_difference_rAlign-2}(b)].

In the main text we therefore focuses on the larger coupling radius, $R_{\theta}=10\,\ell$, where effective interactions between particles are stronger and the effects of nonreciprocity are more pronounced. 

\section{\label{sec:no_repulsion}Effect of repulsion}
In the main text, we focus on systems that include repulsive interactions. As a reference, we here present continuum and BD simulation results for systems without repulsion.

\subsection{Continuum results without repulsion}
On the continuum level, the effect of repulsion is captured through the effective density-dependent velocity $v^{\rm eff} = {\rm Pe} - z\,\rho$, see \ref{app:continuum_model}. In repulsive systems, the self-propulsion velocity of particles is reduced in crowded situations, where free motion is hindered by the presence of neighboring particles. In the absence of repulsion, the particle velocities remain unaffected by the local density. Thus, on the continuum level, the absence of repulsion is modeled by setting the velocity-reduction parameter to $z=0$.

\begin{figure}
	\includegraphics[width=0.5\linewidth]{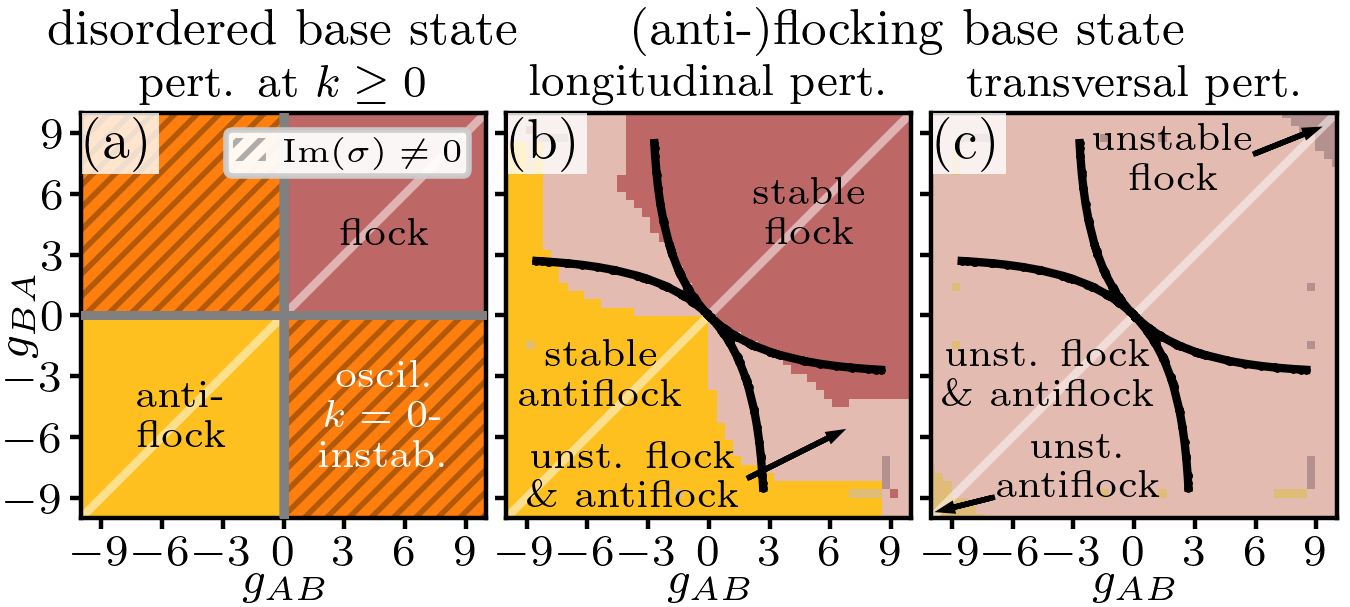}
	\caption{\label{fig:diagrams_arbitrary_k_Rtheta-10_no_repulsion}Stability diagrams for arbitrary $k\geq 0$-instabilities in the absence of repulsion. The diagrams are obtained from linear stability analyses of the (a) isotropic disordered and (b,c) homogeneous (anti)flocking base states of the continuum Eqs.~\eqref{eq:continuum_eq_density}-\eqref{eq:continuum_eq_polarization} with $z=0$. Perturbations to the (anti)flocking base states can be (b) longitudinal or (c) transversal to the base state. Exceptional points of the disordered and (anti)flocking base states are indicated as gray and black lines, respectively. The alignment radius is $R_{\theta}=10\,\ell$. Other parameters are specified in \ref{sapp:parameter_choice_continuum}.}
\end{figure} 

The stability diagrams for finite-wavelength perturbations in the absence of repulsion are shown in Fig.~\ref{fig:diagrams_arbitrary_k_Rtheta-10_no_repulsion}. Overall, these phase diagrams resemble those for repulsive systems (see Fig.~\ref{fig:diagrams_arbitrary_k_Rtheta-10}). In particular, the stability of the disordered base state remains unchanged regardless of whether repulsion is present [compare Figs.~\ref{fig:diagrams_arbitrary_k_Rtheta-10_no_repulsion}(a) and \ref{fig:diagrams_arbitrary_k_Rtheta-10}(a)]. The reason for this is that the stability of the disordered base state is dominated by polarization instabilities, which only occur at $k=0$, whereas density instabilities appear only at finite $k>0$.

The stability of the (anti)flocking base states against longitudinal perturbations is shown in Fig.~\ref{fig:diagrams_arbitrary_k_Rtheta-10_no_repulsion}(b). Similar to the repulsive case [Fig.~\ref{fig:diagrams_arbitrary_k_Rtheta-10}(b)], homogeneous flocking is stable for $g_{AB}, g_{BA} \gtrsim 0$. However, while homogeneous antiflocking is never stable against longitudinal perturbations in systems with repulsion, it is stable in systems without repulsion for $g_{AB}, g_{BA} \lesssim 0$.

The stability of (anti)flocking base states against transversal perturbations is shown in Fig.~\ref{fig:diagrams_arbitrary_k_Rtheta-10_no_repulsion}(c). Here, we find no stable (anti)flocking, similar to the repulsive case [Fig.~\ref{fig:diagrams_arbitrary_k_Rtheta-10}(c)], where antiflocking stability is only observed in small parameter regions.

\begin{figure}
	\includegraphics[width=1\linewidth]{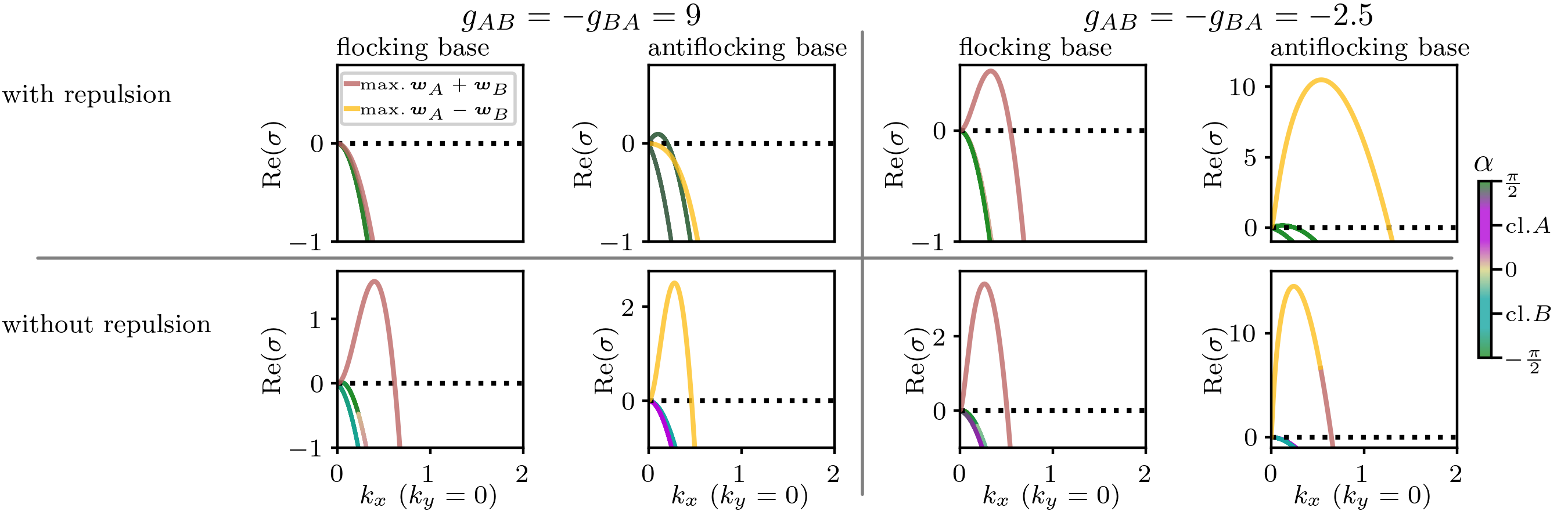}
	\caption{\label{fig:growth_rates_longitudinal}Growth rates of longitudinal perturbations to flocking and antiflocking base states for systems with and without repulsion for two different parameter combinations between the critical exceptional points. The growth rates are shown as functions of the wavenumber $k_x$ (longitudinal to the (anti)flocking direction), while we set $k_y=0$. The colors indicate the respective eigenvector direction. The alignment radius is $R_{\theta}=10\,\ell$. Other parameters are specified in \ref{sapp:parameter_choice_continuum}.}
\end{figure}

\begin{figure}
	\includegraphics[width=1\linewidth]{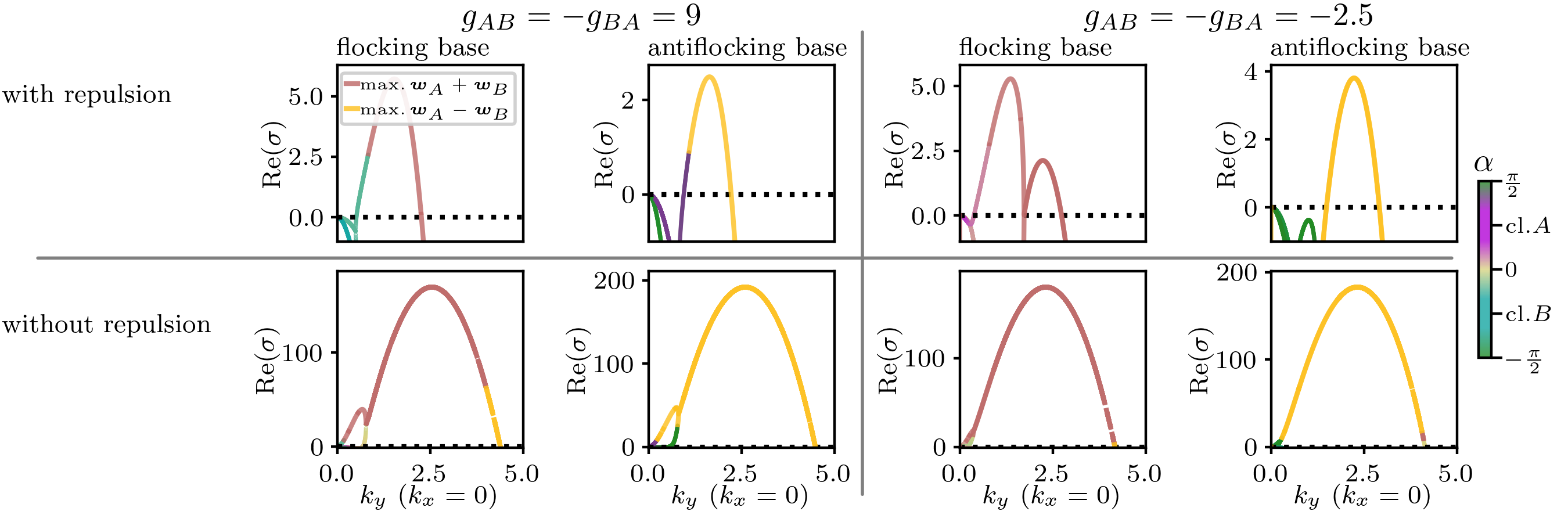}
	\caption{\label{fig:growth_rates_transversal}Growth rates of transversal perturbations to flocking and antiflocking base states for systems with and without repulsion for two different parameter combinations between the critical exceptional points. The growth rates are shown as functions of the wavenumber $k_y$ (transversal to the (anti)flocking direction), while we set $k_x=0$. The colors indicate the respective eigenvector direction. The alignment radius is $R_{\theta}=10\,\ell$. Other parameters are specified in \ref{sapp:parameter_choice_continuum}.}
\end{figure}

To further compare the characteristics of the stability of (anti)flocking states with and without repulsion, we show the full growth rates as functions of the wavenumber of longitudinal ($k_x$) and transversal ($k_y$) perturbations in Figs.~\ref{fig:growth_rates_longitudinal} and \ref{fig:growth_rates_transversal}. We focus on two different parameter sets, chosen between the critical exceptional points.

There are some differences between systems with and without repulsion. In non-repulsive systems, the real part of the growth rates is generally larger than in systems with repulsion. Further, the range of unstable models and the wavenumber corresponding to the maximum real growth rate is typically shifted in the absence of repulsion. One possible implication is that dynamical behaviors occur at a different length scale. However, with one exception for longitudinal perturbations at $g_{AB}=-g_{BA}=9$, the nature of the instability remains unchanged when comparing repulsive and non-repulsive cases.

In this study, we do not perform an in-depth analysis of the various types of instabilities associated with (anti)flocking base states, as the complexity quickly increases in the large parameter space. However, the results presented here already suggest that the fundamental mechanisms driving the instabilities are largely preserved, regardless of the presence or absence of repulsion.

\subsection{BD simulation results without repulsion}
\begin{figure}
	\includegraphics[width=0.6\linewidth]{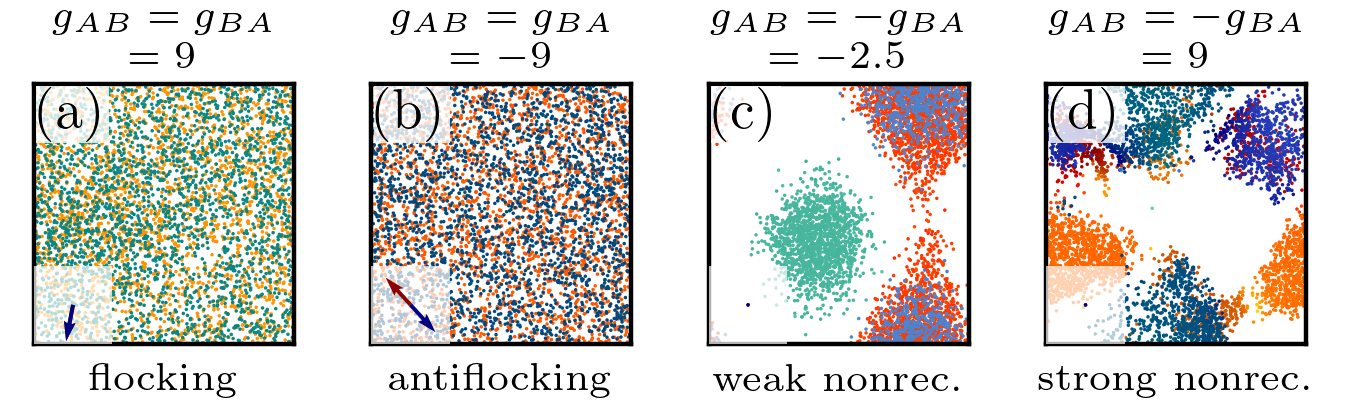}
	\caption{\label{fig:snapshots_no_repulsion}Snapshot of BD simulations without repulsion. Particle species and orientation are colored as in Fig.~1 of the main text.}
\end{figure}

We first consider BD results with reciprocal couplings. Here, we compare the cases of flocking and antiflocking. The snapshots in Figs.~\ref{fig:snapshots_no_repulsion}(a) and (b) show that both flocking and antiflocking also emerge in the absence of repulsion -- with a homogeneous distribution of particles throughout the system. In reciprocal systems with repulsion [snapshots in Fig.~1(c) and (d) in main text], this was only the case for flocking and not for anti-flocking. However, when focusing on polarization dynamics at $k=0$, where density fluctuations are irrelevant, reciprocal flocking and antiflocking exhibit the same characteristics regardless of the presence of absence of repulsion: flocking corresponds to the collective motion of all particles in the same direction, whereas antiflocking is characterized by antiparallel flocks of each species.

Turning now to the nonreciprocal case, we find in the absence of repulsion [Fig.~\ref{fig:snapshots_no_repulsion}(c) and (d)] again a behavior that closely resembles the repulsive case [Fig.~1(e) and (f) in main text]. Specifically, for weak antisymmetric couplings ($g_{AB} = -g_{BA} =-2.5$), particles of the same species form large synchronized rotating clusters. Some $B$-particles become trapped within the $A$-cluster. For strong antisymmetric couplings ($g_{AB} = -g_{BA} =9$), clusters are smaller and no longer fully synchronized. There is, however, one important difference to the case with repulsion. In the absence of repulsion, no minimum separation between particles is enforced and particles can accumulate in very small regions. Over time, this leads to an inhomogeneous density distribution with large distances between particles in different accumulations.

\begin{figure}
	\includegraphics[width=0.5\linewidth]{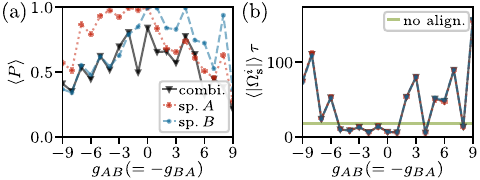}
	\caption{\label{fig:polarisation_chirality_rAlign-10_no_repulsion}Polarization and spontaneous chirality for antisymmetric system without repulsion and $g_{AB}=-g_{BA}$. The data points represent ensemble and time averages of the polarization and spontaneous chirality for species $A$ (red), species $B$ (blue), and all particles combined (black). In (b), the green line shows the spontaneous chirality in a system without alignment couplings.}
\end{figure}

The polarization and spontaneous chirality for an antisymmetric system without repulsion are shown in Fig.~\ref{fig:polarisation_chirality_rAlign-10_no_repulsion}. Qualitatively, the results resemble those obtained in systems with repulsion [Fig.~4 in main text]. However, the distinction between weak anti-symmetric couplings with small spontaneous chirality and stronger antisymmetric couplings with large spontaneous chirality is not as clear as for systems with repulsion. This can be explained by the tendency of non-repulsive, aligning particles to accumulate in small regions, as there is no repulsion to enforce a minimum separation between them. As a result, the zone within the interaction radius around a particle contains less non-aligned particles of the other species. Therefore, although non-reciprocal couplings are large, the effect of non-reciprocity is reduced.

\begin{figure}
	\includegraphics[width=0.5\linewidth]{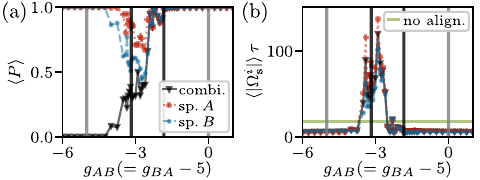}
	\caption{\label{fig:polarisation_chirality_rAlign-10_N-5000_fixed_difference-5_no_repulsion}Polarization and spontaneous chirality when crossing exceptional points in non-reciprocal system without repulsion and $g_{AB}=g_{BA}-5$. The vertical gray and black lines indicate non-critical and critical exceptional points, respectively. The data points represent ensemble and time averages of the polarization and spontaneous chirality for species $A$ (red), species $B$ (blue), and all particles combined (black). In (b), the green line shows the spontaneous chirality in a system without alignment couplings.}
\end{figure}

Fig.~\ref{fig:polarisation_chirality_rAlign-10_N-5000_fixed_difference-5_no_repulsion} shows the polarization and spontaneous chirality as the system without repulsion crosses exceptional points along the path $g_{AB}=g_{BA}-5$. Similar to the case with repulsion (Fig.~5 in main text), the polarization increases from $P=0$ to $P=1$ as $g_{AB}$ increases, and the spontaneous chirality exhibits a significant increase only within a specific range of $g_{AB}$. The increase in spontaneous chirality is accompanied by two peaks. However, unlike the case with repulsion, these peaks do not occur at the coupling strength associated with critical exceptional points.
	
This mismatch is likely related to a change in the effective interaction radius in the absence of repulsion. In systems with repulsion, particles occupy a specific area fraction of the system. Given a fixed alignment radius, this sets a maximum number of particle that any single particle can interact with. At the same time, when the alignment radius is large enough, there is also a minimum number of nearby particles that each particle will always interact with (due to the fixed overall area fraction of particles). However, in the absence of repulsion, particles can accumulate in highly dense regions without being pushed apart. As a result, within such dense accumulations, a single particle may interact with more neighbors than it would in a repulsive system. Conversely, because these dense clusters reduce the overall area fraction occupied by particles, it becomes easier for different clusters to become spatially separated and not interact with each other at all. This leads to a non-trivial change of the effective coupling radius between particles.

For the continuum description, this change in the effective coupling radius has an important consequence: the scaling relation $g_{ab}'=g_{ab}\,R_{\theta}^2\,\rho_0^b/2$, which links the microscopic coupling strength $g_{ab}$ to the coupling-level strength $g_{ab}'$, may no longer hold accurately in the absence of repulsion. This discrepancy affects not only the interspecies couplings ($g'_{AB}$, $g'_{BA}$) but also the intraspecies couplings ($g_{AA}'$, $g_{BB}'$). Determining how to properly rescale $g'_{ab}$ is not straightforward. Consequently, the change in the effective coupling radius implies that the phase diagram and the locations of exceptional points shown in Fig.~1 in the main text are likely different when repulsion is absent. Therefore, the observed mismatch between the coupling strengths at which peaks in the spontaneous chirality appear and those related to critical exceptional points in the repulsive case is not too surprising.

\subsubsection{{\label{ssec:interplay_repulsion_alignment}}Interplay between repulsion and alignment radius}
All results in the main text pertain to systems with steric repulsion and a large interaction radius ($R_{\theta}=10\,\ell$). Importantly, nonreciprocity-induced chirality is also present in systems without repulsion, as well as in those with smaller interaction radii, see \ref{app:dynamics_alignment_radius}. Although the precise interplay of repulsion and alignment radius on the dynamical behavior at the particle level is non-trivial, results obtained in systems with and without repulsion are similar.

\begin{figure}
	\includegraphics[width=0.5\linewidth]{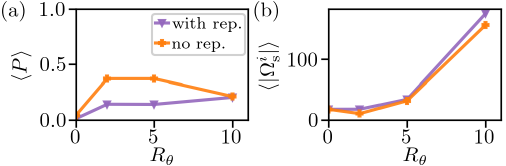}
	\caption{\label{fig:polarisation_chirality_over_rAlign_comparison_no_repulsion}Polarization and mean absolute value of spontaneous chirality as a function of alignment radius $R_{\theta}$ for antisymmetric system with $g_{AB}=-g_{BA}=9$. The average to calculate $\langle P \rangle$ and $\langle \vert \Omega_{\rm s}^i \vert \rangle$ is taken over all particles, times, and noise realizations. The purple line indicates results for repulsive particles, while the orange line indicates results for non-repulsive particles.}
\end{figure}

As an example, we consider in Fig.~\ref{fig:polarisation_chirality_over_rAlign_comparison_no_repulsion}, the dependency of overall polarization and the spontaneous chirality on $R_{\theta}$ for $\delta = 9$.

For systems with repulsion, the polarization remains approximately constant for all $R_{\theta}$ due to nearly $R_{\theta}$-independent asymmetric cluster formation with highly polarized single-species particles [purple line in Fig.~\ref{fig:polarisation_chirality_over_rAlign_comparison_no_repulsion}(a), snapshots in Fig.~1(f) of the main text and Fig.~\ref{fig:stability_diagram_with_snapshots_Rtheta-2_mean-field}(f) of the SM]. Chiral motion, on the other hand, is induced by nonreciprocal couplings with particles of the other species. Thus, it increases with increasing $R_{\theta}$ due to enhanced chances of having opposite-species particles within the interaction radius.

In contrast, for systems without repulsion (orange line), the polarization first increases for intermediate radii ($R_{\theta}\lesssim 5\,\ell$), but then decreases again for larger $R_{\theta}(=10\,\ell)$. The reason is that, when the interaction radius is relatively small, non-repulsive particles of the same species cluster in very small regions. Within these single-species cluster, particles strongly align their motion, which leads to a large polarization. However, at the same time, it becomes unlikely that two tightly accumulated single-species clusters are close enough to interact within the small interaction radius. This limits the occurrence of nonreciprocal interspecies couplings and results in low spontaneous chirality. Yet, with larger coupling radii, particles become more easily influenced by particles of the opposite species, and single-species clustering occurs less. While this leads to decreased polarization, nonreciprocal couplings are increased. The resulting nonreciprocity-induced spontaneous chirality is very large since nonrepulsive particles can move freely without being hindered by other particles.

Thus, while the particle behavior remains qualitatively similar with and without repulsion for both small and large alignment radii, the interplay between intraspecies alignment couplings, which promote single-species accumulations \cite{vicsek_1995_novel_type_phase_transition,elena_2018_velocity_alignment_MIPS,kreienkamp_klapp_2024_dynamical_structures_phase_separating_nonreciprocal_polar_active_mixtures}, and the number of interacting neighbors, which depends both on the alignment radius and the presence of repulsion, affects the system's dynamics qualitatively.


\section*{Supplementary References}
\addtocontents{toc}{\string\tocdepth@munge}
%